\documentclass[12pt,a4paper]{article}
\pdfoutput=1
\usepackage{jheppub,psfrag,slashed,cancel,lscape,caption,array,graphicx,subcaption}
\usepackage[utf8]{inputenc}
\usepackage{amsmath}
\usepackage{mathtools}
\usepackage{mathrsfs}
\usepackage{multirow}
\usepackage{bbold}
\usepackage{booktabs} 
\usepackage{graphicx,relsize} 

\usepackage{xkeyval}

\usepackage{tikz,fp}
\usetikzlibrary{external,fixedpointarithmetic,decorations.pathmorphing,positioning,calc,fadings,decorations.markings,decorations.pathreplacing,patterns,arrows}

\tikzset{
  label distance=-3pt,
  >=stealth,
  inner sep=1.5pt,
  boson/.style={
    decoration={snake, segment length=2mm, amplitude=0.5mm},
    decorate,
  },
  quark/.style={
    postaction={decorate},
    decoration={markings,mark=at position .5 with {\draw[->] (-0.01,0) -- (0.07,0);}}
  },
  aquark/.style={
    postaction={decorate},
    decoration={markings,mark=at position .5 with {\draw[->] (0.01,0) -- (-0.07,0);}}
  },
  gluonOld/.style={
    line width=0.7pt,
    decorate,
    decoration={coil,aspect=-0.5,amplitude=-2pt,segment length=2.2pt}
  },
  gluon/.style={
    line width=0.5pt,
    decorate,
    decoration={coil,aspect=-0.75,amplitude=-1.5pt,segment length=2pt}
  },
  agluon/.style={
    line width=0.5pt,
    decorate,
    decoration={coil,aspect=0.75,amplitude=1.5pt,segment length=2pt}
  },
  scalar/.style={
    dashed
  },
  dscalar/.style={
    dashed,
    postaction={decorate},
    decoration={markings,mark=at position 0.65 with {\draw[->] (-0.01,0) -- (0.07,0);}}
  },
  adscalar/.style={
    dashed,
    postaction={decorate},
    decoration={markings,mark=at position 0.65 with {\draw[->] (-0.01,0) -- (0.07,0);}}
  },
  line/.style={
  },
  doubleline/.style={
    double
  },
  middleArrow/.style={
    draw=white,
    decoration={markings,mark=at position 0.65 with{\arrow[scale=1,black]{>}}},
    decorate
  },
  dot/.style={
  	postaction={decorate},
    decoration={markings,mark=between positions 0.5 and 0.5 step 1 with {\draw[fill] (0,0) circle [radius=1.5pt];}}  
  },
  dots/.style={
  	postaction={decorate},
    decoration={markings,mark=between positions 0.33 and 0.66 step 0.33 with {\draw[fill] (0,0) circle [radius=1.5pt];}}  
  },
  dotss/.style={
  	postaction={decorate},
    decoration={markings,mark=between positions 0.25 and 0.75 step 0.25 with {\draw[fill] (0,0) circle [radius=1.5pt];}}  
  },
  blob/.style={
  	pattern=north west lines,
    draw=black
  },
  blobBlack/.style={
    draw=black,
    fill=black
  },
  blobWhite/.style={
    draw=black,
    fill=white
  }
}

\makeatletter
\def\gSetStdExtTypes#1{\presetkeys{gTypes}{eA=#1,eB=#1,eC=#1,eD=#1,eE=#1}{}}
\def\gSetStdIntTypes#1{\presetkeys{gTypes}{iA=#1,iB=#1,iC=#1,iD=#1,iE=#1,iF=#1,iG=#1,iH=#1,iI=#1,iJ=#1}{}}
\def\gSetStdTypes#1{\gSetStdExtTypes{#1}\gSetStdIntTypes{#1}}

\define@cmdkeys{general}{scale}
\define@cmdkeys{general}{rotate}
\define@cmdkeys{general}{font}
\define@cmdkeys{general}{yshift}
\presetkeys{general}{scale=1, rotate=0, font=\scriptsize, yshift=0pt}{}

\define@cmdkeys{gTypes}{eA,eB,eC,eD,eE,iA,iB,iC,iD,iE,iF,iG,iH,iI,iJ}
\gSetStdTypes{line}
\define@key{pre}{all}{\gSetStdTypes{#1}}
\define@key{pre}{allE}{\gSetStdExtTypes{#1}}
\define@key{pre}{allI}{\gSetStdIntTypes{#1}}

\define@cmdkeys{gLabels}{eLA,eLB,eLC,eLD,eLE,iLA,iLB,iLC,iLD,iLE,iLF,iLG,iLH,iLI,iLJ}
\presetkeys{gLabels}{eLA={},eLB={},eLC={},eLD={},eLE={},iLA={},iLB={},iLC={},iLD={},iLE={},iLF={},iLG={},iLH={},iLI={},iLJ={}}{}

\define@boolkeys{dots}{lDots,rDots}
\presetkeys{dots}{lDots=false, rDots=false}{}

\define@cmdkeys{blobs}{blobA,blobB,blobC}
\presetkeys{blobs}{blobA=blob,blobB=blob,blobC=blob}{}

\def\gTreeTri{\@ifnextchar[\@gTreeTri{\@gTreeTri[]}}
\def\@gTreeTri[#1]#2{%
  \setkeys*{pre}{#1}%
  \setrmkeys{general,gTypes,gLabels}%
  \begin{tikzpicture}
    [line width=1pt,
    baseline={([yshift=-0.5ex+\cmdKV@general@yshift]current bounding box.center)},
    scale=\cmdKV@general@scale,
    rotate=\cmdKV@general@rotate,
    font=\cmdKV@general@font]
    \draw[\cmdKV@gTypes@eA] (0.7,0.3) -- (0.4,0.3) node[label=(-180+\cmdKV@general@rotate):\cmdKV@gLabels@eLA] {};
    \draw[\cmdKV@gTypes@eB] (0.7,0.3) -- (0.9,0.6) node[label=(\cmdKV@general@rotate):\cmdKV@gLabels@eLB] {};
    \draw[\cmdKV@gTypes@eC] (0.7,0.3) -- (0.9,0) node[label=(\cmdKV@general@rotate):\cmdKV@gLabels@eLC] {};
    #2
  \end{tikzpicture}
  \gSetStdTypes{line}
}

\def\gTreeS{\@ifnextchar[\@gTreeS{\@gTreeS[]}}
\def\@gTreeS[#1]#2{%
  \setkeys*{pre}{#1}%
  \setrmkeys{general,gTypes,gLabels}%
  \begin{tikzpicture}
    [line width=1pt,
    baseline={([yshift=-0.5ex+\cmdKV@general@yshift]current bounding box.center)},
    scale=\cmdKV@general@scale,
    rotate=\cmdKV@general@rotate,
    font=\cmdKV@general@font]
    \draw[\cmdKV@gTypes@eA] (0.2,0.3) -- (0,0) node[label=(-180+\cmdKV@general@rotate):\cmdKV@gLabels@eLA] {};
    \draw[\cmdKV@gTypes@eB] (0.2,0.3) -- (0,0.6) node[label=(180+\cmdKV@general@rotate):\cmdKV@gLabels@eLB] {};
    \draw[\cmdKV@gTypes@eC] (0.7,0.3) -- (0.9,0.6) node[label=(\cmdKV@general@rotate):\cmdKV@gLabels@eLC] {};
    \draw[\cmdKV@gTypes@eD] (0.7,0.3) -- (0.9,0) node[label=(\cmdKV@general@rotate):\cmdKV@gLabels@eLD] {};    
    \draw[label distance=0,\cmdKV@gTypes@iA] (0.2,0.3) -- node[label=(90+\cmdKV@general@rotate):\cmdKV@gLabels@iLA] {} (0.7,0.3);
    #2
  \end{tikzpicture}
  \gSetStdTypes{line}
}

\def\gTreeT{\@ifnextchar[\@gTreeT{\@gTreeT[]}}
\def\@gTreeT[#1]#2{%
  \setkeys*{pre}{#1}%
  \setrmkeys{general,gTypes,gLabels}%
  \begin{tikzpicture}
    [line width=1pt,
    baseline={([yshift=-0.5ex+\cmdKV@general@yshift]current bounding box.center)},
    scale=\cmdKV@general@scale,
    rotate=\cmdKV@general@rotate,
    font=\cmdKV@general@font]
    \draw[\cmdKV@gTypes@eA] (0.45,0.1) -- (0,0) node[label=(-180+\cmdKV@general@rotate):\cmdKV@gLabels@eLA] {};
    \draw[\cmdKV@gTypes@eB] (0.45,0.5) -- (0,0.6) node[label=(180+\cmdKV@general@rotate):\cmdKV@gLabels@eLB] {};
    \draw[\cmdKV@gTypes@eC] (0.45,0.5) -- (0.9,0.6) node[label=(\cmdKV@general@rotate):\cmdKV@gLabels@eLC] {};
    \draw[\cmdKV@gTypes@eD] (0.45,0.1) -- (0.9,0) node[label=(\cmdKV@general@rotate):\cmdKV@gLabels@eLD] {};
    \draw[label distance=0,\cmdKV@gTypes@iA] (0.45,0.5) -- node[label=(\cmdKV@general@rotate):\cmdKV@gLabels@iLA] {} (0.45,0.1);
    #2
  \end{tikzpicture}
  \gSetStdTypes{line}
}

\def\gTreeU{\@ifnextchar[\@gTreeU{\@gTreeU[]}}
\def\@gTreeU[#1]#2{%
  \setkeys*{pre}{#1}%
  \setrmkeys{general,gTypes,gLabels}%
  \begin{tikzpicture}
    [line width=1pt,
    baseline={([yshift=-0.5ex+\cmdKV@general@yshift]current bounding box.center)},
    scale=\cmdKV@general@scale,
    rotate=\cmdKV@general@rotate,
    font=\cmdKV@general@font]
    \draw[\cmdKV@gTypes@eA] (0.45,0.5) -- (0,0) node[label=(-180+\cmdKV@general@rotate):\cmdKV@gLabels@eLA] {};
    \foreach \x in {5,...,14} \fill[opacity=0.12,white] (0.27,0.3) circle (\x/100);
    \draw[\cmdKV@gTypes@eB] (0.45,0.1) -- (0,0.6) node[label=(180+\cmdKV@general@rotate):\cmdKV@gLabels@eLB] {};
    \draw[\cmdKV@gTypes@eC] (0.45,0.5) -- (0.9,0.6) node[label=(\cmdKV@general@rotate):\cmdKV@gLabels@eLC] {};
    \draw[\cmdKV@gTypes@eD] (0.45,0.1) -- (0.9,0) node[label=(\cmdKV@general@rotate):\cmdKV@gLabels@eLD] {};
    \draw[label distance=0,\cmdKV@gTypes@iA] (0.45,0.5) -- node[label=(\cmdKV@general@rotate):\cmdKV@gLabels@iLA] {} (0.45,0.1);
    #2
  \end{tikzpicture}
  \gSetStdTypes{line}
}

\def\gTreeFourPt{\@ifnextchar[\@gTreeFourPt{\@gTreeFourPt[]}}
\def\@gTreeFourPt[#1]#2{%
  \setkeys*{pre}{#1}%
  \setrmkeys{general,gTypes,gLabels}%
  \begin{tikzpicture}
    [line width=1pt,
    baseline={([yshift=-0.5ex+\cmdKV@general@yshift]current bounding box.center)},
    scale=\cmdKV@general@scale,
    rotate=\cmdKV@general@rotate,
    font=\cmdKV@general@font]
    \draw[\cmdKV@gTypes@eA] (0,0) -- (-0.4,-0.4) node[label=(-180+\cmdKV@general@rotate):\cmdKV@gLabels@eLA] {};
    \draw[\cmdKV@gTypes@eB] (0,0) -- (-0.4,0.4) node[label=(180+\cmdKV@general@rotate):\cmdKV@gLabels@eLB] {};
    \draw[\cmdKV@gTypes@eC] (0,0) -- (0.4,0.4) node[label=(\cmdKV@general@rotate):\cmdKV@gLabels@eLC] {};
    \draw[\cmdKV@gTypes@eD] (0,0) -- (0.4,-0.4) node[label=(\cmdKV@general@rotate):\cmdKV@gLabels@eLD] {};
    #2
  \end{tikzpicture}
  \gSetStdTypes{line}
}

\def\gTadA{\@ifnextchar[\@gTadA{\@gTadA[]}}
\def\@gTadA[#1]#2{%
  \setkeys*{pre}{#1}%
  \setrmkeys{general,gTypes,gLabels}%
  \begin{tikzpicture}
    [line width=1pt,
    baseline={([yshift=-0.5ex+\cmdKV@general@yshift]current bounding box.center)},
    scale=\cmdKV@general@scale,
    rotate=\cmdKV@general@rotate,
    font=\cmdKV@general@font]
    \draw[\cmdKV@gTypes@eA] (-0.7,0.3) -- (-0.9,0) node[label=(180+\cmdKV@general@rotate):\cmdKV@gLabels@eLA] {};
    \draw[\cmdKV@gTypes@eB] (-0.7,0.3) -- (-0.9,0.6) node[label=(180+\cmdKV@general@rotate):\cmdKV@gLabels@eLB] {};
    \draw[\cmdKV@gTypes@eC] (0.15,0.45) -- (0.5,0.6) node[label=(\cmdKV@general@rotate):\cmdKV@gLabels@eLC] {};
    \draw[\cmdKV@gTypes@eD] (-0.3,0.3) -- (-0.2,0) node[label=(\cmdKV@general@rotate):\cmdKV@gLabels@eLD] {};
    \draw[label distance=0,\cmdKV@gTypes@iA] (-0.7,0.3) -- node[label=(90+\cmdKV@general@rotate):\cmdKV@gLabels@iLA] {} (-0.3,0.3);
    \draw[label distance=0,\cmdKV@gTypes@iB] (-0.3,0.3) -- node[label=(-80+\cmdKV@general@rotate):\cmdKV@gLabels@iLB] {} (0.15,0.45);
    \draw[label distance=0,\cmdKV@gTypes@iC] (0.15,0.45) -- node[label=(45+\cmdKV@general@rotate):\cmdKV@gLabels@iLC] {} (-0.18,0.69);
    \draw[label distance=-0.5,\cmdKV@gTypes@iD] (-0.18,0.69) .. controls ($(-0.18,0.69) + (-0.12,-0.12)$) and ($(-0.35,0.87) + (-0.12,-0.12)$) .. (-0.35,0.87) node[label=(180+\cmdKV@general@rotate):\cmdKV@gLabels@iLD] {} .. controls ($(-0.35,0.87) + (0.12,0.12)$) and ($(-0.18,0.69) + (0.12,0.12)$) .. (-0.18,0.69);
    #2
  \end{tikzpicture}
  \gSetStdTypes{line}
}

\def\gTadB{\@ifnextchar[\@gTadB{\@gTadB[]}}
\def\@gTadB[#1]#2{%
  \setkeys*{pre}{#1}%
  \setrmkeys{general,gTypes,gLabels}%
  \begin{tikzpicture}
    [line width=1pt,
    baseline={([yshift=-0.5ex+\cmdKV@general@yshift]current bounding box.center)},
    scale=\cmdKV@general@scale,
    rotate=\cmdKV@general@rotate,
    font=\cmdKV@general@font]
    \draw[\cmdKV@gTypes@eA] (0,0.3) -- (-0.2,0) node[label=(-180+\cmdKV@general@rotate):\cmdKV@gLabels@eLA] {};
    \draw[\cmdKV@gTypes@eB] (0,0.3) -- (-0.2,0.6) node[label=(180+\cmdKV@general@rotate):\cmdKV@gLabels@eLB] {};
    \draw[\cmdKV@gTypes@eC] (0.7,0.3) -- (0.9,0.6) node[label=(\cmdKV@general@rotate):\cmdKV@gLabels@eLC] {};
    \draw[\cmdKV@gTypes@eD] (0.7,0.3) -- (0.9,0) node[label=(\cmdKV@general@rotate):\cmdKV@gLabels@eLD] {};
    \draw[label distance=0,\cmdKV@gTypes@iA] (0,0.3) -- node[label=(-90+\cmdKV@general@rotate):\cmdKV@gLabels@iLA] {} (0.35,0.3);
    \draw[label distance=0,\cmdKV@gTypes@iB] (0.35,0.3) -- node[label=(\cmdKV@general@rotate):\cmdKV@gLabels@iLB] {} (0.35,0.6);
    \draw[label distance=-0.5,\cmdKV@gTypes@iC] (0.35,0.6) .. controls (0.18,0.6) and (0.18,0.85) .. (0.35,0.85) .. controls (0.52,0.85) and (0.52,0.6) .. node[label=(\cmdKV@general@rotate):\cmdKV@gLabels@iLC] {} (0.35,0.6);
    \draw[label distance=0,\cmdKV@gTypes@iD] (0.35,0.3) -- node[label=(-90+\cmdKV@general@rotate):\cmdKV@gLabels@iLD] {} (0.7,0.3);
    #2
  \end{tikzpicture}
  \gSetStdTypes{line}
}
\def\gBubA{\@ifnextchar[\@gBubA{\@gBubA[]}}
\def\@gBubA[#1]#2{%
  \setkeys*{pre}{#1}%
  \setrmkeys{general,gTypes,gLabels}%
  \begin{tikzpicture}
    [line width=1pt,
    baseline={([yshift=-0.5ex+\cmdKV@general@yshift]current bounding box.center)},
    scale=\cmdKV@general@scale,
    rotate=\cmdKV@general@rotate,
    font=\cmdKV@general@font]
    \draw[\cmdKV@gTypes@eA] (0.5,0.7) -- (0.2,0.7) node[label=(180+\cmdKV@general@rotate):\cmdKV@gLabels@eLA] {};
    \draw[\cmdKV@gTypes@eB] (0.9,0.7) -- (1.2,0.7) node[label=(\cmdKV@general@rotate):\cmdKV@gLabels@eLB] {};
    \draw[label distance=0,\cmdKV@gTypes@iA] (0.5,0.7) .. controls (0.5,0.95) and (0.9,0.95) .. node[label=(90+\cmdKV@general@rotate):\cmdKV@gLabels@iLA] {} (0.9,0.7);
    \draw[label distance=0,\cmdKV@gTypes@iB] (0.9,0.7) .. controls (0.9,0.45) and (0.5,0.45) .. node[label=(-90+\cmdKV@general@rotate):\cmdKV@gLabels@iLB] {} (0.5,0.7);
    #2
  \end{tikzpicture}
  \gSetStdTypes{line}
}
\def\gBubB{\@ifnextchar[\@gBubB{\@gBubB[]}}
\def\@gBubB[#1]#2{%
  \setkeys*{pre}{#1}%
  \setrmkeys{general,gTypes,gLabels}%
  \begin{tikzpicture}
    [line width=1pt,
    baseline={([yshift=-0.5ex+\cmdKV@general@yshift]current bounding box.center)},
    scale=\cmdKV@general@scale,
    rotate=\cmdKV@general@rotate,
    font=\cmdKV@general@font]
    \draw[\cmdKV@gTypes@eA] (0.2,0.7) -- (0,0.4) node[label=(-180+\cmdKV@general@rotate):\cmdKV@gLabels@eLA] {};
    \draw[\cmdKV@gTypes@eB] (0.2,0.7) -- (0,1) node[label=(180+\cmdKV@general@rotate):\cmdKV@gLabels@eLB] {};
    \draw[\cmdKV@gTypes@eC] (1.2,0.7) -- (1.4,1) node[label=(\cmdKV@general@rotate):\cmdKV@gLabels@eLC] {};
    \draw[\cmdKV@gTypes@eD] (1.2,0.7) -- (1.4,0.4) node[label=(\cmdKV@general@rotate):\cmdKV@gLabels@eLD] {};
    \draw[label distance=0,\cmdKV@gTypes@iA] (0.2,0.7) -- node[label=(90+\cmdKV@general@rotate):\cmdKV@gLabels@iLA] {} (0.5,0.7);
    \draw[label distance=0,\cmdKV@gTypes@iB] (0.5,0.7) .. controls (0.5,0.95) and (0.9,0.95) .. node[label=(90+\cmdKV@general@rotate):\cmdKV@gLabels@iLB] {} (0.9,0.7);
    \draw[label distance=0,\cmdKV@gTypes@iC] (0.9,0.7) .. controls (0.9,0.45) and (0.5,0.45) .. node[label=(-90+\cmdKV@general@rotate):\cmdKV@gLabels@iLC] {} (0.5,0.7);
    \draw[label distance=0,\cmdKV@gTypes@iD] (0.9,0.7) -- node[label=(90+\cmdKV@general@rotate):\cmdKV@gLabels@iLD] {} (1.2,0.7);
    #2
  \end{tikzpicture}
  \gSetStdTypes{line}
}

\def\gBubC{\@ifnextchar[\@gBubC{\@gBubC[]}}
\def\@gBubC[#1]#2{%
  \setkeys*{pre}{#1}%
  \setrmkeys{general,gTypes,gLabels}%
  \begin{tikzpicture}
    [line width=1pt,
    baseline={([yshift=-0.5ex+\cmdKV@general@yshift]current bounding box.center)},
    scale=\cmdKV@general@scale,
    rotate=\cmdKV@general@rotate,
    font=\cmdKV@general@font]
    \draw[\cmdKV@gTypes@eA] (-0.7,0.3) -- (-0.9,0) node[label=(180+\cmdKV@general@rotate):\cmdKV@gLabels@eLA] {};
    \draw[\cmdKV@gTypes@eB] (-0.7,0.3) -- (-0.9,0.6) node[label=(180+\cmdKV@general@rotate):\cmdKV@gLabels@eLB] {};
    \draw[\cmdKV@gTypes@eC] (0.3,0.5) -- (0.5,0.6) node[label=(\cmdKV@general@rotate):\cmdKV@gLabels@eLC] {};
    \draw[\cmdKV@gTypes@eD] (-0.3,0.3) -- (-0.2,0) node[label=(\cmdKV@general@rotate):\cmdKV@gLabels@eLD] {};
    \draw[label distance=0,\cmdKV@gTypes@iA] (-0.7,0.3) -- node[label=(90+\cmdKV@general@rotate):\cmdKV@gLabels@iLA] {} (-0.3,0.3);
    \draw[label distance=0,\cmdKV@gTypes@iB] (-0.3,0.3) -- node[label=(90+\cmdKV@general@rotate):\cmdKV@gLabels@iLB] {} (-0,0.4);
    \draw[label distance=0,\cmdKV@gTypes@iC] (0,0.4) .. controls (-0.03,0.6) and (0.22,0.65) .. node[label=(90+\cmdKV@general@rotate):\cmdKV@gLabels@iLC] {} (0.3,0.5);
    \draw[label distance=0,\cmdKV@gTypes@iD] (0.3,0.5) .. controls (0.38,0.3) and (0.08,0.21) .. node[label=(-90+\cmdKV@general@rotate):\cmdKV@gLabels@iLD] {} (0,0.4);
    #2
  \end{tikzpicture}
  \gSetStdTypes{line}
}
\def\gTriA{\@ifnextchar[\@gTriA{\@gTriA[]}}
\def\@gTriA[#1]#2{%
  \setkeys*{pre}{#1}%
  \setrmkeys{general,gTypes,gLabels}%
  \begin{tikzpicture}
    [line width=1pt,
    baseline={([yshift=-0.5ex+\cmdKV@general@yshift]current bounding box.center)},
    scale=\cmdKV@general@scale,
    rotate=\cmdKV@general@rotate,
    font=\cmdKV@general@font]
    \draw[\cmdKV@gTypes@eC] (0:0.3) -- (0:0.55) node[label=(\cmdKV@general@rotate):\cmdKV@gLabels@eLC] {};
    \draw[\cmdKV@gTypes@eA] (-120:0.3) -- (-120:0.55) node[label=(-180+\cmdKV@general@rotate):\cmdKV@gLabels@eLA] {};
    \draw[\cmdKV@gTypes@eB] (120:0.3) -- (120:0.55) node[label=(180+\cmdKV@general@rotate):\cmdKV@gLabels@eLB] {};    
    \draw[label distance=0,\cmdKV@gTypes@iA] (-120:0.3) -- node[label=(180+\cmdKV@general@rotate):\cmdKV@gLabels@iLA] {} (120:0.3);
    \draw[label distance=0,\cmdKV@gTypes@iB] (120:0.3) -- node[label=(60+\cmdKV@general@rotate):\cmdKV@gLabels@iLB] {} (0:0.3);
    \draw[label distance=0,\cmdKV@gTypes@iC] (0:0.3) -- node[label=(-60+\cmdKV@general@rotate):\cmdKV@gLabels@iLC] {} (-120:0.3);
    #2
  \end{tikzpicture}
  \gSetStdTypes{line}
}
\def\gTriB{\@ifnextchar[\@gTriB{\@gTriB[]}}
\def\@gTriB[#1]#2{%
  \setkeys*{pre}{#1}%
  \setrmkeys{general,gTypes,gLabels}%
  \begin{tikzpicture}
    [line width=1pt,
    baseline={([yshift=-0.5ex+\cmdKV@general@yshift]current bounding box.center)},
    scale=\cmdKV@general@scale,
    rotate=\cmdKV@general@rotate,
    font=\cmdKV@general@font]
    \draw[\cmdKV@gTypes@eA] (0.3,0.45) -- (0,0.3) node[label=(-180+\cmdKV@general@rotate):\cmdKV@gLabels@eLA] {};
    \draw[\cmdKV@gTypes@eB] (0.3,0.95) -- (0,1.1) node[label=(180+\cmdKV@general@rotate):\cmdKV@gLabels@eLB] {};
    \draw[\cmdKV@gTypes@eC] (1,0.7) -- (1.3,1.1) node[label=(\cmdKV@general@rotate):\cmdKV@gLabels@eLC] {};
    \draw[\cmdKV@gTypes@eD] (1,0.7) -- (1.3,0.3) node[label=(\cmdKV@general@rotate):\cmdKV@gLabels@eLD] {};
    \draw[label distance=0,\cmdKV@gTypes@iA] (0.3,0.45) -- node[label=(180+\cmdKV@general@rotate):\cmdKV@gLabels@iLA] {} (0.3,0.95);
    \draw[label distance=0,\cmdKV@gTypes@iB] (0.3,0.95) -- node[label=(70+\cmdKV@general@rotate):\cmdKV@gLabels@iLB] {} (0.7,0.7);
    \draw[label distance=0,\cmdKV@gTypes@iC] (0.7,0.7) -- node[label=(-70+\cmdKV@general@rotate):\cmdKV@gLabels@iLC] {} (0.3,0.45);
    \draw[label distance=0,\cmdKV@gTypes@iD] (1,0.7) -- node[label=(90+\cmdKV@general@rotate):\cmdKV@gLabels@iLD] {} (0.7,0.7);
    #2
  \end{tikzpicture}
  \gSetStdTypes{line}
}
\def\gBox{\@ifnextchar[\@gBox{\@gBox[]}}
\def\@gBox[#1]#2{%
  \setkeys*{pre}{#1}%
  \setrmkeys{general,gTypes,gLabels}%
  \begin{tikzpicture}
    [line width=1pt,
    baseline={([yshift=-0.5ex+\cmdKV@general@yshift]current bounding box.center)},
    scale=\cmdKV@general@scale,
    rotate=\cmdKV@general@rotate,
    font=\cmdKV@general@font]
    \draw[\cmdKV@gTypes@eA] (0.25,0.25) -- (0,0) node[label=(-180+\cmdKV@general@rotate):\cmdKV@gLabels@eLA] {};
    \draw[\cmdKV@gTypes@eB] (0.25,0.75) -- (0,1) node[label=(180+\cmdKV@general@rotate):\cmdKV@gLabels@eLB] {};
    \draw[\cmdKV@gTypes@eC] (0.75,0.75) -- (1,1) node[label=(\cmdKV@general@rotate):\cmdKV@gLabels@eLC] {};
    \draw[\cmdKV@gTypes@eD] (0.75,0.25) -- (1,0) node[label=(\cmdKV@general@rotate):\cmdKV@gLabels@eLD] {};
    \draw[label distance=0,\cmdKV@gTypes@iA] (0.25,0.25) -- node[label=(180+\cmdKV@general@rotate):\cmdKV@gLabels@iLA] {} (0.25,0.75);
    \draw[label distance=0,\cmdKV@gTypes@iB] (0.25,0.75) -- node[label=(90+\cmdKV@general@rotate):\cmdKV@gLabels@iLB] {} (0.75,0.75);
    \draw[label distance=0,\cmdKV@gTypes@iC] (0.75,0.75) -- node[label=(\cmdKV@general@rotate):\cmdKV@gLabels@iLC] {} (0.75,0.25);
    \draw[label distance=0,\cmdKV@gTypes@iD] (0.75,0.25) -- node[label=(-90+\cmdKV@general@rotate):\cmdKV@gLabels@iLD] {} (0.25,0.25);
    #2
  \end{tikzpicture}
  \gSetStdTypes{line}
}
\def\gPenta{\@ifnextchar[\@gPenta{\@gPenta[]}}
\def\@gPenta[#1]#2{%
  \setkeys*{pre}{#1}%
  \setrmkeys{general,gTypes,gLabels}%
  \begin{tikzpicture}
    [line width=1pt,
    baseline={([yshift=-0.5ex+\cmdKV@general@yshift]current bounding box.center)},
    scale=\cmdKV@general@scale,
    rotate=\cmdKV@general@rotate,
    font=\cmdKV@general@font]
    \draw[\cmdKV@gTypes@eA] (-144:0.35) -- (-144:0.75) node[label=(-144+\cmdKV@general@rotate):\cmdKV@gLabels@eLA] {};
    \draw[\cmdKV@gTypes@eB] (144:0.35) -- (144:0.75) node[label=(144+\cmdKV@general@rotate):\cmdKV@gLabels@eLB] {};
    \draw[\cmdKV@gTypes@eC] (72:0.35) -- (72:0.75) node[label=(72+\cmdKV@general@rotate):\cmdKV@gLabels@eLC] {};
    \draw[\cmdKV@gTypes@eD] (0:0.35) -- (0:0.75) node[label=(\cmdKV@general@rotate):\cmdKV@gLabels@eLD] {};
    \draw[\cmdKV@gTypes@eE] (-72:0.35) -- (-72:0.75) node[label=(-72+\cmdKV@general@rotate):\cmdKV@gLabels@eLE] {};
    \draw[label distance=0,\cmdKV@gTypes@iA] (-144:0.35) -- node[label=(180+\cmdKV@general@rotate):\cmdKV@gLabels@iLA] {} (144:0.35);
    \draw[label distance=0,\cmdKV@gTypes@iB] (144:0.35) -- node[label=(108+\cmdKV@general@rotate):\cmdKV@gLabels@iLB] {} (72:0.35);
    \draw[label distance=0,\cmdKV@gTypes@iC] (72:0.35) -- node[label=(36+\cmdKV@general@rotate):\cmdKV@gLabels@iLC] {} (0:0.35);
    \draw[label distance=0,\cmdKV@gTypes@iD] (0:0.35) -- node[label=(-36+\cmdKV@general@rotate):\cmdKV@gLabels@iLD] {} (-72:0.35);
    \draw[label distance=0,\cmdKV@gTypes@iE] (-72:0.35) -- node[label=(-108+\cmdKV@general@rotate):\cmdKV@gLabels@iLE] {} (-144:0.35);
    #2
  \end{tikzpicture}
  \gSetStdTypes{line}
}

\def\gBoxBox{\@ifnextchar[\@gBoxBox{\@gBoxBox[]}}
\def\@gBoxBox[#1]#2{%
  \setkeys*{pre}{#1}%
  \setrmkeys{general,gTypes,gLabels}%
  \begin{tikzpicture}
    [line width=1pt,
    baseline={([yshift=-0.5ex+\cmdKV@general@yshift]current bounding box.center)},
    scale=\cmdKV@general@scale,
    rotate=\cmdKV@general@rotate,
    font=\cmdKV@general@font]
    \draw[\cmdKV@gTypes@eA] (0.3,0.3) -- (0,0) node[label=(-180+\cmdKV@general@rotate):\cmdKV@gLabels@eLA] {};
    \draw[\cmdKV@gTypes@eB] (0.3,0.9) -- (0,1.2) node[label=(180+\cmdKV@general@rotate):\cmdKV@gLabels@eLB] {};
    \draw[\cmdKV@gTypes@eC] (1.5,0.9) -- (1.8,1.2) node[label=(\cmdKV@general@rotate):\cmdKV@gLabels@eLC] {};
    \draw[\cmdKV@gTypes@eD] (1.5,0.3) -- (1.8,0) node[label=(\cmdKV@general@rotate):\cmdKV@gLabels@eLD] {};
    \draw[label distance=0,\cmdKV@gTypes@iA] (0.3,0.3) -- node[label=(180+\cmdKV@general@rotate):\cmdKV@gLabels@iLA] {} (0.3,0.9);
    \draw[label distance=0,\cmdKV@gTypes@iB] (0.3,0.9) -- node[label=(90+\cmdKV@general@rotate):\cmdKV@gLabels@iLB] {} (0.9,0.9);
    \draw[label distance=0,\cmdKV@gTypes@iC] (0.9,0.9) -- node[label=(90+\cmdKV@general@rotate):\cmdKV@gLabels@iLC] {} (1.5,0.9);
    \draw[label distance=0,\cmdKV@gTypes@iD] (1.5,0.9) -- node[label=(\cmdKV@general@rotate):\cmdKV@gLabels@iLD] {} (1.5,0.3);
    \draw[label distance=0,\cmdKV@gTypes@iE] (1.5,0.3) -- node[label=(-90+\cmdKV@general@rotate):\cmdKV@gLabels@iLE] {} (0.9,0.3);
    \draw[label distance=0,\cmdKV@gTypes@iF] (0.9,0.3) -- node[label=(-90+\cmdKV@general@rotate):\cmdKV@gLabels@iLF] {} (0.3,0.3);
    \draw[label distance=0,\cmdKV@gTypes@iG] (0.9,0.9) -- node[label=(\cmdKV@general@rotate):\cmdKV@gLabels@iLG] {} (0.9,0.3);
    #2
  \end{tikzpicture}
  \gSetStdTypes{line}
}

\def\gBoxBoxNP{\@ifnextchar[\@gBoxBoxNP{\@gBoxBoxNP[]}}
\def\@gBoxBoxNP[#1]#2{%
  \setkeys*{pre}{#1}%
  \setrmkeys{general,gTypes,gLabels}%
  \begin{tikzpicture}
    [line width=1pt,
    baseline={([yshift=-0.5ex+\cmdKV@general@yshift]current bounding box.center)},
    scale=\cmdKV@general@scale,
    rotate=\cmdKV@general@rotate,
    font=\cmdKV@general@font]
    \draw[\cmdKV@gTypes@eA] (-1.9,0.3) -- (-2.2,0) node[label=(180+\cmdKV@general@rotate):\cmdKV@gLabels@eLA] {};
    \draw[\cmdKV@gTypes@eB] (-1.9,1.1) -- (-2.2,1.4) node[label=(180+\cmdKV@general@rotate):\cmdKV@gLabels@eLB] {};
    \draw[\cmdKV@gTypes@eC] (-0.3,0.7) -- (0,0.7) node[label=(\cmdKV@general@rotate):\cmdKV@gLabels@eLC] {};
    \draw[\cmdKV@gTypes@eD] (-1.1,0.7) -- (-1.4,0.7) node[label=(180+\cmdKV@general@rotate):\cmdKV@gLabels@eLD] {};
    \draw[label distance=0,\cmdKV@gTypes@iA] (-1.9,0.3) -- node[label=(180+\cmdKV@general@rotate):\cmdKV@gLabels@iLA] {} (-1.9,1.1);
    \draw[label distance=0,\cmdKV@gTypes@iB] (-1.9,1.1) -- node[label=(90+\cmdKV@general@rotate):\cmdKV@gLabels@iLB] {} (-0.7,1.1);
    \draw[label distance=0,\cmdKV@gTypes@iC] (-0.7,1.1) -- node[label=(45+\cmdKV@general@rotate):\cmdKV@gLabels@iLC] {} (-0.3,0.7);
    \draw[label distance=0,\cmdKV@gTypes@iD] (-0.3,0.7) -- node[label=(-45+\cmdKV@general@rotate):\cmdKV@gLabels@iLD] {} (-0.7,0.3);
    \draw[label distance=0,\cmdKV@gTypes@iE] (-0.7,0.3) -- node[label=(-90+\cmdKV@general@rotate):\cmdKV@gLabels@iLE] {} (-1.9,0.3);
    \draw[label distance=0,\cmdKV@gTypes@iF] (-0.7,1.1) -- node[label=(180+\cmdKV@general@rotate):\cmdKV@gLabels@iLF] {} (-1.1,0.7);
    \draw[label distance=0,\cmdKV@gTypes@iG] (-1.1,0.7) -- node[label=(-180+\cmdKV@general@rotate):\cmdKV@gLabels@iLG] {} (-0.7,0.3);
    #2
  \end{tikzpicture}
  \gSetStdTypes{line}
}

\def\gBoxBoxNPB{\@ifnextchar[\@gBoxBoxNPB{\@gBoxBoxNPB[]}}
\def\@gBoxBoxNPB[#1]#2{%
  \setkeys*{pre}{#1}%
  \setrmkeys{general,gTypes,gLabels}%
  \begin{tikzpicture}
    [line width=1pt,
    baseline={([yshift=-0.5ex+\cmdKV@general@yshift]current bounding box.center)},
    scale=\cmdKV@general@scale,
    rotate=\cmdKV@general@rotate,
    font=\cmdKV@general@font]
    \draw[\cmdKV@gTypes@eA] (0.3,0.3) -- (0,0) node[label=(-180+\cmdKV@general@rotate):\cmdKV@gLabels@eLA] {};
    \draw[\cmdKV@gTypes@eB] (0.3,0.9) -- (0,1.2) node[label=(180+\cmdKV@general@rotate):\cmdKV@gLabels@eLB] {};
    \draw[\cmdKV@gTypes@eC] (1.5,0.9) -- (1.8,1.2) node[label=(\cmdKV@general@rotate):\cmdKV@gLabels@eLC] {};
    \draw[\cmdKV@gTypes@eD] (1.5,0.3) -- (1.8,0) node[label=(\cmdKV@general@rotate):\cmdKV@gLabels@eLD] {};
    \draw[label distance=0,\cmdKV@gTypes@iA] (0.3,0.3) -- node[label=(180+\cmdKV@general@rotate):\cmdKV@gLabels@iLA] {} (0.3,0.9);
    \draw[label distance=0,\cmdKV@gTypes@iB] (0.3,0.9) -- node[label=(90+\cmdKV@general@rotate):\cmdKV@gLabels@iLB] {} (0.9,0.9);
    \draw[label distance=0,\cmdKV@gTypes@iC] (0.9,0.9) -- node[label=(90+\cmdKV@general@rotate):\cmdKV@gLabels@iLC] {} (1.5,0.3);
    \foreach \x in {5,...,14} \fill[opacity=0.12,white] (1.2,0.6) circle (\x/100);
    \draw[label distance=0,\cmdKV@gTypes@iD] (1.5,0.3) -- node[label=(\cmdKV@general@rotate):\cmdKV@gLabels@iLD] {} (1.5,0.9);
    \draw[label distance=0,\cmdKV@gTypes@iE] (1.5,0.9) -- node[label=(-90+\cmdKV@general@rotate):\cmdKV@gLabels@iLE] {} (0.9,0.3);
    \draw[label distance=0,\cmdKV@gTypes@iF] (0.9,0.3) -- node[label=(-90+\cmdKV@general@rotate):\cmdKV@gLabels@iLF] {} (0.3,0.3);
    \draw[label distance=-1,\cmdKV@gTypes@iG] (0.9,0.9) -- node[label=(180+\cmdKV@general@rotate):\cmdKV@gLabels@iLG] {} (0.9,0.3);
    #2
  \end{tikzpicture}
  \gSetStdTypes{line}
}

\def\gBoxBoxNPC{\@ifnextchar[\@gBoxBoxNPC{\@gBoxBoxNPC[]}}
\def\@gBoxBoxNPC[#1]#2{%
  \setkeys*{pre}{#1}%
  \setrmkeys{general,gTypes,gLabels}%
  \begin{tikzpicture}
    [line width=1pt,
    baseline={([yshift=-0.5ex+\cmdKV@general@yshift]current bounding box.center)},
    scale=\cmdKV@general@scale,
    rotate=\cmdKV@general@rotate,
    font=\cmdKV@general@font]
    \draw[\cmdKV@gTypes@eA] (0.3,0.3) -- (0,0) node[label=(-180+\cmdKV@general@rotate):\cmdKV@gLabels@eLA] {};
    \draw[\cmdKV@gTypes@eB] (0.3,0.9) -- (0,1.2) node[label=(180+\cmdKV@general@rotate):\cmdKV@gLabels@eLB] {};
    \draw[\cmdKV@gTypes@eC] (1.5,0.9) -- (1.8,1.2) node[label=(\cmdKV@general@rotate):\cmdKV@gLabels@eLC] {};
    \draw[\cmdKV@gTypes@eD] (1.5,0.3) -- (1.8,0) node[label=(\cmdKV@general@rotate):\cmdKV@gLabels@eLD] {};
    \draw[label distance=0,\cmdKV@gTypes@iA] (0.3,0.3) -- node[label=(180+\cmdKV@general@rotate):\cmdKV@gLabels@iLA] {} (0.3,0.9);
    \draw[label distance=0,\cmdKV@gTypes@iB] (0.3,0.9) -- node[label=(90+\cmdKV@general@rotate):\cmdKV@gLabels@iLB] {} (0.9,0.9);
    \draw[label distance=0,\cmdKV@gTypes@iC] (0.9,0.9) -- node[label=(90+\cmdKV@general@rotate):\cmdKV@gLabels@iLC] {} (1.5,0.9);
    \draw[label distance=0,\cmdKV@gTypes@iD] (1.5,0.9) -- node[label=(\cmdKV@general@rotate):\cmdKV@gLabels@iLD] {} (0.9,0.3);
    \foreach \x in {5,...,14} \fill[opacity=0.12,white] (1.2,0.6) circle (\x/100);
    \draw[label distance=0,\cmdKV@gTypes@iE] (1.5,0.3) -- node[label=(-90+\cmdKV@general@rotate):\cmdKV@gLabels@iLE] {} (0.9,0.3);
    \draw[label distance=0,\cmdKV@gTypes@iF] (0.9,0.3) -- node[label=(-90+\cmdKV@general@rotate):\cmdKV@gLabels@iLF] {} (0.3,0.3);
    \draw[label distance=-1,\cmdKV@gTypes@iG] (0.9,0.9) -- node[label=(180+\cmdKV@general@rotate):\cmdKV@gLabels@iLG] {} (1.5,0.3);
    #2
  \end{tikzpicture}
  \gSetStdTypes{line}
}

\def\gBoxTriNPA{\@ifnextchar[\@gBoxTriNPA{\@gBoxTriNPA[]}}
\def\@gBoxTriNPA[#1]#2{%
  \setkeys*{pre}{#1}%
  \setrmkeys{general,gTypes,gLabels}%
  \begin{tikzpicture}
    [line width=1pt,
    baseline={([yshift=-0.5ex+\cmdKV@general@yshift]current bounding box.center)},
    scale=\cmdKV@general@scale,
    rotate=\cmdKV@general@rotate,
    font=\cmdKV@general@font]
    \draw[\cmdKV@gTypes@eA] (0.3,0.6) -- (0,0) node[label=(-180+\cmdKV@general@rotate):\cmdKV@gLabels@eLA] {};
    \draw[\cmdKV@gTypes@eB] (0.3,0.6) -- (0,1.2) node[label=(180+\cmdKV@general@rotate):\cmdKV@gLabels@eLB] {};
    \draw[\cmdKV@gTypes@eC] (1.5,0.9) -- (1.8,1.2) node[label=(\cmdKV@general@rotate):\cmdKV@gLabels@eLC] {};
    \draw[\cmdKV@gTypes@eD] (1.5,0.3) -- (1.8,0) node[label=(\cmdKV@general@rotate):\cmdKV@gLabels@eLD] {};
    \draw[label distance=0,\cmdKV@gTypes@iA] (0.3,0.6) -- node[label=(90+\cmdKV@general@rotate):\cmdKV@gLabels@iLA] {} (0.9,0.9);
    \draw[label distance=0,\cmdKV@gTypes@iB] (0.9,0.9) -- node[label=(90+\cmdKV@general@rotate):\cmdKV@gLabels@iLB] {} (1.5,0.9);
    \draw[label distance=-1,\cmdKV@gTypes@iC] (0.9,0.9) -- node[label=(180+\cmdKV@general@rotate):\cmdKV@gLabels@iLC] {} (1.5,0.3);
    \foreach \x in {5,...,14} \fill[opacity=0.12,white] (1.2,0.6) circle (\x/100);
    \draw[label distance=0,\cmdKV@gTypes@iD] (1.5,0.3) -- node[label=(-90+\cmdKV@general@rotate):\cmdKV@gLabels@iLD] {} (0.9,0.3);
    \draw[label distance=0,\cmdKV@gTypes@iE] (0.9,0.3) -- node[label=(-90+\cmdKV@general@rotate):\cmdKV@gLabels@iLE] {} (0.3,0.6);
    \draw[label distance=-1,\cmdKV@gTypes@iF] (1.5,0.9) -- node[label=(180+\cmdKV@general@rotate):\cmdKV@gLabels@iLF] {} (0.9,0.3);
    #2
  \end{tikzpicture}
  \gSetStdTypes{line}
}

\def\gBoxTriNPB{\@ifnextchar[\@gBoxTriNPB{\@gBoxTriNPB[]}}
\def\@gBoxTriNPB[#1]#2{%
  \setkeys*{pre}{#1}%
  \setrmkeys{general,gTypes,gLabels}%
  \begin{tikzpicture}
    [line width=1pt,
    baseline={([yshift=-0.5ex+\cmdKV@general@yshift]current bounding box.center)},
    scale=\cmdKV@general@scale,
    rotate=\cmdKV@general@rotate,
    font=\cmdKV@general@font]
    \draw[\cmdKV@gTypes@eA] (-1.9,0.7) -- (-2.2,0) node[label=(180+\cmdKV@general@rotate):\cmdKV@gLabels@eLA] {};
    \draw[\cmdKV@gTypes@eB] (-1.9,0.7) -- (-2.2,1.4) node[label=(180+\cmdKV@general@rotate):\cmdKV@gLabels@eLB] {};
    \draw[\cmdKV@gTypes@eC] (-0.3,0.7) -- (0,0.7) node[label=(\cmdKV@general@rotate):\cmdKV@gLabels@eLC] {};
    \draw[\cmdKV@gTypes@eD] (-1.1,0.7) -- (-0.7,0.7) node[label=(\cmdKV@general@rotate):\cmdKV@gLabels@eLD] {};
    \draw[label distance=0,\cmdKV@gTypes@iA] (-1.9,0.7) -- node[label=(90+\cmdKV@general@rotate):\cmdKV@gLabels@iLA] {} (-1.6,0.7);
    \draw[label distance=0,\cmdKV@gTypes@iB] (-1.6,0.7) -- node[label=(90+\cmdKV@general@rotate):\cmdKV@gLabels@iLB] {} (-0.7,1.1);
    \draw[label distance=0,\cmdKV@gTypes@iC] (-0.7,1.1) -- node[label=(45+\cmdKV@general@rotate):\cmdKV@gLabels@iLC] {} (-0.3,0.7);
    \draw[label distance=0,\cmdKV@gTypes@iD] (-0.3,0.7) -- node[label=(-45+\cmdKV@general@rotate):\cmdKV@gLabels@iLD] {} (-0.7,0.3);
    \draw[label distance=0,\cmdKV@gTypes@iE] (-0.7,0.3) -- node[label=(-90+\cmdKV@general@rotate):\cmdKV@gLabels@iLE] {} (-1.6,0.7);
    \draw[label distance=0,\cmdKV@gTypes@iF] (-0.7,1.1) -- node[label=(180+\cmdKV@general@rotate):\cmdKV@gLabels@iLF] {} (-1.1,0.7);
    \draw[label distance=0,\cmdKV@gTypes@iG] (-1.1,0.7) -- node[label=(-180+\cmdKV@general@rotate):\cmdKV@gLabels@iLG] {} (-0.7,0.3);
    #2
  \end{tikzpicture}
  \gSetStdTypes{line}
}

\def\gTriPenta{\@ifnextchar[\@gTriPenta{\@gTriPenta[]}}
\def\@gTriPenta[#1]#2{%
  \setkeys*{pre}{#1}%
  \setrmkeys{general,gTypes,gLabels}%
  \begin{tikzpicture}
    [line width=1pt,
    baseline={([yshift=-0.5ex+\cmdKV@general@yshift]current bounding box.center)},
    scale=\cmdKV@general@scale,
    rotate=\cmdKV@general@rotate,
    font=\cmdKV@general@font]
    \draw[\cmdKV@gTypes@eA] (-108:0.5) -- (-135:1) node[label=(145+\cmdKV@general@rotate):\cmdKV@gLabels@eLA] {};
    \draw[\cmdKV@gTypes@eB] (180:0.5) -- (180:1) node[label=(180+\cmdKV@general@rotate):\cmdKV@gLabels@eLB] {};
    \draw[\cmdKV@gTypes@eC] (108:0.5) -- (135:1) node[label=(-145+\cmdKV@general@rotate):\cmdKV@gLabels@eLC] {};
    \draw[\cmdKV@gTypes@eD] (0:1.2) -- (0:1.5) node[label=(\cmdKV@general@rotate):\cmdKV@gLabels@eLD] {};
    \draw[label distance=0,\cmdKV@gTypes@iA] (-108:0.5) -- node[label=(-135+\cmdKV@general@rotate):\cmdKV@gLabels@iLA] {} (-180:0.5);
    \draw[label distance=0,\cmdKV@gTypes@iB] (180:0.5) -- node[label=(135+\cmdKV@general@rotate):\cmdKV@gLabels@iLB] {} (108:0.5);
    \draw[label distance=0,\cmdKV@gTypes@iC] (108:0.5) -- node[label=(72+\cmdKV@general@rotate):\cmdKV@gLabels@iLC] {} (36:0.5);
    \draw[label distance=0,\cmdKV@gTypes@iD] (36:0.5) -- node[label=(72+\cmdKV@general@rotate):\cmdKV@gLabels@iLD] {} (0:1.2);
    \draw[label distance=2,\cmdKV@gTypes@iE] (0:1.2) -- node[label=(-90+\cmdKV@general@rotate):\cmdKV@gLabels@iLE] {} (-36:0.5);
    \draw[label distance=2,\cmdKV@gTypes@iF] (-36:0.5) -- node[label=(-90+\cmdKV@general@rotate):\cmdKV@gLabels@iLF] {} (-108:0.5);
    \draw[label distance=0,\cmdKV@gTypes@iG] (36:0.5) -- node[label=(180+\cmdKV@general@rotate):\cmdKV@gLabels@iLG] {} (-36:0.5);
    #2
  \end{tikzpicture}
  \gSetStdTypes{line}
}

\def\gTriTri{\@ifnextchar[\@gTriTri{\@gTriTri[]}}
\def\@gTriTri[#1]#2{%
  \setkeys*{pre}{#1}%
  \setrmkeys{general,gTypes,gLabels}%
  \begin{tikzpicture}
    [line width=1pt,
    baseline={([yshift=-0.5ex+\cmdKV@general@yshift]current bounding box.center)},
    scale=\cmdKV@general@scale,
    rotate=\cmdKV@general@rotate,
    font=\cmdKV@general@font]
    \draw[\cmdKV@gTypes@eA] (0.3,0.3) -- (0,0) node[label=(-180+\cmdKV@general@rotate):\cmdKV@gLabels@eLA] {};
    \draw[\cmdKV@gTypes@eB] (0.3,1.1) -- (0,1.4) node[label=(180+\cmdKV@general@rotate):\cmdKV@gLabels@eLB] {};
    \draw[\cmdKV@gTypes@eC] (1.9,1.1) -- (2.2,1.4) node[label=(\cmdKV@general@rotate):\cmdKV@gLabels@eLC] {};
    \draw[\cmdKV@gTypes@eD] (1.9,0.3) -- (2.2,0) node[label=(\cmdKV@general@rotate):\cmdKV@gLabels@eLD] {};
    \draw[label distance=0,\cmdKV@gTypes@iA] (0.3,0.3) -- node[label=(180+\cmdKV@general@rotate):\cmdKV@gLabels@iLA] {} (0.3,1.1);
    \draw[label distance=-2,\cmdKV@gTypes@iB] (0.3,1.1) -- node[label=(45+\cmdKV@general@rotate):\cmdKV@gLabels@iLB] {} (0.9,0.7);
    \draw[label distance=-2,\cmdKV@gTypes@iC] (0.9,0.7) -- node[label=(-45+\cmdKV@general@rotate):\cmdKV@gLabels@iLC] {} (0.3,0.3);
    \draw[label distance=0,\cmdKV@gTypes@iD] (0.9,0.7) -- node[label=(90+\cmdKV@general@rotate):\cmdKV@gLabels@iLD] {} (1.3,0.7);
    \draw[label distance=-2,\cmdKV@gTypes@iE] (1.3,0.7) -- node[label=(135+\cmdKV@general@rotate):\cmdKV@gLabels@iLE] {} (1.9,1.1);
    \draw[label distance=0,\cmdKV@gTypes@iF] (1.9,1.1) -- node[label=(\cmdKV@general@rotate):\cmdKV@gLabels@iLF] {} (1.9,0.3);
    \draw[label distance=-2,\cmdKV@gTypes@iG] (1.9,0.3) -- node[label=(-135+\cmdKV@general@rotate):\cmdKV@gLabels@iLG] {} (1.3,0.7);
    #2
  \end{tikzpicture}
  \gSetStdTypes{line}
}

\def\gTriTriB{\@ifnextchar[\@gTriTriB{\@gTriTriB[]}}
\def\@gTriTriB[#1]#2{%
  \setkeys*{pre}{#1}%
  \setrmkeys{general,gTypes,gLabels}%
  \begin{tikzpicture}
    [line width=1pt,
    baseline={([yshift=-0.5ex+\cmdKV@general@yshift]current bounding box.center)},
    scale=\cmdKV@general@scale,
    rotate=\cmdKV@general@rotate,
    font=\cmdKV@general@font]
    \draw[\cmdKV@gTypes@eA] (0.3,0.6) -- (0,0) node[label=(-180+\cmdKV@general@rotate):\cmdKV@gLabels@eLA] {};
    \draw[\cmdKV@gTypes@eB] (0.3,0.6) -- (0,1.2) node[label=(180+\cmdKV@general@rotate):\cmdKV@gLabels@eLB] {};
    \draw[\cmdKV@gTypes@eC] (2.1,0.6) -- (2.4,1.2) node[label=(\cmdKV@general@rotate):\cmdKV@gLabels@eLC] {};
    \draw[\cmdKV@gTypes@eD] (2.1,0.6) -- (2.4,0) node[label=(\cmdKV@general@rotate):\cmdKV@gLabels@eLD] {};
    \draw[label distance=0,\cmdKV@gTypes@iA] (0.3,0.6) -- node[label=(90+\cmdKV@general@rotate):\cmdKV@gLabels@iLA] {} (0.6,0.6);
    \draw[label distance=-2,\cmdKV@gTypes@iB] (0.6,0.6) -- node[label=(135+\cmdKV@general@rotate):\cmdKV@gLabels@iLB] {} (1.2,0.9);
    \draw[label distance=-2,\cmdKV@gTypes@iC] (1.2,0.9) -- node[label=(45+\cmdKV@general@rotate):\cmdKV@gLabels@iLC] {} (1.8,0.6);
    \draw[label distance=-2,\cmdKV@gTypes@iD] (1.8,0.6) -- node[label=(-48+\cmdKV@general@rotate):\cmdKV@gLabels@iLD] {} (1.2,0.3);
    \draw[label distance=-2,\cmdKV@gTypes@iE] (1.2,0.3) -- node[label=(-135+\cmdKV@general@rotate):\cmdKV@gLabels@iLE] {} (0.6,0.6);
    \draw[label distance=-2,\cmdKV@gTypes@iF] (1.2,0.9) -- node[label=(\cmdKV@general@rotate):\cmdKV@gLabels@iLF] {} (1.2,0.3);
    \draw[label distance=0,\cmdKV@gTypes@iG] (1.8,0.6) -- node[label=(90+\cmdKV@general@rotate):\cmdKV@gLabels@iLG] {} (2.1,0.6);
    #2
  \end{tikzpicture}
  \gSetStdTypes{line}
}

\def\gBoxTri{\@ifnextchar[\@gBoxTri{\@gBoxTri[]}}
\def\@gBoxTri[#1]#2{%
  \setkeys*{pre}{#1}%
  \setrmkeys{general,gTypes,gLabels}%
  \begin{tikzpicture}
    [line width=1pt,
    baseline={([yshift=-0.5ex+\cmdKV@general@yshift]current bounding box.center)},
    scale=\cmdKV@general@scale,
    rotate=\cmdKV@general@rotate,
    font=\cmdKV@general@font]
    \draw[\cmdKV@gTypes@eA] (0.3,0.3) -- (0,0) node[label=(-180+\cmdKV@general@rotate):\cmdKV@gLabels@eLA] {};
    \draw[\cmdKV@gTypes@eB] (0.3,0.9) -- (0,1.2) node[label=(180+\cmdKV@general@rotate):\cmdKV@gLabels@eLB] {};
    \draw[\cmdKV@gTypes@eC] (1.8,0.6) -- (2.1,1.2) node[label=(\cmdKV@general@rotate):\cmdKV@gLabels@eLC] {};
    \draw[\cmdKV@gTypes@eD] (1.8,0.6) -- (2.1,0) node[label=(\cmdKV@general@rotate):\cmdKV@gLabels@eLD] {};
    \draw[label distance=0,\cmdKV@gTypes@iA] (0.3,0.3) -- node[label=(180+\cmdKV@general@rotate):\cmdKV@gLabels@iLA] {} (0.3,0.9);
    \draw[label distance=0,\cmdKV@gTypes@iB] (0.3,0.9) -- node[label=(90+\cmdKV@general@rotate):\cmdKV@gLabels@iLB] {} (0.9,0.9);
    \draw[label distance=0,\cmdKV@gTypes@iC] (0.9,0.9) -- node[label=(90+\cmdKV@general@rotate):\cmdKV@gLabels@iLC] {} (1.5,0.6);
    \draw[label distance=0,\cmdKV@gTypes@iD] (1.5,0.6) -- node[label=(90+\cmdKV@general@rotate):\cmdKV@gLabels@iLD] {} (1.8,0.6);
    \draw[label distance=0,\cmdKV@gTypes@iE] (1.5,0.6) -- node[label=(-90+\cmdKV@general@rotate):\cmdKV@gLabels@iLE] {} (0.9,0.3);
    \draw[label distance=0,\cmdKV@gTypes@iF] (0.9,0.3) -- node[label=(-90+\cmdKV@general@rotate):\cmdKV@gLabels@iLF] {} (0.3,0.3);
    \draw[label distance=0,\cmdKV@gTypes@iG] (0.9,0.9) -- node[label=(180+\cmdKV@general@rotate):\cmdKV@gLabels@iLG] {} (0.9,0.3);
    #2
  \end{tikzpicture}
  \gSetStdTypes{line}
}

\def\gBoxTriB{\@ifnextchar[\@gBoxTriB{\@gBoxTriB[]}}
\def\@gBoxTriB[#1]#2{%
  \setkeys*{pre}{#1}%
  \setrmkeys{general,gTypes,gLabels}%
  \begin{tikzpicture}
    [line width=1pt,
    baseline={([yshift=-0.5ex+\cmdKV@general@yshift]current bounding box.center)},
    scale=\cmdKV@general@scale,
    rotate=\cmdKV@general@rotate,
    font=\cmdKV@general@font]
    \draw[\cmdKV@gTypes@eA] (0.3,0.3) -- (0,0) node[label=(-180+\cmdKV@general@rotate):\cmdKV@gLabels@eLA] {};
    \draw[\cmdKV@gTypes@eB] (0.1,1.1) -- (-0.4,1.1) node[label=(180+\cmdKV@general@rotate):\cmdKV@gLabels@eLB] {};
    \draw[\cmdKV@gTypes@eC] (0.1,1.1) -- (0.1,1.6) node[label=(\cmdKV@general@rotate):\cmdKV@gLabels@eLC] {};
    \draw[\cmdKV@gTypes@eD] (1.5,0.6) -- (2,0.6) node[label=(\cmdKV@general@rotate):\cmdKV@gLabels@eLD] {};
    \draw[label distance=0,\cmdKV@gTypes@iA] (0.3,0.3) -- node[label=(180+\cmdKV@general@rotate):\cmdKV@gLabels@iLA] {} (0.3,0.9);
    \draw[label distance=-2,\cmdKV@gTypes@iB] (0.3,0.9) -- node[label=(45+\cmdKV@general@rotate):\cmdKV@gLabels@iLB] {} (0.1,1.1);
    \draw[label distance=0,\cmdKV@gTypes@iC] (0.3,0.9) -- node[label=(90+\cmdKV@general@rotate):\cmdKV@gLabels@iLC] {} (0.9,0.9);
    \draw[label distance=0,\cmdKV@gTypes@iD] (0.9,0.9) -- node[label=(90+\cmdKV@general@rotate):\cmdKV@gLabels@iLD] {} (1.5,0.6);  
    \draw[label distance=0,\cmdKV@gTypes@iE] (1.5,0.6) -- node[label=(-90+\cmdKV@general@rotate):\cmdKV@gLabels@iLE] {} (0.9,0.3);
    \draw[label distance=0,\cmdKV@gTypes@iF] (0.9,0.3) -- node[label=(-90+\cmdKV@general@rotate):\cmdKV@gLabels@iLF] {} (0.3,0.3);
    \draw[label distance=0,\cmdKV@gTypes@iG] (0.9,0.9) -- node[label=(180+\cmdKV@general@rotate):\cmdKV@gLabels@iLG] {} (0.9,0.3);
    #2
  \end{tikzpicture}
  \gSetStdTypes{line}
}

\def\gBoxBubA{\@ifnextchar[\@gBoxBubA{\@gBoxBubA[]}}
\def\@gBoxBubA[#1]#2{%
  \setkeys*{pre}{#1}%
  \setrmkeys{general,gTypes,gLabels}%
  \begin{tikzpicture}
    [line width=1pt,
    baseline={([yshift=-0.5ex+\cmdKV@general@yshift]current bounding box.center)},
    scale=\cmdKV@general@scale,
    rotate=\cmdKV@general@rotate,
    font=\cmdKV@general@font]
    \draw[\cmdKV@gTypes@eA] (0.1,0.1) -- (-0.1,-0.1) node[label=(-180+\cmdKV@general@rotate):\cmdKV@gLabels@eLA] {};
    \draw[\cmdKV@gTypes@eB] (0.1,0.9) -- (-0.1,1.1) node[label=(180+\cmdKV@general@rotate):\cmdKV@gLabels@eLB] {};
    \draw[\cmdKV@gTypes@eC] (0.9,0.9) -- (1.1,1.1) node[label=(\cmdKV@general@rotate):\cmdKV@gLabels@eLC] {};
    \draw[\cmdKV@gTypes@eD] (0.9,0.1) -- (1.1,-0.1) node[label=(\cmdKV@general@rotate):\cmdKV@gLabels@eLD] {};
    \draw[label distance=0,\cmdKV@gTypes@iA] (0.1,0.1) -- node[label=(180+\cmdKV@general@rotate):\cmdKV@gLabels@iLA] {} (0.1,0.9);
    \draw[label distance=0,\cmdKV@gTypes@iB] (0.1,0.9) -- node[label=(90+\cmdKV@general@rotate):\cmdKV@gLabels@iLB] {} (0.32,0.9);
    \draw[label distance=0,\cmdKV@gTypes@iC] (0.32,0.9) .. controls (0.32,1.15) and (0.68,1.15) .. node[label=(90+\cmdKV@general@rotate):\cmdKV@gLabels@iLC] {} (0.68,0.9);
    \draw[label distance=0,\cmdKV@gTypes@iD] (0.68,0.9) .. controls (0.68,0.65) and (0.32,0.65) .. node[label=(-90+\cmdKV@general@rotate):\cmdKV@gLabels@iLD] {} (0.32,0.9);
    \draw[label distance=0,\cmdKV@gTypes@iE] (0.68,0.9) -- node[label=(90+\cmdKV@general@rotate):\cmdKV@gLabels@iLE] {} (0.9,0.9);
    \draw[label distance=0,\cmdKV@gTypes@iF] (0.9,0.9) -- node[label=(\cmdKV@general@rotate):\cmdKV@gLabels@iLF] {} (0.9,0.1);
    \draw[label distance=0,\cmdKV@gTypes@iG] (0.9,0.1) -- node[label=(-90+\cmdKV@general@rotate):\cmdKV@gLabels@iLG] {} (0.1,0.1);
    #2
  \end{tikzpicture}
  \gSetStdTypes{line}
}

\def\gBoxBubB{\@ifnextchar[\@gBoxBubB{\@gBoxBubB[]}}
\def\@gBoxBubB[#1]#2{%
  \setkeys*{pre}{#1}%
  \setrmkeys{general,gTypes,gLabels}%
  \begin{tikzpicture}
    [line width=1pt,
    baseline={([yshift=-0.5ex+\cmdKV@general@yshift]current bounding box.center)},
    scale=\cmdKV@general@scale,
    rotate=\cmdKV@general@rotate,
    font=\cmdKV@general@font]
    \draw[label distance=-1,\cmdKV@gTypes@eA] (0,-0.4) -- (0,-0.7) node[label=(180+\cmdKV@general@rotate):\cmdKV@gLabels@eLA] {};
    \draw[\cmdKV@gTypes@eB] (-0.4,0) -- (-0.7,0) node[label=(180+\cmdKV@general@rotate):\cmdKV@gLabels@eLB] {};
    \draw[label distance=-1,\cmdKV@gTypes@eC] (0,0.4) -- (0,0.7) node[label=(180+\cmdKV@general@rotate):\cmdKV@gLabels@eLC] {};
    \draw[\cmdKV@gTypes@eD] (1.1,0) -- (1.4,0) node[label=(\cmdKV@general@rotate):\cmdKV@gLabels@eLD] {};
    \draw[label distance=0,\cmdKV@gTypes@iA] (0,-0.4) -- node[label=(-135+\cmdKV@general@rotate):\cmdKV@gLabels@iLA] {} (-0.4,0);
    \draw[label distance=0,\cmdKV@gTypes@iB] (-0.4,0) -- node[label=(135+\cmdKV@general@rotate):\cmdKV@gLabels@iLB] {} (0,0.4);
    \draw[label distance=0,\cmdKV@gTypes@iC] (0,0.4) -- node[label=(45+\cmdKV@general@rotate):\cmdKV@gLabels@iLC] {} (0.4,0);
    \draw[label distance=0,\cmdKV@gTypes@iD] (0.4,0) -- node[label=(-45+\cmdKV@general@rotate):\cmdKV@gLabels@iLD] {} (0,-0.4);
    \draw[label distance=0,\cmdKV@gTypes@iE] (0.4,0) --  node[label=(90+\cmdKV@general@rotate):\cmdKV@gLabels@iLE] {} (0.7,0);
    \draw[label distance=0,\cmdKV@gTypes@iF] (0.7,0) .. controls (0.7,0.26) and (1.1,0.26) .. node[label=(90+\cmdKV@general@rotate):\cmdKV@gLabels@iLF] {} (1.1,0);
    \draw[label distance=0,\cmdKV@gTypes@iG] (1.1,0) .. controls (1.1,-0.26) and (0.7,-0.26) ..  node[label=(-90+\cmdKV@general@rotate):\cmdKV@gLabels@iLG] {} (0.7,0);
    #2
  \end{tikzpicture}
  \gSetStdTypes{line}
}

\def\gBoxBubC{\@ifnextchar[\@gBoxBubC{\@gBoxBubC[]}}
\def\@gBoxBubC[#1]#2{%
  \setkeys*{pre}{#1}%
  \setrmkeys{general,gTypes,gLabels}%
  \begin{tikzpicture}
    [line width=1pt,
    baseline={([yshift=-0.5ex+\cmdKV@general@yshift]current bounding box.center)},
    scale=\cmdKV@general@scale,
    rotate=\cmdKV@general@rotate,
    font=\cmdKV@general@font]
    \draw[\cmdKV@gTypes@eA] (0.1,0.1) -- (-0.1,-0.1) node[label=(-180+\cmdKV@general@rotate):\cmdKV@gLabels@eLA] {};
    \draw[\cmdKV@gTypes@eB] (0.1,0.9) -- (-0.1,1.1) node[label=(180+\cmdKV@general@rotate):\cmdKV@gLabels@eLB] {};
    \draw[\cmdKV@gTypes@eC] (0.9,0.9) -- (1.1,1.1) node[label=(\cmdKV@general@rotate):\cmdKV@gLabels@eLC] {};
    \draw[\cmdKV@gTypes@eD] (0.9,0.1) -- (1.1,-0.1) node[label=(\cmdKV@general@rotate):\cmdKV@gLabels@eLD] {};
    \draw[label distance=0,\cmdKV@gTypes@iA] (0.1,0.1) -- node[label=(180+\cmdKV@general@rotate):\cmdKV@gLabels@iLA] {} (0.1,0.9);
    \draw[label distance=0,\cmdKV@gTypes@iB] (0.1,0.9) to[bend left] node[label=(90+\cmdKV@general@rotate):\cmdKV@gLabels@iLB] {} (0.9,0.9);
    \draw[label distance=0,\cmdKV@gTypes@iC] (0.9,0.9) to[bend left] node[label=(-90+\cmdKV@general@rotate):\cmdKV@gLabels@iLC] {} (0.1,0.9);
    \draw[label distance=0,\cmdKV@gTypes@iD] (0.9,0.9) -- node[label=(\cmdKV@general@rotate):\cmdKV@gLabels@iLD] {} (0.9,0.1);
    \draw[label distance=0,\cmdKV@gTypes@iE] (0.9,0.1) -- node[label=(-90+\cmdKV@general@rotate):\cmdKV@gLabels@iLE] {} (0.1,0.1);
    #2
  \end{tikzpicture}
  \gSetStdTypes{line}
}

\def\gPentaTad{\@ifnextchar[\@gPentaTad{\@gPentaTad[]}}
\def\@gPentaTad[#1]#2{%
  \setkeys*{pre}{#1}%
  \setrmkeys{general,gTypes,gLabels}%
  \begin{tikzpicture}
    [line width=1pt,
    baseline={([yshift=-0.5ex+\cmdKV@general@yshift]current bounding box.center)},
    scale=\cmdKV@general@scale,
    rotate=\cmdKV@general@rotate,
    font=\cmdKV@general@font]
    \draw[label distance=-0.5,\cmdKV@gTypes@eA] (108:0.35) -- (108:0.75) node[label=(180+\cmdKV@general@rotate):\cmdKV@gLabels@eLA] {};
    \draw[\cmdKV@gTypes@eB] (36:0.35) -- (36:0.75) node[label=(36+\cmdKV@general@rotate):\cmdKV@gLabels@eLB] {};
    \draw[\cmdKV@gTypes@eC] (-36:0.35) -- (-36:0.75) node[label=(-36+\cmdKV@general@rotate):\cmdKV@gLabels@eLC] {};
    \draw[label distance=-0.5,\cmdKV@gTypes@eD] (-108:0.35) -- (-108:0.75) node[label=(-180+\cmdKV@general@rotate):\cmdKV@gLabels@eLD] {};
    \draw[label distance=0,\cmdKV@gTypes@iA] (108:0.35) -- node[label=(72+\cmdKV@general@rotate):\cmdKV@gLabels@iLA] {} (36:0.35);
    \draw[label distance=0,\cmdKV@gTypes@iB] (36:0.35) -- node[label=(\cmdKV@general@rotate):\cmdKV@gLabels@iLB] {} (-36:0.35);
    \draw[label distance=0,\cmdKV@gTypes@iC] (-36:0.35) -- node[label=(-72+\cmdKV@general@rotate):\cmdKV@gLabels@iLC] {} (-108:0.35);
    \draw[label distance=0,\cmdKV@gTypes@iD] (-108:0.35) -- node[label=(-144+\cmdKV@general@rotate):\cmdKV@gLabels@iLD] {} (-180:0.35);
    \draw[label distance=0,\cmdKV@gTypes@iE] (180:0.35) -- node[label=(144+\cmdKV@general@rotate):\cmdKV@gLabels@iLE] {} (108:0.35);
    \draw[label distance=0,\cmdKV@gTypes@iF] (180:0.35) -- node[label=(90+\cmdKV@general@rotate):\cmdKV@gLabels@iLF] {} (180:0.75);
    \draw[label distance=-0.5,\cmdKV@gTypes@iG] (180:0.75) .. controls ($(180:0.75)+(0,-0.18)$) and ($(180:1)+(0,-0.18)$) .. (180:1) node[label=(180+\cmdKV@general@rotate):\cmdKV@gLabels@iLG] {} .. controls ($(180:1)+(0,0.18)$) and ($(180:0.75)+(0,0.18)$) .. (180:0.75);
    #2
  \end{tikzpicture}
  \gSetStdTypes{line}
}

\def\gBoxTadA{\@ifnextchar[\@gBoxTadA{\@gBoxTadA[]}}
\def\@gBoxTadA[#1]#2{%
  \setkeys*{pre}{#1}%
  \setrmkeys{general,gTypes,gLabels}%
  \begin{tikzpicture}
    [line width=1pt,
    baseline={([yshift=-0.5ex+\cmdKV@general@yshift]current bounding box.center)},
    scale=\cmdKV@general@scale,
    rotate=\cmdKV@general@rotate,
    font=\cmdKV@general@font]
    \draw[label distance=-0.5,\cmdKV@gTypes@eA] (0,0.4) -- (0,0.7) node[label=(\cmdKV@general@rotate):\cmdKV@gLabels@eLA] {};
    \draw[\cmdKV@gTypes@eB] (0.4,0) -- (0.7,0) node[label=(\cmdKV@general@rotate):\cmdKV@gLabels@eLB] {};
    \draw[label distance=-0.5,\cmdKV@gTypes@eC] (0,-0.4) -- (0,-0.7) node[label=(\cmdKV@general@rotate):\cmdKV@gLabels@eLC] {};
    \draw[\cmdKV@gTypes@eD] (-0.7,0) -- (-0.7,-0.3) node[label=(-90+\cmdKV@general@rotate):\cmdKV@gLabels@eLD] {};
    \draw[label distance=0,\cmdKV@gTypes@iA] (0,0.4) -- node[label=(45+\cmdKV@general@rotate):\cmdKV@gLabels@iLA] {} (0.4,0);
    \draw[label distance=0,\cmdKV@gTypes@iB] (0.4,0) -- node[label=(-45+\cmdKV@general@rotate):\cmdKV@gLabels@iLB] {} (0,-0.4);
    \draw[label distance=0,\cmdKV@gTypes@iC] (0,-0.4) -- node[label=(-135+\cmdKV@general@rotate):\cmdKV@gLabels@iLC] {} (-0.4,0);
    \draw[label distance=0,\cmdKV@gTypes@iD] (-0.4,0) -- node[label=(135+\cmdKV@general@rotate):\cmdKV@gLabels@iLD] {} (0,0.4);
    \draw[label distance=0,\cmdKV@gTypes@iE] (-0.4,0) --  node[label=(90+\cmdKV@general@rotate):\cmdKV@gLabels@iLE] {} (-0.7,0);
    \draw[label distance=0,\cmdKV@gTypes@iF] (-0.7,0) -- node[label=(90+\cmdKV@general@rotate):\cmdKV@gLabels@iLF] {} (-1.1,0);
    \draw[label distance=-0.5,\cmdKV@gTypes@iG] (-1.1,0) .. controls (-1.1,-0.18) and (-1.35,-0.18) .. (-1.35,0) node[label=(180+\cmdKV@general@rotate):\cmdKV@gLabels@iLG] {} .. controls (-1.35,0.18) and (-1.1,0.18) .. (-1.1,0);
    #2
  \end{tikzpicture}
  \gSetStdTypes{line}
}

\def\gSunsetA{\@ifnextchar[\@gSunsetA{\@gSunsetA[]}}
\def\@gSunsetA[#1]#2{%
  \setkeys*{pre}{#1}%
  \setrmkeys{general,gTypes,gLabels}%
  \begin{tikzpicture}
    [line width=1pt,
    baseline={([yshift=-0.5ex+\cmdKV@general@yshift]current bounding box.center)},
    scale=\cmdKV@general@scale,
    rotate=\cmdKV@general@rotate,
    font=\cmdKV@general@font]
    \draw[\cmdKV@gTypes@eA] (0.3,0.6) -- (0,0) node[label=(-180+\cmdKV@general@rotate):\cmdKV@gLabels@eLA] {};
    \draw[\cmdKV@gTypes@eB] (0.3,0.6) -- (0,1.2) node[label=(180+\cmdKV@general@rotate):\cmdKV@gLabels@eLB] {};
    \draw[\cmdKV@gTypes@eC] (1.5,0.6) -- (1.8,1.2) node[label=(\cmdKV@general@rotate):\cmdKV@gLabels@eLC] {};
    \draw[\cmdKV@gTypes@eD] (1.5,0.6) -- (1.8,0) node[label=(\cmdKV@general@rotate):\cmdKV@gLabels@eLD] {};
    \draw[label distance=0,\cmdKV@gTypes@iA] (0.3,0.6) to[bend left=90] node[label=(90+\cmdKV@general@rotate):\cmdKV@gLabels@iLA] {} (1.5,0.6);
    \draw[label distance=0,\cmdKV@gTypes@iB] (1.5,0.6) to[bend left=90] node[label=(-90+\cmdKV@general@rotate):\cmdKV@gLabels@iLB] {} (0.3,0.6);
    \draw[label distance=-2,\cmdKV@gTypes@iC] (0.3,0.6) -- node[label=(90+\cmdKV@general@rotate):\cmdKV@gLabels@iLC] {} (1.5,0.6);
    #2
  \end{tikzpicture}
  \gSetStdTypes{line}
}

\def\gTriBubA{\@ifnextchar[\@gTriBubA{\@gTriBubA[]}}
\def\@gTriBubA[#1]#2{%
  \setkeys*{pre}{#1}%
  \setrmkeys{general,gTypes,gLabels}%
  \begin{tikzpicture}
    [line width=1pt,
    baseline={([yshift=-0.5ex+\cmdKV@general@yshift]current bounding box.center)},
    scale=\cmdKV@general@scale,
    rotate=\cmdKV@general@rotate,
    font=\cmdKV@general@font]
    \draw[\cmdKV@gTypes@eA] (0.4,0.3) -- (0,0) node[label=(-180+\cmdKV@general@rotate):\cmdKV@gLabels@eLA] {};
    \draw[\cmdKV@gTypes@eB] (0.4,0.9) -- (0,1.2) node[label=(180+\cmdKV@general@rotate):\cmdKV@gLabels@eLB] {};
    \draw[\cmdKV@gTypes@eC] (1.4,0.6) -- (1.8,1.2) node[label=(\cmdKV@general@rotate):\cmdKV@gLabels@eLC] {};
    \draw[\cmdKV@gTypes@eD] (1.4,0.6) -- (1.8,0) node[label=(\cmdKV@general@rotate):\cmdKV@gLabels@eLD] {};
    \draw[label distance=0,\cmdKV@gTypes@iA] (0.4,0.3) to[bend left=40] node[label=(180+\cmdKV@general@rotate):\cmdKV@gLabels@iLA] {} (0.4,0.9);
    \draw[label distance=0,\cmdKV@gTypes@iB] (0.4,0.9) -- node[label=(90+\cmdKV@general@rotate):\cmdKV@gLabels@iLB] {} (1.4,0.6);
    \draw[label distance=0,\cmdKV@gTypes@iC] (1.4,0.6) -- node[label=(-90+\cmdKV@general@rotate):\cmdKV@gLabels@iLC] {} (0.4,0.3);
    \draw[label distance=-1,\cmdKV@gTypes@iD] (0.4,0.9) to[bend left=40] node[label=(\cmdKV@general@rotate):\cmdKV@gLabels@iLD] {} (0.4,0.3);
    #2
  \end{tikzpicture}
  \gSetStdTypes{line}
}

\def\gTriBubB{\@ifnextchar[\@gTriBubB{\@gTriBubB[]}}
\def\@gTriBubB[#1]#2{%
  \setkeys*{pre}{#1}%
  \setrmkeys{general,gTypes,gLabels}%
  \begin{tikzpicture}
    [line width=1pt,
    baseline={([yshift=-0.5ex+\cmdKV@general@yshift]current bounding box.center)},
    scale=\cmdKV@general@scale,
    rotate=\cmdKV@general@rotate,
    font=\cmdKV@general@font]
    \draw[\cmdKV@gTypes@eA] (0.3,0.6) -- (0,0) node[label=(-180+\cmdKV@general@rotate):\cmdKV@gLabels@eLA] {};
    \draw[\cmdKV@gTypes@eB] (0.3,0.6) -- (0,1.2) node[label=(180+\cmdKV@general@rotate):\cmdKV@gLabels@eLB] {};
    \draw[\cmdKV@gTypes@eC] (1.05,0.8) -- (2.1,1.2) node[label=(\cmdKV@general@rotate):\cmdKV@gLabels@eLC] {};
    \draw[\cmdKV@gTypes@eD] (1.5,0.6) -- (2.1,0) node[label=(\cmdKV@general@rotate):\cmdKV@gLabels@eLD] {};
    \draw[label distance=0,\cmdKV@gTypes@iA] (0.3,0.6) -- node[label=(90+\cmdKV@general@rotate):\cmdKV@gLabels@iLA] {} (0.6,0.6);
    \draw[label distance=-2,\cmdKV@gTypes@iB] (0.6,0.6) -- node[label=(135+\cmdKV@general@rotate):\cmdKV@gLabels@iLB] {} (1.05,0.8);
    \draw[label distance=-2,\cmdKV@gTypes@iC] (1.05,0.8) -- node[label=(45+\cmdKV@general@rotate):\cmdKV@gLabels@iLC] {} (1.5,0.6);
    \draw[label distance=-3,\cmdKV@gTypes@iD] (1.5,0.6) to[bend left=40] node[label=(-45+\cmdKV@general@rotate):\cmdKV@gLabels@iLD] {} (1.4,0.3);
    \draw[label distance=0,\cmdKV@gTypes@iE] (1.4,0.3) to[bend left=40] node[label=(-90+\cmdKV@general@rotate):\cmdKV@gLabels@iLE] {} (0.7,0.3);
    \draw[label distance=0,\cmdKV@gTypes@iF] (0.7,0.3) to[bend left=40] node[label=(+90+\cmdKV@general@rotate):\cmdKV@gLabels@iLF] {} (1.4,0.3);
    \draw[label distance=-3,\cmdKV@gTypes@iG] (0.7,0.3) to[bend left=40] node[label=(-135+\cmdKV@general@rotate):\cmdKV@gLabels@iLG] {} (0.6,0.6);
    #2
  \end{tikzpicture}
  \gSetStdTypes{line}
}

\def\gTriBubC{\@ifnextchar[\@gTriBubC{\@gTriBubC[]}}
\def\@gTriBubC[#1]#2{%
  \setkeys*{pre}{#1}%
  \setrmkeys{general,gTypes,gLabels}%
  \begin{tikzpicture}
    [line width=1pt,
    baseline={([yshift=-0.5ex+\cmdKV@general@yshift]current bounding box.center)},
    scale=\cmdKV@general@scale,
    rotate=\cmdKV@general@rotate,
    font=\cmdKV@general@font]
    \draw[\cmdKV@gTypes@eA] (0.1,0.1) -- (-0.1,-0.1) node[label=(-180+\cmdKV@general@rotate):\cmdKV@gLabels@eLA] {};
    \draw[\cmdKV@gTypes@eB] (0.1,0.9) -- (-0.1,1.1) node[label=(180+\cmdKV@general@rotate):\cmdKV@gLabels@eLB] {};
    \draw[\cmdKV@gTypes@eC] (1.1,0.5) -- (1.4,1.1) node[label=(\cmdKV@general@rotate):\cmdKV@gLabels@eLC] {};
    \draw[\cmdKV@gTypes@eD] (1.1,0.5) -- (1.4,-0.1) node[label=(\cmdKV@general@rotate):\cmdKV@gLabels@eLD] {};
    \draw[label distance=0,\cmdKV@gTypes@iA] (0.1,0.1) -- node[label=(180+\cmdKV@general@rotate):\cmdKV@gLabels@iLA] {} (0.1,0.32);
    \draw[label distance=0,\cmdKV@gTypes@iB] (0.1,0.32) .. controls (-0.15,0.32) and (-0.15,0.68) .. node[label=(180+\cmdKV@general@rotate):\cmdKV@gLabels@iLB] {} (0.1,0.68);
    \draw[label distance=-1,\cmdKV@gTypes@iC] (0.1,0.68) .. controls (0.35,0.68) and (0.35,0.32) .. node[label=(\cmdKV@general@rotate):\cmdKV@gLabels@iLC] {} (0.1,0.32);
    \draw[label distance=0,\cmdKV@gTypes@iD] (0.1,0.68) -- node[label=(180+\cmdKV@general@rotate):\cmdKV@gLabels@iLD] {} (0.1,0.9);
    \draw[label distance=-2,\cmdKV@gTypes@iE] (0.1,0.9) -- node[label=(60+\cmdKV@general@rotate):\cmdKV@gLabels@iLE] {} (0.8,0.5);
    \draw[label distance=0,\cmdKV@gTypes@iF] (0.8,0.5) -- node[label=(90+\cmdKV@general@rotate):\cmdKV@gLabels@iLF] {} (1.1,0.5);
    \draw[label distance=-2,\cmdKV@gTypes@iG] (0.8,0.5) -- node[label=(-60+\cmdKV@general@rotate):\cmdKV@gLabels@iLG] {} (0.1,0.1);
    #2
  \end{tikzpicture}
  \gSetStdTypes{line}
}

\def\gTriBubD{\@ifnextchar[\@gTriBubD{\@gTriBubD[]}}
\def\@gTriBubD[#1]#2{%
  \setkeys*{pre}{#1}%
  \setrmkeys{general,gTypes,gLabels}%
  \begin{tikzpicture}
    [line width=1pt,
    baseline={([yshift=-0.5ex+\cmdKV@general@yshift]current bounding box.center)},
    scale=\cmdKV@general@scale,
    rotate=\cmdKV@general@rotate,
    font=\cmdKV@general@font]
    \draw[\cmdKV@gTypes@eA] (0.3,0.3) -- (0,0) node[label=(-180+\cmdKV@general@rotate):\cmdKV@gLabels@eLA] {};
    \draw[\cmdKV@gTypes@eB] (0.3,0.9) -- (0,1.2) node[label=(180+\cmdKV@general@rotate):\cmdKV@gLabels@eLB] {};
    \draw[\cmdKV@gTypes@eC] (1.8,0.6) -- (2.1,1.2) node[label=(\cmdKV@general@rotate):\cmdKV@gLabels@eLC] {};
    \draw[\cmdKV@gTypes@eD] (1.8,0.6) -- (2.1,0) node[label=(\cmdKV@general@rotate):\cmdKV@gLabels@eLD] {};
    \draw[label distance=0,\cmdKV@gTypes@iA] (0.3,0.3) -- node[label=(180+\cmdKV@general@rotate):\cmdKV@gLabels@iLA] {} (0.3,0.9);
    \draw[label distance=-4,\cmdKV@gTypes@iB] (0.3,0.9) -- node[label=(45+\cmdKV@general@rotate):\cmdKV@gLabels@iLB] {} (0.8,0.6);
    \draw[label distance=-4,\cmdKV@gTypes@iC] (0.8,0.6) -- node[label=(-45+\cmdKV@general@rotate):\cmdKV@gLabels@iLC] {} (0.3,0.3);
    \draw[label distance=0,\cmdKV@gTypes@iD] (0.8,0.6) -- node[label=(90+\cmdKV@general@rotate):\cmdKV@gLabels@iLD] {} (1.1,0.6);
    \draw[label distance=0,\cmdKV@gTypes@iE] (1.1,0.6) to[bend left=90] node[label=(90+\cmdKV@general@rotate):\cmdKV@gLabels@iLE] {} (1.5,0.6);
    \draw[label distance=0,\cmdKV@gTypes@iF] (1.5,0.6) to[bend left=90] node[label=(-90+\cmdKV@general@rotate):\cmdKV@gLabels@iLF] {} (1.1,0.6);
    \draw[label distance=0,\cmdKV@gTypes@iG] (1.5,0.6) -- node[label=(90+\cmdKV@general@rotate):\cmdKV@gLabels@iLG] {} (1.8,0.6);
    #2
  \end{tikzpicture}
  \gSetStdTypes{line}
}

\def\gTriBubE{\@ifnextchar[\@gTriBubE{\@gTriBubE[]}}
\def\@gTriBubE[#1]#2{%
  \setkeys*{pre}{#1}%
  \setrmkeys{general,gTypes,gLabels}%
  \begin{tikzpicture}
    [line width=1pt,
    baseline={([yshift=-0.5ex+\cmdKV@general@yshift]current bounding box.center)},
    scale=\cmdKV@general@scale,
    rotate=\cmdKV@general@rotate,
    font=\cmdKV@general@font]
    \draw[\cmdKV@gTypes@eA] (0.3,0.3) -- (0,0) node[label=(-180+\cmdKV@general@rotate):\cmdKV@gLabels@eLA] {};
    \draw[\cmdKV@gTypes@eB] (0.3,0.9) -- (0,1.2) node[label=(180+\cmdKV@general@rotate):\cmdKV@gLabels@eLB] {};
    \draw[\cmdKV@gTypes@eC] (1.2,0.6) -- (1.2,1.1) node[label=(90+\cmdKV@general@rotate):\cmdKV@gLabels@eLC] {};
    \draw[\cmdKV@gTypes@eD] (2,0.6) -- (2.4,0.6) node[label=(\cmdKV@general@rotate):\cmdKV@gLabels@eLD] {};
    \draw[label distance=0,\cmdKV@gTypes@iA] (0.3,0.3) -- node[label=(180+\cmdKV@general@rotate):\cmdKV@gLabels@iLA] {} (0.3,0.9);
    \draw[label distance=-4,\cmdKV@gTypes@iB] (0.3,0.9) -- node[label=(45+\cmdKV@general@rotate):\cmdKV@gLabels@iLB] {} (0.8,0.6);
    \draw[label distance=-4,\cmdKV@gTypes@iC] (0.8,0.6) -- node[label=(-45+\cmdKV@general@rotate):\cmdKV@gLabels@iLC] {} (0.3,0.3);
    \draw[label distance=0,\cmdKV@gTypes@iD] (0.8,0.6) -- node[label=(90+\cmdKV@general@rotate):\cmdKV@gLabels@iLD] {} (1.2,0.6);
    \draw[label distance=0,\cmdKV@gTypes@iE] (1.6,0.6) to[bend left=90] node[label=(90+\cmdKV@general@rotate):\cmdKV@gLabels@iLE] {} (2,0.6);
    \draw[label distance=0,\cmdKV@gTypes@iF] (2,0.6) to[bend left=90] node[label=(-90+\cmdKV@general@rotate):\cmdKV@gLabels@iLF] {} (1.6,0.6);
    \draw[label distance=0,\cmdKV@gTypes@iG] (1.2,0.6) -- node[label=(90+\cmdKV@general@rotate):\cmdKV@gLabels@iLG] {} (1.6,0.6);
    #2
  \end{tikzpicture}
  \gSetStdTypes{line}
}

\def\gBubBubA{\@ifnextchar[\@gBubBubA{\@gBubBubA[]}}
\def\@gBubBubA[#1]#2{%
  \setkeys*{pre}{#1}%
  \setrmkeys{general,gTypes,gLabels}%
  \begin{tikzpicture}
    [line width=1pt,
    baseline={([yshift=-0.5ex+\cmdKV@general@yshift]current bounding box.center)},
    scale=\cmdKV@general@scale,
    rotate=\cmdKV@general@rotate,
    font=\cmdKV@general@font]
    \draw[\cmdKV@gTypes@eA] (0.3,0.6) -- (0,0) node[label=(-180+\cmdKV@general@rotate):\cmdKV@gLabels@eLA] {};
    \draw[\cmdKV@gTypes@eB] (0.3,0.6) -- (0,1.2) node[label=(180+\cmdKV@general@rotate):\cmdKV@gLabels@eLB] {};
    \draw[\cmdKV@gTypes@eC] (1.5,0.6) -- (1.8,1.2) node[label=(\cmdKV@general@rotate):\cmdKV@gLabels@eLC] {};
    \draw[\cmdKV@gTypes@eD] (1.5,0.6) -- (1.8,0) node[label=(\cmdKV@general@rotate):\cmdKV@gLabels@eLD] {};
    \draw[label distance=0,\cmdKV@gTypes@iA] (0.3,0.6) to[bend left=90] node[label=(90+\cmdKV@general@rotate):\cmdKV@gLabels@iLA] {} (0.9,0.6);
    \draw[label distance=0,\cmdKV@gTypes@iB] (0.9,0.6) to[bend left=90] node[label=(90+\cmdKV@general@rotate):\cmdKV@gLabels@iLB] {} (1.5,0.6);
    \draw[label distance=0,\cmdKV@gTypes@iC] (1.5,0.6) to[bend left=90] node[label=(-90+\cmdKV@general@rotate):\cmdKV@gLabels@iLC] {} (0.9,0.6);
    \draw[label distance=0,\cmdKV@gTypes@iD] (0.9,0.6) to[bend left=90] node[label=(-90+\cmdKV@general@rotate):\cmdKV@gLabels@iLD] {} (0.3,0.6);
    #2
  \end{tikzpicture}
  \gSetStdTypes{line}
}

\def\gBubBubB{\@ifnextchar[\@gBubBubB{\@gBubBubB[]}}
\def\@gBubBubB[#1]#2{%
  \setkeys*{pre}{#1}%
  \setrmkeys{general,gTypes,gLabels}%
  \begin{tikzpicture}
    [line width=1pt,
    baseline={([yshift=-0.5ex+\cmdKV@general@yshift]current bounding box.center)},
    scale=\cmdKV@general@scale,
    rotate=\cmdKV@general@rotate,
    font=\cmdKV@general@font]
    \draw[\cmdKV@gTypes@eA] (0.3,0.6) -- (0,0) node[label=(-180+\cmdKV@general@rotate):\cmdKV@gLabels@eLA] {};
    \draw[\cmdKV@gTypes@eB] (0.3,0.6) -- (0,1.2) node[label=(180+\cmdKV@general@rotate):\cmdKV@gLabels@eLB] {};
    \draw[\cmdKV@gTypes@eC] (1.8,0.6) -- (2.1,1.2) node[label=(\cmdKV@general@rotate):\cmdKV@gLabels@eLC] {};
    \draw[\cmdKV@gTypes@eD] (1.8,0.6) -- (2.1,0) node[label=(\cmdKV@general@rotate):\cmdKV@gLabels@eLD] {};
    \draw[label distance=0,\cmdKV@gTypes@iA] (0.3,0.6) -- node[label=(90+\cmdKV@general@rotate):\cmdKV@gLabels@iLA] {} (0.6,0.6);
    \draw[label distance=0,\cmdKV@gTypes@iB] (0.6,0.6) to[bend left=90] node[label=(90+\cmdKV@general@rotate):\cmdKV@gLabels@iLB] {} (1.5,0.6);
    \draw[label distance=0,\cmdKV@gTypes@iC] (1.5,0.6) -- node[label=(90+\cmdKV@general@rotate):\cmdKV@gLabels@iLC] {} (1.8,0.6);
    \draw[label distance=-3,\cmdKV@gTypes@iD] (1.5,0.6) to[bend left=40] node[label=(-45+\cmdKV@general@rotate):\cmdKV@gLabels@iLD] {} (1.4,0.3);
    \draw[label distance=0,\cmdKV@gTypes@iE] (1.4,0.3) to[bend left=40] node[label=(-90+\cmdKV@general@rotate):\cmdKV@gLabels@iLE] {} (0.7,0.3);
    \draw[label distance=0,\cmdKV@gTypes@iF] (0.7,0.3) to[bend left=40] node[label=(+90+\cmdKV@general@rotate):\cmdKV@gLabels@iLF] {} (1.4,0.3);
    \draw[label distance=-3,\cmdKV@gTypes@iG] (0.7,0.3) to[bend left=40] node[label=(-135+\cmdKV@general@rotate):\cmdKV@gLabels@iLG] {} (0.6,0.6);
    #2
  \end{tikzpicture}
  \gSetStdTypes{line}
}

\def\gBubBubC{\@ifnextchar[\@gBubBubC{\@gBubBubC[]}}
\def\@gBubBubC[#1]#2{%
  \setkeys*{pre}{#1}%
  \setrmkeys{general,gTypes,gLabels}%
  \begin{tikzpicture}
    [line width=1pt,
    baseline={([yshift=-0.5ex+\cmdKV@general@yshift]current bounding box.center)},
    scale=\cmdKV@general@scale,
    rotate=\cmdKV@general@rotate,
    font=\cmdKV@general@font]
    \draw[\cmdKV@gTypes@eA] (0.3,0.6) -- (0,0) node[label=(-180+\cmdKV@general@rotate):\cmdKV@gLabels@eLA] {};
    \draw[\cmdKV@gTypes@eB] (0.3,0.6) -- (0,1.2) node[label=(180+\cmdKV@general@rotate):\cmdKV@gLabels@eLB] {};
    \draw[\cmdKV@gTypes@eC] (2,0.6) -- (2.3,1.2) node[label=(\cmdKV@general@rotate):\cmdKV@gLabels@eLC] {};
    \draw[\cmdKV@gTypes@eD] (2,0.6) -- (2.3,0) node[label=(\cmdKV@general@rotate):\cmdKV@gLabels@eLD] {};
    \draw[label distance=0,\cmdKV@gTypes@iA] (0.3,0.6) -- node[label=(90+\cmdKV@general@rotate):\cmdKV@gLabels@iLA] {} (0.6,0.6);
    \draw[label distance=0,\cmdKV@gTypes@iB] (0.6,0.6) to[bend left=90] node[label=(90+\cmdKV@general@rotate):\cmdKV@gLabels@iLB] {} (1,0.6);
    \draw[label distance=0,\cmdKV@gTypes@iC] (1,0.6) to[bend left=90] node[label=(-90+\cmdKV@general@rotate):\cmdKV@gLabels@iLC] {} (0.6,0.6);
    \draw[label distance=0,\cmdKV@gTypes@iD] (1,0.6) -- node[label=(90+\cmdKV@general@rotate):\cmdKV@gLabels@iLD] {} (1.3,0.6);
    \draw[label distance=0,\cmdKV@gTypes@iE] (1.3,0.6) to[bend left=90] node[label=(90+\cmdKV@general@rotate):\cmdKV@gLabels@iLE] {} (1.7,0.6);
    \draw[label distance=0,\cmdKV@gTypes@iF] (1.7,0.6) to[bend left=90] node[label=(-90+\cmdKV@general@rotate):\cmdKV@gLabels@iLF] {} (1.3,0.6);
    \draw[label distance=0,\cmdKV@gTypes@iG] (1.7,0.6) -- node[label=(90+\cmdKV@general@rotate):\cmdKV@gLabels@iLG] {} (2,0.6);
    #2
  \end{tikzpicture}
  \gSetStdTypes{line}
}

\def\gBoxZ{\@ifnextchar[\@gBoxZ{\@gBoxZ[]}}
\def\@gBoxZ[#1]#2{%
  \setkeys*{pre}{#1}%
  \setrmkeys{general,gTypes,gLabels}%
  \begin{tikzpicture}
    [line width=1pt,
    baseline={([yshift=-0.5ex+\cmdKV@general@yshift]current bounding box.center)},
    scale=\cmdKV@general@scale,
    rotate=\cmdKV@general@rotate,
    font=\cmdKV@general@font]
    \draw[\cmdKV@gTypes@eA] (0.15,0.15) -- (-0.1,-0.1) node[label=(-180+\cmdKV@general@rotate):\cmdKV@gLabels@eLA] {};
    \draw[\cmdKV@gTypes@eB] (0.15,0.85) -- (-0.1,1.1) node[label=(180+\cmdKV@general@rotate):\cmdKV@gLabels@eLB] {};
    \draw[\cmdKV@gTypes@eC] (0.85,0.85) -- (1.1,1.1) node[label=(\cmdKV@general@rotate):\cmdKV@gLabels@eLC] {};
    \draw[\cmdKV@gTypes@eD] (0.85,0.15) -- (1.1,-0.1) node[label=(\cmdKV@general@rotate):\cmdKV@gLabels@eLD] {};
    \draw[label distance=0,\cmdKV@gTypes@iA] (0.15,0.15) -- node[label=(180+\cmdKV@general@rotate):\cmdKV@gLabels@iLA] {} (0.15,0.85);
    \draw[label distance=0,\cmdKV@gTypes@iB] (0.15,0.85) -- node[label=(90+\cmdKV@general@rotate):\cmdKV@gLabels@iLB] {} (0.85,0.85);
    \draw[label distance=0,\cmdKV@gTypes@iC] (0.85,0.85) -- node[label=(\cmdKV@general@rotate):\cmdKV@gLabels@iLC] {} (0.85,0.15);
    \draw[label distance=0,\cmdKV@gTypes@iD] (0.85,0.15) -- node[label=(-90+\cmdKV@general@rotate):\cmdKV@gLabels@iLD] {} (0.15,0.15);
    \draw[label distance=-4,\cmdKV@gTypes@iE] (0.15,0.15) -- node[label=(135+\cmdKV@general@rotate):\cmdKV@gLabels@iLE] {} (0.85,0.85);
    #2
  \end{tikzpicture}
  \gSetStdTypes{line}
}

\def\gBoxBoxBoxA{\@ifnextchar[\@gBoxBoxBoxA{\@gBoxBoxBoxA[]}}
\def\@gBoxBoxBoxA[#1]#2{%
  \setkeys*{pre}{#1}%
  \setrmkeys{general,gTypes,gLabels}%
  \begin{tikzpicture}
    [line width=1pt,
    baseline={([yshift=-0.5ex+\cmdKV@general@yshift]current bounding box.center)},
    scale=\cmdKV@general@scale,
    rotate=\cmdKV@general@rotate,
    font=\cmdKV@general@font]
    \draw[\cmdKV@gTypes@eA] (-0.3,0.3) -- (-0.6,0) node[label=(-180+\cmdKV@general@rotate):\cmdKV@gLabels@eLA] {};
    \draw[\cmdKV@gTypes@eB] (-0.3,0.9) -- (-0.6,1.2) node[label=(180+\cmdKV@general@rotate):\cmdKV@gLabels@eLB] {};
    \draw[\cmdKV@gTypes@eC] (1.5,0.9) -- (1.8,1.2) node[label=(\cmdKV@general@rotate):\cmdKV@gLabels@eLC] {};
    \draw[\cmdKV@gTypes@eD] (1.5,0.3) -- (1.8,0) node[label=(\cmdKV@general@rotate):\cmdKV@gLabels@eLD] {};
    \draw[label distance=0,\cmdKV@gTypes@iA] (-0.3,0.3) -- node[label=(180+\cmdKV@general@rotate):\cmdKV@gLabels@iLA] {} (-0.3,0.9);
    \draw[label distance=0,\cmdKV@gTypes@iB] (-0.3,0.9) -- node[label=(90+\cmdKV@general@rotate):\cmdKV@gLabels@iLB] {} (0.3,0.9);
    \draw[label distance=0,\cmdKV@gTypes@iC] (0.3,0.9) -- node[label=(90+\cmdKV@general@rotate):\cmdKV@gLabels@iLC] {} (0.9,0.9);
    \draw[label distance=0,\cmdKV@gTypes@iD] (0.9,0.9) -- node[label=(90+\cmdKV@general@rotate):\cmdKV@gLabels@iLD] {} (1.5,0.9);
    \draw[label distance=0,\cmdKV@gTypes@iE] (1.5,0.9) -- node[label=(\cmdKV@general@rotate):\cmdKV@gLabels@iLE] {} (1.5,0.3);
    \draw[label distance=0,\cmdKV@gTypes@iF] (1.5,0.3) -- node[label=(-90+\cmdKV@general@rotate):\cmdKV@gLabels@iLF] {} (0.9,0.3);
    \draw[label distance=0,\cmdKV@gTypes@iG] (0.9,0.3) -- node[label=(-90+\cmdKV@general@rotate):\cmdKV@gLabels@iLG] {} (0.3,0.3);
    \draw[label distance=0,\cmdKV@gTypes@iH] (0.3,0.3) -- node[label=(-90+\cmdKV@general@rotate):\cmdKV@gLabels@iLH] {} (-0.3,0.3);
    \draw[label distance=0,\cmdKV@gTypes@iI] (0.3,0.9) -- node[label=(\cmdKV@general@rotate):\cmdKV@gLabels@iLI] {} (0.3,0.3);
    \draw[label distance=0,\cmdKV@gTypes@iJ] (0.9,0.9) -- node[label=(\cmdKV@general@rotate):\cmdKV@gLabels@iLJ] {} (0.9,0.3);
    #2
  \end{tikzpicture}
  \gSetStdTypes{line}
}

\def\gBoxBoxBoxB{\@ifnextchar[\@gBoxBoxBoxB{\@gBoxBoxBoxB[]}}
\def\@gBoxBoxBoxB[#1]#2{%
  \setkeys*{pre}{#1}%
  \setrmkeys{general,gTypes,gLabels}%
  \begin{tikzpicture}
    [line width=1pt,
    baseline={([yshift=-0.5ex+\cmdKV@general@yshift]current bounding box.center)},
    scale=\cmdKV@general@scale,
    rotate=\cmdKV@general@rotate,
    font=\cmdKV@general@font]
    \draw[\cmdKV@gTypes@eA] (0.3,0.3) -- (0,0) node[label=(-180+\cmdKV@general@rotate):\cmdKV@gLabels@eLA] {};
    \draw[\cmdKV@gTypes@eB] (0.3,1.5) -- (0,1.8) node[label=(180+\cmdKV@general@rotate):\cmdKV@gLabels@eLB] {};
    \draw[\cmdKV@gTypes@eC] (1.5,1.5) -- (1.8,1.8) node[label=(\cmdKV@general@rotate):\cmdKV@gLabels@eLC] {};
    \draw[\cmdKV@gTypes@eD] (1.5,0.3) -- (1.8,0) node[label=(\cmdKV@general@rotate):\cmdKV@gLabels@eLD] {};
    \draw[label distance=0,\cmdKV@gTypes@iA] (0.3,0.3) -- node[label=(180+\cmdKV@general@rotate):\cmdKV@gLabels@iLA] {} (0.3,0.9);
    \draw[label distance=0,\cmdKV@gTypes@iB] (0.3,0.9) -- node[label=(180+\cmdKV@general@rotate):\cmdKV@gLabels@iLB] {} (0.3,1.5);
    \draw[label distance=0,\cmdKV@gTypes@iC] (0.3,1.5) -- node[label=(90+\cmdKV@general@rotate):\cmdKV@gLabels@iLC] {} (0.9,1.5);
    \draw[label distance=0,\cmdKV@gTypes@iD] (0.9,1.5) -- node[label=(90+\cmdKV@general@rotate):\cmdKV@gLabels@iLD] {} (1.5,1.5);
    \draw[label distance=0,\cmdKV@gTypes@iE] (1.5,1.5) -- node[label=(\cmdKV@general@rotate):\cmdKV@gLabels@iLE] {} (1.5,0.3);
    \draw[label distance=0,\cmdKV@gTypes@iF] (1.5,0.3) -- node[label=(-90+\cmdKV@general@rotate):\cmdKV@gLabels@iLF] {} (0.9,0.3);
    \draw[label distance=0,\cmdKV@gTypes@iG] (0.9,0.3) -- node[label=(-90+\cmdKV@general@rotate):\cmdKV@gLabels@iLG] {} (0.3,0.3);
    \draw[label distance=0,\cmdKV@gTypes@iH] (0.9,1.5) -- node[label=(\cmdKV@general@rotate):\cmdKV@gLabels@iLH] {} (0.9,0.9);
    \draw[label distance=0,\cmdKV@gTypes@iI] (0.9,0.9) -- node[label=(\cmdKV@general@rotate):\cmdKV@gLabels@iLI] {} (0.9,0.3);
    \draw[label distance=0,\cmdKV@gTypes@iJ] (0.9,0.9) -- node[label=(90+\cmdKV@general@rotate):\cmdKV@gLabels@iLJ] {} (0.3,0.9);
    #2
  \end{tikzpicture}
  \gSetStdTypes{line}
}

\def\gEFTTree{\@ifnextchar[\@gEFTTree{\@gEFTTree[]}}
\def\@gEFTTree[#1]#2{%
  \setkeys*{pre}{#1}%
  \setrmkeys{general,gTypes,gLabels}%
  \begin{tikzpicture}
    [line width=1pt,
    baseline={([yshift=-0.5ex+\cmdKV@general@yshift]current bounding box.center)},
    scale=\cmdKV@general@scale,
    rotate=\cmdKV@general@rotate,
    font=\cmdKV@general@font]
    \draw[\cmdKV@gTypes@iA] (0,0.5) node[label=(90+\cmdKV@general@rotate):\cmdKV@gLabels@eLA] {} -- node[label=(\cmdKV@general@rotate):\cmdKV@gLabels@iLA] {} (0,-0.5) node[label=(-90+\cmdKV@general@rotate):\cmdKV@gLabels@eLB] {};
    \draw[fill] (0,0.5) circle (0.02);
    \draw[fill] (0,-0.5) circle (0.02);
    #2
  \end{tikzpicture}
  \gSetStdTypes{line}
}

\def\gEFTV{\@ifnextchar[\@gEFTV{\@gEFTV[]}}
\def\@gEFTV[#1]#2{%
  \setkeys*{pre}{#1}%
  \setrmkeys{general,gTypes,gLabels}%
  \begin{tikzpicture}
    [line width=1pt,
    baseline={([yshift=-0.5ex+\cmdKV@general@yshift]current bounding box.center)},
    scale=\cmdKV@general@scale,
    rotate=\cmdKV@general@rotate,
    font=\cmdKV@general@font]
    \draw[\cmdKV@gTypes@iA] (0,0.5) node[label=(90+\cmdKV@general@rotate):\cmdKV@gLabels@eLA] {} -- node[label=(200+\cmdKV@general@rotate):\cmdKV@gLabels@iLA] {} (0.3,-0.5) node[label=(-90+\cmdKV@general@rotate):\cmdKV@gLabels@eLC] {};
    \draw[\cmdKV@gTypes@iB] (0.6,0.5) node[label=(90+\cmdKV@general@rotate):\cmdKV@gLabels@eLB] {} -- node[label=(-20+\cmdKV@general@rotate):\cmdKV@gLabels@iLB] {} (0.3,-0.5);
    \draw[fill] (0,0.5) circle (0.02);
    \draw[fill] (0.3,-0.5) circle (0.02);
    \draw[fill] (0.6,0.5) circle (0.02);
    #2
  \end{tikzpicture}
  \gSetStdTypes{line}
}

\def\gEFTY{\@ifnextchar[\@gEFTY{\@gEFTY[]}}
\def\@gEFTY[#1]#2{%
  \setkeys*{pre}{#1}%
  \setrmkeys{general,gTypes,gLabels}%
  \begin{tikzpicture}
    [line width=1pt,
    baseline={([yshift=-0.5ex+\cmdKV@general@yshift]current bounding box.center)},
    scale=\cmdKV@general@scale,
    rotate=\cmdKV@general@rotate,
    font=\cmdKV@general@font]
    \draw[\cmdKV@gTypes@iA] (0,0.5) node[label=(90+\cmdKV@general@rotate):\cmdKV@gLabels@eLA] {} -- node[label=(225+\cmdKV@general@rotate):\cmdKV@gLabels@iLA] {} (0.3,0);
    \draw[\cmdKV@gTypes@iB] (0.6,0.5) node[label=(90+\cmdKV@general@rotate):\cmdKV@gLabels@eLB] {} -- node[label=(-45+\cmdKV@general@rotate):\cmdKV@gLabels@iLB] {} (0.3,0);
    \draw[\cmdKV@gTypes@iC] (0.3,0) -- node[label=(\cmdKV@general@rotate):\cmdKV@gLabels@iLC] {} (0.3,-0.5) node[label=(-90+\cmdKV@general@rotate):\cmdKV@gLabels@eLC] {};
    \draw[fill] (0,0.5) circle (0.02);
    \draw[fill] (0.6,0.5) circle (0.02);
    \draw[fill] (0.3,-0.5) circle (0.02);
    #2
  \end{tikzpicture}
  \gSetStdTypes{line}
}

\def\gEFTPsi{\@ifnextchar[\@gEFTPsi{\@gEFTPsi[]}}
\def\@gEFTPsi[#1]#2{%
  \setkeys*{pre}{#1}%
  \setrmkeys{general,gTypes,gLabels}%
  \begin{tikzpicture}
    [line width=1pt,
    baseline={([yshift=-0.5ex+\cmdKV@general@yshift]current bounding box.center)},
    scale=\cmdKV@general@scale,
    rotate=\cmdKV@general@rotate,
    font=\cmdKV@general@font]
    \draw[\cmdKV@gTypes@iA] (0,0.5) node[label=(90+\cmdKV@general@rotate):\cmdKV@gLabels@eLA] {} -- node[label=(225+\cmdKV@general@rotate):\cmdKV@gLabels@iLA] {} (0.5,0);
    \draw[\cmdKV@gTypes@iB] (0.5,0.5) node[label=(90+\cmdKV@general@rotate):\cmdKV@gLabels@eLB] {} -- node[label=(\cmdKV@general@rotate):\cmdKV@gLabels@iLB] {} (0.5,0);
    \draw[\cmdKV@gTypes@iC] (1,0.5) node[label=(90+\cmdKV@general@rotate):\cmdKV@gLabels@eLC] {} -- node[label=(-45+\cmdKV@general@rotate):\cmdKV@gLabels@iLC] {} (0.5,0);
    \draw[\cmdKV@gTypes@iD] (0.5,0) -- node[label=(\cmdKV@general@rotate):\cmdKV@gLabels@iLD] {} (0.5,-0.5) node[label=(-90+\cmdKV@general@rotate):\cmdKV@gLabels@eLD] {};
    \draw[fill] (0,0.5) circle (0.02);
    \draw[fill] (0.5,0.5) circle (0.02);
    \draw[fill] (1,0.5) circle (0.02);
    \draw[fill] (0.5,-0.5) circle (0.02);
    #2
  \end{tikzpicture}
  \gSetStdTypes{line}
}

\def\gEFTVinY{\@ifnextchar[\@gEFTVinY{\@gEFTVinY[]}}
\def\@gEFTVinY[#1]#2{%
  \setkeys*{pre}{#1}%
  \setrmkeys{general,gTypes,gLabels}%
  \begin{tikzpicture}
    [line width=1pt,
    baseline={([yshift=-0.5ex+\cmdKV@general@yshift]current bounding box.center)},
    scale=\cmdKV@general@scale,
    rotate=\cmdKV@general@rotate,
    font=\cmdKV@general@font]
    \draw[\cmdKV@gTypes@iA] (0,0.5) node[label=(90+\cmdKV@general@rotate):\cmdKV@gLabels@eLA] {} -- node[label=(225+\cmdKV@general@rotate):\cmdKV@gLabels@iLA] {} (0.25,0.25);
    \draw[\cmdKV@gTypes@iE] (0.25,0.25) -- node[label=(225+\cmdKV@general@rotate):\cmdKV@gLabels@iLE] {} (0.5,0);
    \draw[\cmdKV@gTypes@iB] (0.5,0.5) node[label=(90+\cmdKV@general@rotate):\cmdKV@gLabels@eLB] {} -- node[label=(\cmdKV@general@rotate):\cmdKV@gLabels@iLB] {} (0.25,0.25);
    \draw[\cmdKV@gTypes@iC] (1,0.5) node[label=(90+\cmdKV@general@rotate):\cmdKV@gLabels@eLC] {} -- node[label=(-45+\cmdKV@general@rotate):\cmdKV@gLabels@iLC] {} (0.5,0);
    \draw[\cmdKV@gTypes@iD] (0.5,0) -- node[label=(\cmdKV@general@rotate):\cmdKV@gLabels@iLD] {} (0.5,-0.5) node[label=(-90+\cmdKV@general@rotate):\cmdKV@gLabels@eLD] {};
    \draw[fill] (0,0.5) circle (0.02);
    \draw[fill] (0.5,0.5) circle (0.02);
    \draw[fill] (1,0.5) circle (0.02);
    \draw[fill] (0.5,-0.5) circle (0.02);
    #2
  \end{tikzpicture}
  \gSetStdTypes{line}
}

\def\gEFTYinvY{\@ifnextchar[\@gEFTYinvY{\@gEFTYinvY[]}}
\def\@gEFTYinvY[#1]#2{%
  \setkeys*{pre}{#1}%
  \setrmkeys{general,gTypes,gLabels}%
  \begin{tikzpicture}
    [line width=1pt,
    baseline={([yshift=-0.5ex+\cmdKV@general@yshift]current bounding box.center)},
    scale=\cmdKV@general@scale,
    rotate=\cmdKV@general@rotate,
    font=\cmdKV@general@font]
    \draw[\cmdKV@gTypes@iA] (0,0.5) node[label=(90+\cmdKV@general@rotate):\cmdKV@gLabels@eLA] {} -- node[label=(225+\cmdKV@general@rotate):\cmdKV@gLabels@iLA] {} (0.3,0.166);
    \draw[\cmdKV@gTypes@iB] (0.6,0.5) node[label=(90+\cmdKV@general@rotate):\cmdKV@gLabels@eLB] {} -- node[label=(-45+\cmdKV@general@rotate):\cmdKV@gLabels@iLB] {} (0.3,0.166);
    \draw[\cmdKV@gTypes@iC] (0.3,-0.166) -- node[label=(135+\cmdKV@general@rotate):\cmdKV@gLabels@iLC] {} (0,-0.5) node[label=(-90+\cmdKV@general@rotate):\cmdKV@gLabels@eLC] {}; 
    \draw[\cmdKV@gTypes@iD] (0.3,-0.166) -- node[label=(45+\cmdKV@general@rotate):\cmdKV@gLabels@iLD] {} (0.6,-0.5) node[label=(-90+\cmdKV@general@rotate):\cmdKV@gLabels@eLD] {};
    \draw[\cmdKV@gTypes@iE] (0.3,0.166) -- node[label=(\cmdKV@general@rotate):\cmdKV@gLabels@iLE] {} (0.3,-0.166);
    \draw[fill] (0,0.5) circle (0.02);
    \draw[fill] (0.6,0.5) circle (0.02);
    \draw[fill] (0,-0.5) circle (0.02);
    \draw[fill] (0.6,-0.5) circle (0.02);
    #2
  \end{tikzpicture}
  \gSetStdTypes{line}
}

\def\gEFTX{\@ifnextchar[\@gEFTX{\@gEFTX[]}}
\def\@gEFTX[#1]#2{%
  \setkeys*{pre}{#1}%
  \setrmkeys{general,gTypes,gLabels}%
  \begin{tikzpicture}
    [line width=1pt,
    baseline={([yshift=-0.5ex+\cmdKV@general@yshift]current bounding box.center)},
    scale=\cmdKV@general@scale,
    rotate=\cmdKV@general@rotate,
    font=\cmdKV@general@font]
    \draw[\cmdKV@gTypes@iA] (0,0.5) node[label=(90+\cmdKV@general@rotate):\cmdKV@gLabels@eLA] {} -- node[label=(225+\cmdKV@general@rotate):\cmdKV@gLabels@iLA] {} (0.3,0);
    \draw[\cmdKV@gTypes@iB] (0.6,0.5) node[label=(90+\cmdKV@general@rotate):\cmdKV@gLabels@eLB] {} -- node[label=(-45+\cmdKV@general@rotate):\cmdKV@gLabels@iLB] {} (0.3,0);
    \draw[\cmdKV@gTypes@iC] (0.3,0) -- node[label=(135+\cmdKV@general@rotate):\cmdKV@gLabels@iLC] {} (0,-0.5) node[label=(-90+\cmdKV@general@rotate):\cmdKV@gLabels@eLC] {}; 
    \draw[\cmdKV@gTypes@iD] (0.3,0) -- node[label=(45+\cmdKV@general@rotate):\cmdKV@gLabels@iLD] {} (0.6,-0.5) node[label=(-90+\cmdKV@general@rotate):\cmdKV@gLabels@eLD] {};
    \draw[fill] (0,0.5) circle (0.02);
    \draw[fill] (0.6,0.5) circle (0.02);
    \draw[fill] (0,-0.5) circle (0.02);
    \draw[fill] (0.6,-0.5) circle (0.02);
    #2
  \end{tikzpicture}
  \gSetStdTypes{line}
}

\def\gEFTH{\@ifnextchar[\@gEFTH{\@gEFTH[]}}
\def\@gEFTH[#1]#2{%
  \setkeys*{pre}{#1}%
  \setrmkeys{general,gTypes,gLabels}%
  \begin{tikzpicture}
    [line width=1pt,
    baseline={([yshift=-0.5ex+\cmdKV@general@yshift]current bounding box.center)},
    scale=\cmdKV@general@scale,
    rotate=\cmdKV@general@rotate,
    font=\cmdKV@general@font]
    \draw[\cmdKV@gTypes@iA] (0,0.5) node[label=(90+\cmdKV@general@rotate):\cmdKV@gLabels@eLA] {} -- node[label=(180+\cmdKV@general@rotate):\cmdKV@gLabels@iLA] {} (0,0);
    \draw[\cmdKV@gTypes@iB] (0.6,0.5) node[label=(90+\cmdKV@general@rotate):\cmdKV@gLabels@eLB] {} -- node[label=(\cmdKV@general@rotate):\cmdKV@gLabels@iLB] {} (0.6,0);
    \draw[\cmdKV@gTypes@iC] (0.0,0) -- node[label=(180+\cmdKV@general@rotate):\cmdKV@gLabels@iLC] {} (0,-0.5) node[label=(-90+\cmdKV@general@rotate):\cmdKV@gLabels@eLC] {}; 
    \draw[\cmdKV@gTypes@iD] (0.6,0) -- node[label=(\cmdKV@general@rotate):\cmdKV@gLabels@iLD] {} (0.6,-0.5) node[label=(-90+\cmdKV@general@rotate):\cmdKV@gLabels@eLD] {};
    \draw[\cmdKV@gTypes@iE] (0,0) -- node[label=(90+\cmdKV@general@rotate):\cmdKV@gLabels@iLE] {} (0.6,0);
    \draw[fill] (0,0.5) circle (0.02);
    \draw[fill] (0.6,0.5) circle (0.02);
    \draw[fill] (0,-0.5) circle (0.02);
    \draw[fill] (0.6,-0.5) circle (0.02);
    #2
  \end{tikzpicture}
  \gSetStdTypes{line}
}

\def\cTree{\@ifnextchar[\@cTree{\@cTree[]}}
\def\@cTree[#1]#2{%
  \setkeys*{pre}{#1}%
  \setrmkeys{general,gTypes,gLabels,blobs}%
  \begin{tikzpicture}
    [line width=1pt,
    baseline={([yshift=-0.5ex+\cmdKV@general@yshift]current bounding box.center)},
    scale=\cmdKV@general@scale,
    rotate=\cmdKV@general@rotate,
    font=\cmdKV@general@font]
    \fill[\cmdKV@blobs@blobA] (0,0) ellipse (0.25 and 0.25);
    \draw[\cmdKV@gTypes@eA] ($ (0,0) + (-135:0.25) $) -- ($ (0,0) + (-135:0.6) $) node[label=(-180+\cmdKV@general@rotate):\cmdKV@gLabels@eLA] {};
    \draw[\cmdKV@gTypes@eB] ($ (0,0) + (135:0.25) $) -- ($ (0,0) + (135:0.6) $) node[label=(180+\cmdKV@general@rotate):\cmdKV@gLabels@eLB] {};
    \draw[\cmdKV@gTypes@eC] ($ (0,0) + (45:0.25) $) -- ($ (0,0) + (45:0.6) $) node[label=(\cmdKV@general@rotate):\cmdKV@gLabels@eLC] {};
    \draw[\cmdKV@gTypes@eD] ($ (0,0) + (-45:0.25) $) -- ($ (0,0) + (-45:0.6) $) node[label=(\cmdKV@general@rotate):\cmdKV@gLabels@eLD] {};
    \node[label=(90+\cmdKV@general@rotate):\cmdKV@gLabels@iLA] at (0,0.3) {};
    \node[label=(\cmdKV@general@rotate):\cmdKV@gLabels@iLB] at (0.3,0) {};
    #2
  \end{tikzpicture}
  \gSetStdTypes{line}
}

\def\cBoxA{\@ifnextchar[\@cBoxA{\@cBoxA[]}}
\def\@cBoxA[#1]#2{%
  \setkeys*{pre}{#1}%
  \setrmkeys{general,gTypes,gLabels,dots,blobs}%
  \begin{tikzpicture}
    [line width=1pt,
    baseline={([yshift=-0.5ex+\cmdKV@general@yshift]current bounding box.center)},
    scale=\cmdKV@general@scale,
    rotate=\cmdKV@general@rotate,
    font=\cmdKV@general@font]
    \fill[\cmdKV@blobs@blobA] (-0.4,0) ellipse (0.2 and 0.4);
    \fill[\cmdKV@blobs@blobB] (0.4,0) ellipse (0.2 and 0.4);
    \draw[\cmdKV@gTypes@eA] ($ (-0.4,0) + (-135:0.26) $) -- ($ (-0.4,0) + (-135:0.7) $) node[label=(-180+\cmdKV@general@rotate):\cmdKV@gLabels@eLA] {};
    \draw[\cmdKV@gTypes@eB] ($ (-0.4,0) + (135:0.26) $) -- ($ (-0.4,0) + (135:0.7) $) node[label=(180+\cmdKV@general@rotate):\cmdKV@gLabels@eLB] {};
    \draw[\cmdKV@gTypes@eC] ($ (0.4,0) + (45:0.26) $) -- ($ (0.4,0) + (45:0.7) $) node[label=(\cmdKV@general@rotate):\cmdKV@gLabels@eLC] {};
    \draw[\cmdKV@gTypes@eD] ($ (0.4,0) + (-45:0.26) $) -- ($ (0.4,0) + (-45:0.7) $) node[label=(\cmdKV@general@rotate):\cmdKV@gLabels@eLD] {};
    \draw[label distance=1,\cmdKV@gTypes@iA] ($ (-0.4,0) + (45:0.26) $) -- node[label=(90+\cmdKV@general@rotate):\cmdKV@gLabels@iLA] {}($ (0.4,0) + (135:0.26) $);
    \draw[label distance=1,\cmdKV@gTypes@iB] ($ (0.4,0) + (-135:0.26) $) -- node[label=(-90+\cmdKV@general@rotate):\cmdKV@gLabels@iLB] {}($ (-0.4,0) + (-45:0.26) $);
    \node[label=(180+\cmdKV@general@rotate):\cmdKV@gLabels@iLC] at (-0.6,0) {};
    \node[label=(\cmdKV@general@rotate):\cmdKV@gLabels@iLD] at (0.6,0) {};
    \ifKV@dots@lDots
      \draw[dotted] ($(-0.4,0) + (-150:0.7)$) to[bend left] ($(-0.4,0) + (150:0.7)$);%
    \fi
    \ifKV@dots@rDots
      \draw[dotted] ($(0.4,0) + (30:0.7)$) to[bend left] ($(0.4,0) + (-30:0.7)$);%
    \fi
    #2
  \end{tikzpicture}
  \gSetStdTypes{line}
}

\def\cBoxAA{\@ifnextchar[\@cBoxAA{\@cBoxAA[]}}
\def\@cBoxAA[#1]#2{%
  \setkeys*{pre}{#1}%
  \setrmkeys{general,gTypes,gLabels,dots,blobs}%
  \begin{tikzpicture}
    [line width=1pt,
    baseline={([yshift=-0.5ex+\cmdKV@general@yshift]current bounding box.center)},
    scale=\cmdKV@general@scale,
    rotate=\cmdKV@general@rotate,
    font=\cmdKV@general@font]
    \fill[\cmdKV@blobs@blobA] (-0.4,0) ellipse (0.2 and 0.4);
    \fill[\cmdKV@blobs@blobB] (0.4,0) ellipse (0.2 and 0.4);
    \draw[\cmdKV@gTypes@eA] ($ (-0.4,0) + (-135:0.26) $) -- ($ (-0.4,0) + (-135:0.7) $) node[label=(-180+\cmdKV@general@rotate):\cmdKV@gLabels@eLA] {};
    \draw[\cmdKV@gTypes@eB] ($ (-0.4,0) + (135:0.26) $) -- ($ (-0.4,0) + (135:0.7) $) node[label=(180+\cmdKV@general@rotate):\cmdKV@gLabels@eLB] {};
    \draw[\cmdKV@gTypes@eC] ($ (0.4,0) + (45:0.26) $) -- ($ (0.4,0) + (45:0.7) $) node[label=(\cmdKV@general@rotate):\cmdKV@gLabels@eLC] {};
    \draw[\cmdKV@gTypes@eD] ($ (0.4,0) + (-45:0.26) $) -- ($ (0.4,0) + (-45:0.7) $) node[label=(\cmdKV@general@rotate):\cmdKV@gLabels@eLD] {};
    \draw[label distance=1,\cmdKV@gTypes@iA] ($ (-0.4,0) + (60:0.32) $) to[bend left] node[label=(90+\cmdKV@general@rotate):\cmdKV@gLabels@iLA] {}($ (0.4,0) + (120:0.32) $);
    \draw[label distance=1,\cmdKV@gTypes@iB] ($ (0.4,0) + (-120:0.32) $) to[bend left]  node[label=(-90+\cmdKV@general@rotate):\cmdKV@gLabels@iLB] {}($ (-0.4,0) + (-60:0.32) $);
    \draw[dotted] (0,0.25) -- (0,-0.25);
    \node[label=(180+\cmdKV@general@rotate):\cmdKV@gLabels@iLC] at (-0.6,0) {};
    \node[label=(\cmdKV@general@rotate):\cmdKV@gLabels@iLD] at (0.6,0) {};
    \ifKV@dots@lDots
      \draw[dotted] ($(-0.4,0) + (-150:0.7)$) to[bend left] ($(-0.4,0) + (150:0.7)$);%
    \fi
    \ifKV@dots@rDots
      \draw[dotted] ($(0.4,0) + (30:0.7)$) to[bend left] ($(0.4,0) + (-30:0.7)$);%
    \fi
    #2
  \end{tikzpicture}
  \gSetStdTypes{line}
}

\def\cBoxB{\@ifnextchar[\@cBoxB{\@cBoxB[]}}
\def\@cBoxB[#1]#2{%
  \setkeys*{pre}{#1}%
  \setrmkeys{general,gTypes,gLabels,dots,blobs}%
  \begin{tikzpicture}
    [line width=1pt,
    baseline={([yshift=-0.5ex+\cmdKV@general@yshift]current bounding box.center)},
    scale=\cmdKV@general@scale,
    rotate=\cmdKV@general@rotate,
    font=\cmdKV@general@font]
    \fill[\cmdKV@blobs@blobA] (0,-0.3) ellipse (0.4 and 0.2);
    \fill[\cmdKV@blobs@blobB] (0,0.3) ellipse (0.4 and 0.2);
    \draw[\cmdKV@gTypes@eA] ($ (0,-0.3) + (180:0.4) $) -- ($ (0,-0.3) + (-170:0.7) $) node[label=(180+\cmdKV@general@rotate):\cmdKV@gLabels@eLA] {};
    \draw[\cmdKV@gTypes@eB] ($ (0,0.3) + (180:0.4) $) -- ($ (0,0.3) + (170:0.7) $) node[label=(180+\cmdKV@general@rotate):\cmdKV@gLabels@eLB] {};
    \draw[\cmdKV@gTypes@eC] ($ (0,0.3) + (0:0.4) $) -- ($ (0,0.3) + (10:0.7) $) node[label=(\cmdKV@general@rotate):\cmdKV@gLabels@eLC] {};
    \draw[\cmdKV@gTypes@eD] ($ (0,-0.3) + (0:0.4) $) -- ($ (0,-0.3) + (-10:0.7) $) node[label=(\cmdKV@general@rotate):\cmdKV@gLabels@eLD] {};
    \draw[label distance=1,\cmdKV@gTypes@iA] ($ (0,-0.3) + (135:0.26) $) -- node[label=(-180+\cmdKV@general@rotate):\cmdKV@gLabels@iLA] {}($ (0,0.3) + (-135:0.26) $);
    \draw[label distance=1,\cmdKV@gTypes@iB] ($ (0,0.3) + (-45:0.26) $) -- node[label=(\cmdKV@general@rotate):\cmdKV@gLabels@iLB] {}($ (0,-0.3) + (45:0.26) $);
    \node[label=(90+\cmdKV@general@rotate):\cmdKV@gLabels@iLC] at (0,0.5) {};
    \node[label=(-90+\cmdKV@general@rotate):\cmdKV@gLabels@iLD] at (0,-0.5) {};
    #2
  \end{tikzpicture}
  \gSetStdTypes{line}
}

\def\cBoxBoxA{\@ifnextchar[\@cBoxBoxA{\@cBoxBoxA[]}}
\def\@cBoxBoxA[#1]#2{%
  \setkeys*{pre}{#1}%
  \setrmkeys{general,gTypes,gLabels,dots,blobs}%
  \begin{tikzpicture}
    [line width=1pt,
    baseline={([yshift=-0.5ex+\cmdKV@general@yshift]current bounding box.center)},
    scale=\cmdKV@general@scale,
    rotate=\cmdKV@general@rotate,
    font=\cmdKV@general@font]
    \fill[\cmdKV@blobs@blobA] (-0.4,0) ellipse (0.2 and 0.4);
    \fill[\cmdKV@blobs@blobB] (0.4,0) ellipse (0.2 and 0.4);
    \fill[\cmdKV@blobs@blobC] (1.2,0) ellipse (0.2 and 0.4);
    \draw[\cmdKV@gTypes@eA] ($ (-0.4,0) + (-135:0.26) $) -- ($ (-0.4,0) + (-135:0.7) $) node[label=(-180+\cmdKV@general@rotate):\cmdKV@gLabels@eLA] {};
    \draw[\cmdKV@gTypes@eB] ($ (-0.4,0) + (135:0.26) $) -- ($ (-0.4,0) + (135:0.7) $) node[label=(180+\cmdKV@general@rotate):\cmdKV@gLabels@eLB] {};
    \draw[\cmdKV@gTypes@eC] ($ (1.2,0) + (45:0.26) $) -- ($ (1.2,0) + (45:0.7) $) node[label=(\cmdKV@general@rotate):\cmdKV@gLabels@eLC] {};
    \draw[\cmdKV@gTypes@eD] ($ (1.2,0) + (-45:0.26) $) -- ($ (1.2,0) + (-45:0.7) $) node[label=(\cmdKV@general@rotate):\cmdKV@gLabels@eLD] {};
    \draw[label distance=1,\cmdKV@gTypes@iA] ($ (-0.4,0) + (45:0.26) $) -- node[label=(90+\cmdKV@general@rotate):\cmdKV@gLabels@iLA] {}($ (0.4,0) + (135:0.26) $);
    \draw[label distance=1,\cmdKV@gTypes@iB] ($ (0.4,0) + (45:0.26) $) -- node[label=(90+\cmdKV@general@rotate):\cmdKV@gLabels@iLB] {}($ (1.2,0) + (135:0.26) $);
    \draw[label distance=1,\cmdKV@gTypes@iC] ($ (1.2,0) + (-135:0.26) $) -- node[label=(-90+\cmdKV@general@rotate):\cmdKV@gLabels@iLC] {}($ (0.4,0) + (-45:0.26) $);
    \draw[label distance=1,\cmdKV@gTypes@iD] ($ (0.4,0) + (-135:0.26) $) -- node[label=(-90+\cmdKV@general@rotate):\cmdKV@gLabels@iLD] {}($ (-0.4,0) + (-45:0.26) $);
    \node[label=(180+\cmdKV@general@rotate):\cmdKV@gLabels@iLE] at (-0.6,0) {};
    \node[label=(\cmdKV@general@rotate):\cmdKV@gLabels@iLF] at (0.6,0) {};
    \node[label=(\cmdKV@general@rotate):\cmdKV@gLabels@iLG] at (1.4,0) {};
    #2
  \end{tikzpicture}
  \gSetStdTypes{line}
}

\def\cBoxBoxB{\@ifnextchar[\@cBoxBoxB{\@cBoxBoxB[]}}
\def\@cBoxBoxB[#1]#2{%
  \setkeys*{pre}{#1}%
  \setrmkeys{general,gTypes,gLabels,dots,blobs}%
  \begin{tikzpicture}
    [line width=1pt,
    baseline={([yshift=-0.5ex+\cmdKV@general@yshift]current bounding box.center)},
    scale=\cmdKV@general@scale,
    rotate=\cmdKV@general@rotate,
    font=\cmdKV@general@font]
    \fill[\cmdKV@blobs@blobA] (0,0.3) ellipse (0.4 and 0.2);
    \fill[\cmdKV@blobs@blobB] (0,-0.3) ellipse (0.4 and 0.2);
    \fill[\cmdKV@blobs@blobC] (1,0) ellipse (0.2 and 0.4);
    \draw[\cmdKV@gTypes@eA] ($ (0,-0.3) + (180:0.4) $) -- ($ (0,-0.3) + (-170:0.7) $) node[label=(180+\cmdKV@general@rotate):\cmdKV@gLabels@eLA] {};
    \draw[\cmdKV@gTypes@eB] ($ (0,0.3) + (180:0.4) $) -- ($ (0,0.3) + (170:0.7) $) node[label=(180+\cmdKV@general@rotate):\cmdKV@gLabels@eLB] {};
    \draw[\cmdKV@gTypes@eC] ($ (1,0) + (45:0.26) $) -- ($ (1,0) + (45:0.7) $) node[label=(\cmdKV@general@rotate):\cmdKV@gLabels@eLC] {};
    \draw[\cmdKV@gTypes@eD] ($ (1,0) + (-45:0.26) $) -- ($ (1,0) + (-45:0.7) $) node[label=(\cmdKV@general@rotate):\cmdKV@gLabels@eLD] {};
    \draw[label distance=1,\cmdKV@gTypes@iA] ($ (0,-0.3) + (135:0.26) $) -- node[label=(-180+\cmdKV@general@rotate):\cmdKV@gLabels@iLA] {}($ (0,0.3) + (-135:0.26) $);
    \draw[label distance=1,\cmdKV@gTypes@iB] ($ (0,0.3) + (0:0.4) $) -- node[label=(90+\cmdKV@general@rotate):\cmdKV@gLabels@iLB] {} ($ (1,0) + (135:0.26) $);
    \draw[label distance=1,\cmdKV@gTypes@iC] ($ (1,0) + (-135:0.26) $) -- node[label=(-90+\cmdKV@general@rotate):\cmdKV@gLabels@iLC] {} ($ (0,-0.3) + (0:0.4) $);
    \draw[label distance=1,\cmdKV@gTypes@iD] ($ (0,0.3) + (-45:0.26) $) -- node[label=(\cmdKV@general@rotate):\cmdKV@gLabels@iLD] {}($ (0,-0.3) + (45:0.26) $);
    \node[label=(90+\cmdKV@general@rotate):\cmdKV@gLabels@iLE] at (0,0.5) {};
    \node[label=(\cmdKV@general@rotate):\cmdKV@gLabels@iLF] at (1.2,0) {};
    \node[label=(-90+\cmdKV@general@rotate):\cmdKV@gLabels@iLG] at (0,-0.5) {};
    #2
  \end{tikzpicture}
  \gSetStdTypes{line}
}

\def\cBoxBoxC{\@ifnextchar[\@cBoxBoxC{\@cBoxBoxC[]}}
\def\@cBoxBoxC[#1]#2{%
  \setkeys*{pre}{#1}%
  \setrmkeys{general,gTypes,gLabels,dots,blobs}%
  \begin{tikzpicture}
    [line width=1pt,
    baseline={([yshift=-0.5ex+\cmdKV@general@yshift]current bounding box.center)},
    scale=\cmdKV@general@scale,
    rotate=\cmdKV@general@rotate,
    font=\cmdKV@general@font]
    \fill[\cmdKV@blobs@blobA] (-0.4,0) ellipse (0.2 and 0.4);
    \fill[\cmdKV@blobs@blobB] (0.4,0) ellipse (0.2 and 0.4);
    \draw[\cmdKV@gTypes@eA] ($ (-0.4,0) + (-135:0.26) $) -- ($ (-0.4,0) + (-135:0.7) $) node[label=(-180+\cmdKV@general@rotate):\cmdKV@gLabels@eLA] {};
    \draw[\cmdKV@gTypes@eB] ($ (-0.4,0) + (135:0.26) $) -- ($ (-0.4,0) + (135:0.7) $) node[label=(180+\cmdKV@general@rotate):\cmdKV@gLabels@eLB] {};
    \draw[\cmdKV@gTypes@eC] ($ (0.4,0) + (45:0.26) $) -- ($ (0.4,0) + (45:0.7) $) node[label=(\cmdKV@general@rotate):\cmdKV@gLabels@eLC] {};
    \draw[\cmdKV@gTypes@eD] ($ (0.4,0) + (-45:0.26) $) -- ($ (0.4,0) + (-45:0.7) $) node[label=(\cmdKV@general@rotate):\cmdKV@gLabels@eLD] {};
    \draw[label distance=1,\cmdKV@gTypes@iA] ($ (-0.4,0) + (60:0.32) $) to[bend left] node[label=(90+\cmdKV@general@rotate):\cmdKV@gLabels@iLA] {}($ (0.4,0) + (120:0.32) $);
    \draw[label distance=1,\cmdKV@gTypes@iB] ($ (0.4,0) + (-120:0.32) $) to[bend left]  node[label=(-90+\cmdKV@general@rotate):\cmdKV@gLabels@iLB] {}($ (-0.4,0) + (-60:0.32) $);
    \draw[label distance=-2,\cmdKV@gTypes@iC] ($ (-0.4,0) + (0:0.2) $) -- node[label=(90+\cmdKV@general@rotate):\cmdKV@gLabels@iLC] {}($ (0.4,0) + (180:0.2) $);
    \node[label=(180+\cmdKV@general@rotate):\cmdKV@gLabels@iLC] at (-0.6,0) {};
    \node[label=(\cmdKV@general@rotate):\cmdKV@gLabels@iLD] at (0.6,0) {};
    \ifKV@dots@lDots
      \draw[dotted] ($(-0.4,0) + (-150:0.7)$) to[bend left] ($(-0.4,0) + (150:0.7)$);%
    \fi
    \ifKV@dots@rDots
      \draw[dotted] ($(0.4,0) + (30:0.7)$) to[bend left] ($(0.4,0) + (-30:0.7)$);%
    \fi
    #2
  \end{tikzpicture}
  \gSetStdTypes{line}
}

\def\cNPBox{\@ifnextchar[\@cNPBox{\@cNPBox[]}}
\def\@cNPBox[#1]#2{%
  \setkeys*{pre}{#1}%
  \setrmkeys{general,gTypes,gLabels,dots,blobs}%
  \begin{tikzpicture}
    [line width=1pt,
    baseline={([yshift=-0.5ex+\cmdKV@general@yshift]current bounding box.center)},
    scale=\cmdKV@general@scale,
    rotate=\cmdKV@general@rotate,
    font=\cmdKV@general@font]
    \fill[\cmdKV@blobs@blobA] (-0.6,0) ellipse (0.2 and 0.4);
    \fill[\cmdKV@blobs@blobB] (0.4,0) ellipse (0.2 and 0.4);
    \draw[\cmdKV@gTypes@eA] ($ (-0.6,0) + (-180:0.2) $) -- ($ (-0.6,0) + (-180:0.6) $) node[label=(-180+\cmdKV@general@rotate):\cmdKV@gLabels@eLA] {};
    \draw[\cmdKV@gTypes@eB] ($ (-0.6,0) + (0:0.2) $) -- (-0.18,0) node[label=(\cmdKV@general@rotate):\cmdKV@gLabels@eLB] {};
    \draw[\cmdKV@gTypes@eC] ($ (0.4,0) + (45:0.26) $) -- ($ (0.4,0) + (45:0.7) $) node[label=(\cmdKV@general@rotate):\cmdKV@gLabels@eLC] {};
    \draw[\cmdKV@gTypes@eD] ($ (0.4,0) + (-45:0.26) $) -- ($ (0.4,0) + (-45:0.7) $) node[label=(\cmdKV@general@rotate):\cmdKV@gLabels@eLD] {};
    \draw[label distance=1,\cmdKV@gTypes@iA] ($ (-0.6,0) + (60:0.3) $) -- node[label=(90+\cmdKV@general@rotate):\cmdKV@gLabels@iLA] {}($ (0.4,0) + (135:0.26) $);
    \draw[label distance=1,\cmdKV@gTypes@iB] ($ (0.4,0) + (-135:0.26) $) -- node[label=(-90+\cmdKV@general@rotate):\cmdKV@gLabels@iLB] {}($ (-0.6,0) + (-60:0.3) $);
    \node[label=(\cmdKV@general@rotate):\cmdKV@gLabels@iLD] at (0.6,0) {};
    #2
  \end{tikzpicture}
  \gSetStdTypes{line}
}

\def\cFourPtRec{\@ifnextchar[\@cFourPtRec{\@cFourPtRec[]}}
\def\@cFourPtRec[#1]#2{%
  \setkeys*{pre}{#1}%
  \setrmkeys{general,gTypes,gLabels,dots,blobs}%
  \begin{tikzpicture}
    [line width=1pt,
    baseline={([yshift=-0.5ex+\cmdKV@general@yshift]current bounding box.center)},
    scale=\cmdKV@general@scale,
    rotate=\cmdKV@general@rotate,
    font=\cmdKV@general@font]
    \fill[\cmdKV@blobs@blobA] (-0.5,0) ellipse (0.2 and 0.3);
    \fill[\cmdKV@blobs@blobB] (0.5,0) ellipse (0.3 and 0.4);
    \filldraw[fill=white,draw=black] (0.55,0.15) circle (0.1);
    \filldraw[fill=white,draw=black] (0.45,-0.1) circle (0.1);
    \draw[\cmdKV@gTypes@eA] ($(-0.5,0) + (135:0.23)$) -- ($(-0.5,0) + (135:0.5)$) node[label=(180+\cmdKV@general@rotate):\cmdKV@gLabels@eLA] {};
    \draw[\cmdKV@gTypes@eB] ($(-0.5,0) + (-135:0.23)$) -- ($(-0.5,0) + (-135:0.5)$) node[label=(180+\cmdKV@general@rotate):\cmdKV@gLabels@eLB] {};
    \draw[\cmdKV@gTypes@eC] ($(0.5,0) + (45:0.35)$) -- ($(0.5,0) + (45:0.7)$) node[label=(\cmdKV@general@rotate):\cmdKV@gLabels@eLC] {};
    \draw[\cmdKV@gTypes@eD] ($(0.5,0) + (-45:0.35)$) -- ($(0.5,0) + (-45:0.7)$) node[label=(\cmdKV@general@rotate):\cmdKV@gLabels@eLD] {};
    \draw[label distance=1,\cmdKV@gTypes@iA] ($(-0.5,0) + (30:0.23)$) -- node[label=(90+\cmdKV@general@rotate):\cmdKV@gLabels@iLA] {} ($(0.5,0) + (157:0.3)$);
    \draw[label distance=1,\cmdKV@gTypes@iB] ($(0.5,0) + (-157:0.3)$) -- node[label=(-90+\cmdKV@general@rotate):\cmdKV@gLabels@iLB] {} ($(-0.5,0) + (-30:0.23)$);
    #2
 \end{tikzpicture}
  \gSetStdTypes{line}
}

\makeatother

\usetikzlibrary{arrows.meta}
\usepackage{tikzexternal}
\tikzsetexternalprefix{figures/}
\tikzset{
  rquark/.style={
    draw=dred,
    postaction={decorate},
    decoration={markings,mark=at position .5 with {\draw[->] (-0.01,0) -- (0.07,0);}}
  },
  raquark/.style={
    draw=dred,
    postaction={decorate},
    decoration={markings,mark=at position .5 with {\draw[->] (0.01,0) -- (-0.07,0);}}
  },
  gquark/.style={
    draw=dgreen,
    postaction={decorate},
    decoration={markings,mark=at position .5 with {\draw[->] (-0.01,0) -- (0.07,0);}}
  },
  gaquark/.style={
    draw=dgreen,
    postaction={decorate},
    decoration={markings,mark=at position .5 with {\draw[->] (0.01,0) -- (-0.07,0);}}
  },
  bquark/.style={
    draw=dblue,
    postaction={decorate},
    decoration={markings,mark=at position .5 with {\draw[->] (-0.01,0) -- (0.07,0);}}
  },
  baquark/.style={
    draw=dblue,
    postaction={decorate},
    decoration={markings,mark=at position .5 with {\draw[->] (0.01,0) -- (-0.07,0);}}
  },
  rline/.style={
    draw=dred
  },
  gline/.style={
    draw=dgreen
  },
  bline/.style={
    draw=dblue
  },
  rquarkA/.style={
    draw=dred,
    postaction={decorate},
    decoration={markings,mark=at position .65 with {
      \draw[{-Stealth[left]}] (-0.01,0) -- (0.07,0);}}
  },
  raquarkA/.style={
    draw=dred,
    postaction={decorate},
    decoration={markings,mark=at position .65 with {
      \draw[{-Stealth[left]}] (0.01,0) -- (-0.07,0);}}
  },
  gquarkA/.style={
    draw=dgreen,
    postaction={decorate},
    decoration={markings,mark=at position .65 with {
      \draw[{-Stealth[left]}] (-0.01,0) -- (0.07,0);}}
  },
  gaquarkA/.style={
    draw=dgreen,
    postaction={decorate},
    decoration={markings,mark=at position .65 with {
      \draw[{-Stealth[left]}] (0.01,0) -- (-0.07,0);}}
  },
  bquarkA/.style={
    draw=dblue,
    postaction={decorate},
    decoration={markings,mark=at position .65 with {
      \draw[{-Stealth[left]}] (-0.01,0) -- (0.07,0);}}
  },
  baquarkA/.style={
    draw=dblue,
    postaction={decorate},
    decoration={markings,mark=at position .65 with {
      \draw[{-Stealth[left]}] (0.01,0) -- (-0.07,0);}}
  }
}

\makeatletter
\DeclareRobustCommand\Cpp{{\em C\raisebox{2pt}{{\relsize{-2}++}}}}
\makeatother

\DeclareFontFamily{OT1}{pzc}{}
\DeclareFontShape{OT1}{pzc}{m}{it}{<-> s * [1.350] pzcmi7t}{}
\DeclareMathAlphabet{\mathpzc}{OT1}{pzc}{m}{it}

\definecolor{myblue}{rgb}{0,0,0.7}

\definecolor{mygreen}{rgb}{0,0.4,0}

\definecolor{myred}{rgb}{0.4,0,0}

\newcommand\varmp{\mathbin{\vcenter{\hbox{%
  \oalign{\hfil$\scriptstyle-$\hfil\cr
          \noalign{\kern-.3ex}
          $\scriptscriptstyle({+})$\cr}%
}}}}

\newcommand\cN{\mathcal{N}}

\newcommand\cV{\mathcal{V}}

\newcommand\cM{\mathcal{M}}
\newcommand\cA{\mathcal{A}}
\newcommand\cB{\mathcal{B}}

\def\d{\mathrm{d}}
\def\Neta{\tilde \eta}

\newcommand\bb{\mathbf{b}}
\newcommand\bc{\mathbf{c}}

\newcommand\nn{\nonumber}

\newcommand{\widebar}{\overline}

\DeclareMathOperator{\tr}{\rm tr}

\newcommand\eps{\epsilon}

\newcommand\braket[1]{\langle #1 \rangle}

\newcommand\spaq[1]{\langle #1\rangle}
\newcommand\spbq[1]{[#1]}

\def\spa#1.#2{\left\langle#1\,#2\right\rangle}
\def\spb#1.#2{\left[#1\,#2\right]}
\def\eqn#1{eq.~(\ref{#1})}
\def\eqns#1#2{eqs.~\eqref{#1} and~\eqref{#2}}

\def\be{\begin{equation}}
\def\ee{\end{equation}}
\def\bea{\begin{eqnarray}}
\def\eea{\end{eqnarray}}
\def\fq{\mathfrak{q}}

\definecolor{dred}{rgb}{0.6,0,0}
\definecolor{dgreen}{rgb}{0,0.5,0}
\definecolor{dblue}{rgb}{0,0,0.6}

\def\ie{i.e. }
\def\eg{e.g. }

\def\eqn#1{eq.~\eqref{#1}}
\def\eqns#1#2{eqs.~\eqref{#1} and~\eqref{#2}}

\def\Eqn#1{Eq.~\eqref{#1}}

\def\Fig#1{Fig.~{\ref{#1}}}

\def\Sec#1{Section~{\ref{#1}}}
\def\Secs#1#2{Sections~{\ref{#1}} and~{\ref{#2}}}

\def\App#1{Appendix~{\ref{#1}}}
\def\rcite#1{ref.~\cite{#1}}
\def\rcites#1{refs.~\cite{#1}}

\title{Two-loop ${\cal N}=1$ SYM Amplitudes via SUSY Decomposition and Massive Spinor-Helicity}
\preprint{UUITP-42/23 \\ \phantom{~} \hfill HU-EP-23/73-RTG}

\author[a,b]{Henrik Johansson,}
\author[c,d]{Gregor K\"{a}lin,}
\author[e,f]{Gustav Mogull,}
\author[a]{and Bram Verbeek}
\affiliation[a]{Department of Physics and Astronomy, Uppsala University, \\ Box 516, 75120 Uppsala, Sweden}
\affiliation[b]{Nordita, Stockholm University and KTH Royal Institute of Technology,\\  Hannes Alfv\'{e}ns  v\"{a}g 12, 10691 Stockholm, Sweden}
\affiliation[c]{SLAC National Accelerator Laboratory, Stanford University, Stanford, CA 94309, USA}
\affiliation[d]{Deutsches Elektronen-Synchrotron DESY, Notkestr. 85, 22607 Hamburg, Germany}
\affiliation[e]{Institut f\"ur Physik und IRIS Adlershof, Humboldt Universit\"at zu Berlin,\\ Zum Gro{\ss}en Windkanal 6, 12489 Berlin, Germany}
\affiliation[f]{Max Planck Institut f\"ur Gravitationsphysik
(Albert Einstein Institut),\\ Am M\"uhlenberg 1, 14476 Potsdam, Germany}

\emailAdd{henrik.johansson@physics.uu.se}
\emailAdd{gregor.kaelin@desy.de}
\emailAdd{gustav.mogull@aei.mpg.de}
\emailAdd{bram.verbeek@physics.uu.se}

\abstract{We obtain a color-kinematics-dual representation of the two-loop four-vector amplitude in a general renormalizable massless $\cN=1$ SYM theory, including internal matter as chiral supermultiplets.
The integrand is constructed to be compatible with dimensional regularization and supersymmetry by employing two strategies (implicitly defining our regularization scheme): supersymmetric decomposition and matching to massive spinor-helicity amplitudes.  All internal vector components inherit their $D$-dimensional properties by relating them to the previously constructed $D\leq6$, $\cN=2$ SQCD amplitude using supersymmetric decomposition identities of individual diagrams. This leaves only diagrams with internal matter lines as unknown masters, which are in turn constrained on $D$-dimensional unitarity cuts by reinterpreting the extra-dimensional momentum components as masses for the chiral supermultiplets.
We rely on the massive spinor-helicity formalism and massive on-shell $\cN=1$ superspace, generalized here to complex masses. 
Finally, we extend the kinematic numerator algebra to include three-term identities that are dual to color identities linear in the matter Clebsch-Gordan coefficients, as well as two new optional identities satisfied by mass-deformed $\cN=4$ and $\cN=2$ SYM theories that preserve $\cN=1$ supersymmetry. Altogether, these identities makes it possible to completely reduce the two-loop integrand to only two master numerators.}


\begin{document}
\maketitle
\flushbottom


\section{Introduction}

To push the frontier of scattering amplitudes computations, $\cN=4$ super Yang-Mills theory (SYM) is often considered a toy model for quantum chromodynamics (QCD), due to its enhanced mathematical structure and the theories' shared footing as $\text{SU}(N_c)$ gauge theories. The connection between these theories can be made more explicit: decomposing the $\cN=4$ supermultiplet, and allowing for representation change and mass deformation, one naturally encounters the field content of QCD. Packaging the leftover component fields into lower-supersymmetric multiplets allows QCD amplitudes to be decomposed into individually simpler parts with, and without, supersymmety. The process of recycling all higher-supersymmetric information in this way when computing an amplitude is referred to as supersymmetric decomposition and has been applied for QCD ---
in particular at tree level~\cite{Dixon:1996wi,Dixon:2010ik,Melia:2013epa} and for one loop amplitudes~\cite{Bern:1993mq,Bern:1994zx,Bern:1996je}. 

With this decomposition in mind, rather than going straight from the maximally supersymmetric toy model to phenomenologically-relevant QCD, it is interesting to have access to results from $\cN=1$ and $\cN=2$ super-QCD (SQCD) with $N_f$ matter multiplets; thus repackaging as much of the eventual QCD result as possible into supersymmetric contributions. These lower-supersymmetric theories are not only relevant for learning about gauge theory, but also appear in the study of gravity. When brought into color-kinematics-dual form~\cite{Bern:2008qj,Bern:2010ue,Johansson:2015oia, Bern:2019prr}, the integrands of SQCD amplitudes are essential ingredients to construct various supergravity amplitudes via the double copy~\cite{Bern:2010ue,Mafra:2011kj, Bern:2011rj, Carrasco:2011mn, Carrasco:2012ca, Chiodaroli:2013upa, Chiodaroli:2015wal, Mafra:2015mja, He:2015wgf, Chiodaroli:2017ehv,He:2017spx, Bern:2017yxu, Chiodaroli:2018dbu,Ben-Shahar:2018uie}, in particular pure $\cN = 0,1,2,3$ supergravities in four dimensions~\cite{Johansson:2014zca,Johansson:2017bfl}. A central application of the double copy is to explore the ultraviolet properties of supergravity multiloop amplitudes~\cite{Bern:2007hh, Bern:2008pv, Bern:2009kd,Bern:2012uf, Bern:2012cd, Bern:2012gh, Bern:2013qca, Bern:2013uka, Bern:2018jmv}. 
More recently, massive 2-to-2 scattering, and related high-energy scattering, in supergravity have found applications in the program of applying modern amplitudes techniques~\cite{Damour:2016gwp,Bern:2019nnu} to the study of bound binary systems and gravitational waves~\cite{DiVecchia:2020ymx,DiVecchia:2021bdo,Herrmann:2021tct}, including spin-effects originating from supersymmetric formulations~\cite{Guevara:2018wpp, Johansson:2019dnu, Bautista:2019evw, Jakobsen:2021zvh,Chiodaroli:2021eug,Cangemi:2022bew}. 

Lower-supersymmetric gauge-theory amplitudes are mathematically interesting objects in their own right. Though they are relatively simpler than non-supersymmetric amplitudes, the lower degree of symmetry leaves such theories less constrained, allowing for features which do not appear in the maximally-supersymmetric case. For example, it is a well-known observation that $\cN=4$ SYM amplitudes contain only polylogarithmic terms of uniform transcendental weight (see~\eg \rcites{Bern:1997nh,Anastasiou:2003kj,Bern:2005iz,Naculich:2008ys,DelDuca:2010zg,Goncharov:2010jf,Golden:2014xqa,Henn:2016jdu,Drummond:2018caf,Abreu:2018aqd,Chicherin:2018yne,Caron-Huot:2019vjl,Dixon:2020cnr}). The next-to-simplest gauge theory is $\cN=2$ SQCD theory at the superconformal point, $N_f=2N_c$, which exhibit properties similar to $\cN=4$ SYM. However, the uniform transcendental-weight structure is broken at two loops already for this superconformal theory~\cite{Dixon2008talk,Leoni:2015zxa}, but the breaking displays a remarkable simplicity and has a non-trivial interplay with the infrared structure of the amplitude~\cite{Duhr:2019ywc,Kalin:2019vjc}. This begs the question if similar mathematical structures can be found when further reducing the degree of supersymmetry.

\begin{figure}[t]
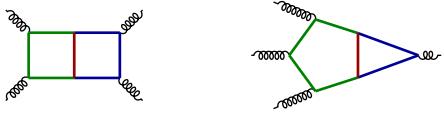

  \centering
  \tikzsetnextfilename{mastersBoxBox}
  \gBoxBox[all=gluon,iA=gline,iB=gline,iC=bline,iD=bline,iE=bline,iF=gline,iG=rline]{}
  \hspace{0.6cm}
  \tikzsetnextfilename{mastersPentaTri}
  \gTriPenta[all=gluon,iA=gline,iB=gline,iC=gline,iD=bline,iE=bline,iF=gline,iG=rline]{}
  \caption{Diagramatic representation of the $\cN=1$ SYM master numerators at the two-loop level.
    External lines are vector multiplets and internal lines are long chiral multiplets of different flavors.}
  \label{fig:masters}
\end{figure}

In this paper we obtain a color-kinematics-dual integrand for the massless two-loop four-vector amplitude of a generic renormalizable $\cN=1$ SYM theory, coupled to chiral multiplets in generic gauge-group representations\footnote{We work with totally antisymmetric three-flavor couplings in the superpotential; however, the general case is straightforward to extract from the given two-loop integrand.} -- of which $\cN=1$ SQCD is a particular example.
Using color-kinematics duality and supersymmetric decomposition, with the $\cN=2$ SQCD results from~\rcite{Johansson:2017bfl} as input, allows us to get to a result of considerable simplicity.  The kinematic Jacobi relations allows us relate BCJ numerators~\cite{Bern:2008qj} of different cubic diagrams, retaining a small subset of independent master diagram in each sector with different number of internal matter multiplets. Finally, supersymmetric decomposition allows us to relate the different matter sectors, such that the full $\cN=1$ SYM amplitude is completely determined by two master numerators with the maximum amount of internal matter, as highlighted by Figure~\ref{fig:masters}. 

A key challenge to our two-loop computation is the tension between manifesting four-dimensional minimal supersymmetry and using dimensional regularisation -- shifting the dimension $D>4$ will break minimal supersymmetry according to conventional wisdom. In contradistinction, extended supersymmetric theories allow for a higher-dimensional uplift which avoids this problem by naturally generating higher-dimensional contributions consistent with supersymmetry. For example, for the two-loop $\cN=2$ SQCD amplitudes obtained in~\rcites{Johansson:2017bfl,Kalin:2018thp,Duhr:2019ywc}, the 6D uplift of the theory to chiral $\cN=(1,0)$ SYM -- along with the use of 6D spinor helicity methods~\cite{Cheung:2009dc} -- was essential in regulating the theory.
For $\cN=1$ theories such an uplift does not exist; nevertheless, we may take a roundabout approach circumventing this no-go result. We consider $\cN=1$ chiral muliplets with complex masses, interpreting the mass as two extra-dimensional components of the momentum {\it post hoc}. Combined with the fact that color-kinematics duality and supersymmetric decomposition allows us to express all unknown diagrams in terms of the pure-matter masters in Figure~\ref{fig:masters}, as well as known $\cN=2$ SQCD contributions~\cite{Johansson:2017bfl}, we thus implicitly define all $\cN=1$ diagrams for extra-dimensional momenta.

For the unitarity-cut computations~\cite{Bern:1994zx,Bern:1994cg,Britto:2004nc,Bern:2007ct} involving matter, we hinge on the massive spinor-helicity formalism of \rcite{Arkani-Hamed:2017jhn}, as well as the associated $\cN=1$ on-shell superspace~\cite{Herderschee:2019ofc,Herderschee:2019dmc}. The use of these formalisms allows for the application of massive on-shell methods in a manifestly supersymmetric way. As a byproduct, we explore in great detail the features of color-kinematics duality for massive $\cN=1$ SYM tree amplitudes, finding both that it is non-trivially satisfied and that there exist new simplifying properties such as bonus kinematic Jacobi relations for pure matter diagrams in special theories. While our two-loop integrand incorporates the information for massive chiral matter multiplets, some care should be taken before integrating it in the massive case, since massive external bubble diagrams contribute, yet these are not guaranteed to be correct when naively obtained from the unitarity method~\cite{Ellis:2008ir,Britto:2011cr}. Further work is needed for exploring the massive external bubble diagrams, and the presented integrand is meant to be integrated in the massless case where such scaleless diagrams vanish. We defer this integration to upcoming work. See previous work on loop amplitudes in $\cN=1$ SYM~\cite{Bidder:2004tx,Bedford:2004py,Quigley:2004pw,Ochirov:2013oca,Chicherin:2018rpz,Chicherin:2018ubl}.

The structure of this paper is as follows. In Section~\ref{sec:neq1Setup} we generalize the established work on massive spinor-helicity variables and on-shell ${{\cal N}=1}$ superspace to include complex masses. We apply this formalism to the three- and four-point tree amplitudes in $\cN=1$ SYM with matter in Section~\ref{sec:susyAmps}, giving the necessary ingredients to compute generalized unitary cuts. In Section~\ref{sec:diagIden} we turn our attention to the contributing diagrams at two loops and show how supersymmetry decomposition and color-kinematics duality imposes constraints on the result before discussing the explicit construction and solution in Section~\ref{sec:numCons}. Finally, in Section~\ref{sec:conclu} we conclude by reviewing the most interesting features of the result and outline the future directions in which this work could be taken.

\section{Massive spinor helicity and on-shell ${{\cal N}=1}$ superspace}
\label{sec:neq1Setup}

In this section we introduce the massive spinor-helicity formalism
and associated on-shell superspace for amplitudes with $\cN=1$ supersymmetries.
The former is based on Arkani-Hamed, Huang and Huang's massive spinor-helicity
formalism \cite{Arkani-Hamed:2017jhn};
the latter on \rcites{Herderschee:2019ofc,Herderschee:2019dmc} by Herderschee, Koren and Trott.
In both cases we generalize to complex masses.

\subsection{Massive spinor helicity with a complex mass}

Given a four-dimensional momentum $p^\mu$ obeying the mass-shell condition $p^2=m \overline m$,
for a complex mass $m$,
we can decompose it using two null momenta $k^\mu,q^\mu$:
\be \label{massiveMom}
p^\mu = k^\mu+ \frac{m \overline m}{2 p \cdot q}q^\mu\,,
\ee
where $q^\mu$ is an arbitrary reference null vector
and $k^\mu$ is defined by the above relation.
Since $k^\mu$ and $q^\mu$ are massless we can use the massless spinor-helicity formalism to convert \eqn{massiveMom} into bi-spinors after contracting $p^\mu$ with the $\sigma^\mu_{\alpha \dot \alpha}$ matrices.\footnote{An explicit parametrization of the massless spinors can be given by:
$|k\rangle =
\left(\begin{matrix}
\sqrt{k_+}  \\
k_\perp/\sqrt{k_+}
\end{matrix}\right)$
and
 $|k] =
\left(\begin{matrix}
\,\overline{k}_\perp/\sqrt{k_+} \, \\
-\sqrt{k_+}
\end{matrix}\right)$,
 where
 $\sigma \cdot k = |k\rangle [k|=
\left(\begin{matrix}
k_+ & \overline{k}_\perp \\
k_\perp & k_-
\end{matrix}\right)=
\left(\begin{matrix}
k^0{+}k^3 & k^1{-} i k^2  \\
k^1{+}i k^2  & k^0{-}k^3
\end{matrix}\right)$.}
We get
\begin{equation}
\sigma \cdot p = |k \rangle [k| +  \frac{m \overline m}{2 p \cdot q}  |q \rangle [q| \equiv | p^{a} \rangle  [ p_a |\,,
\end{equation}
where we have suppressed the $(\alpha, \dot \alpha)$ spinor indices\footnote{For real $SO(1,3)$ momenta the spinor indices belongs to $SL(2, \mathbb{C})$; however, in general we consider complex momenta for which the signature of the Lorentz group is irrelevant.} of $SU(2)_L\times SU(2)_R\sim SO(1,3)$, and on the right-hand side we have implicitly defined massive spinors that carry $a,b, \ldots$ little group indices of $SU(2) \sim SO(3)$.
The massive spinors can be chosen explicitly as
\begin{align}\label{massiveWeyl}
|p^a\rangle=\left(\begin{matrix}
~ |q \rangle   \frac{m}{\braket{kq}}  \\ |k\rangle~
\end{matrix}\right)\,, &&
|p^a]=\left(\begin{matrix}
  |k]~\\ ~ |q] \frac{\overline m}{[kq]}
\end{matrix}\right)\,,
\end{align}
and the transposed spinors $\langle p^a|$ and $[ p^a|$ are obtained by mirroring the massless spinors, $ |k \rangle {\rightarrow} \langle k | $, as is naturally implied by the notation.
Little-group indices are lowered and raised as
$|p_a\rangle=\epsilon_{ab}|p^b\rangle$,
$|p^a\rangle=\epsilon^{ab}|p_b\rangle$ respectively,
and analogously for the square spinors
(we use $\epsilon^{12}=\epsilon_{21}=1$).
For real momenta the two spinors are complex conjugates of each other
up to a similarity transform.

One of the main advantages of the parametrization (\ref{massiveWeyl}) is that the massless limit is non-singular and transparent: $|p^a\rangle \rightarrow \big(0,|p\rangle\big)$ and $|p^a] \rightarrow \big(|p],0\big)$. While the appearance of an unspecified reference vector may seem worrisome, we note that different choices of $q$ are related by a little-group rotation acting on the spinors, and have no physical consequence. Indeed, since the little group is distinct for each on-shell particle, one can in general pick a set of $n$ independent $q_i$, one for each external state.

We may contract a massive spinor with itself in various ways, giving
\bea
\spa{p_a}.{p_b} &=& m \epsilon_{ab}\,, \hskip1.1cm                 \spb{p^a}.{p^b} = \overline m \epsilon^{ab}\,,  \nn \\
|p_a\rangle \langle p^a | &=& \mathbb{1}\, m  \,,\hskip1.07cm   |p^a][p_a | = \mathbb{1}\, \overline{m} \,, \nn \\
|p^a\rangle [p_a | &=& \sigma \cdot p  \,,\hskip0.93cm               |p_a] \langle p^a | = \bar \sigma \cdot p \,,
\eea
where the $\mathbb{1}$'s are identity matrices in the left and right $SU(2)$ groups respectively. From this we can see that the spinors (\ref{massiveWeyl}) correspond to on-shell states/wavefunctions, meaning that they satisfy the equations
\be
(\sigma \cdot p) |p^a]=\overline{m} |p^a \rangle\,,~~~~(\bar \sigma \cdot p) |p^a \rangle =m |p^a]\,.
\ee
These can be interpreted as the two Weyl equations for a decomposed Majorana fermion with complex mass.\footnote{The masses can be made real after rescaling the spinors by a complex phase, which gives the standard Weyl equations. Our conventions then follow those in \rcite{Ochirov:2018uyq}.}

Considering two distinct states, we may contract the Lorentz spinor indices for two spinors of momenta $p_i$ and $p_j$, which we often abbreviate as
\be \label{spinorcontr}
\spa{i^a}.{j^b} \equiv \spa{p_i^a}.{p_j^b}\,,~~~~ \spb{i^a}.{j^b} \equiv \spb{p_i^a}.{p_j^b}\,.
\ee
where $i$ and $j$ are integers that label the particles. Similar abbreviations are used for the individual spinors $|i^a\rangle = |p^a_i\rangle$, etc. Note that the spinor contractions are odd under exchange of particles: $\spa{i^a}.{j^b}= - \spa{j^b}.{i^a}$ and  $\spb{i^a}.{j^b}= - \spb{j^b}.{i^a}$.

While the contractions in \eqn{spinorcontr} are Lorentz invariants,  they are also covariant objects with respect to the little groups of two different particles. Is there some way that we can soak up the little group indices such that we get fully invariant objects? One obvious way is to contract a string of spinor products such that they form a Dirac trace,~\eg
\begin{align}
\begin{aligned}
 \spa{1^a}.{2^b} \spb{2_b}.{1_a} &= \langle1^a| 2 | 1_a ] = \tr(p_1 p_2)= 2 p_1 \cdot p_2\,, \\
 \spa{1^a}.{2_b} \spa{2^b}.{1_a} &= 2 m_1 m_2 \,,~~~~~\spb{1_a}.{2^b} \spb{2_b}.{1^a} = 2 \overline m_1  \overline m_2\,,
\end{aligned}
\end{align}
where we for simplicity avoid explicit Feynman-slash notation,
but it is assumed in appropriate places.

\subsection{Massive on-shell superspace}

A more interesting way to soak up the little group indices is to introduce a basis of
Grassmann-odd auxiliary objects that transform as spinors in the little group,
and as scalars in the Lorenz group.
For the $i$'th particle we call them $\eta_{i,a}$,
where again the little-group index is raised using $\eta_i^a=\eps^{ab}\eta_{i,b}$.
Through them we can define massive Grassmann-odd spinors that have no free little-group indices:
\begin{equation}\label{eq:grassmannSpinors}
|\fq_i \rangle   \equiv  |i^a \rangle \eta_{i,a} \,, ~~~~~~ |\fq_i ]   \equiv  |i_a ] \eta_{i}^a
\end{equation}
The notation will become clear shortly as we start discussing supersymmetry in Section~\ref{sec:OnShellSuperspace}. However, note that these spinors are simply a convenient notation for dealing with the bookkeeping of little-group indices, and they can be used for any four-dimensional theory that has massive fermions. For example, one can simplify some of the above discussed properties. The spinor products are now even under permutations:
\be
\spaq{\fq_1 \fq_2} = \spaq{\fq_2 \fq_1}\,,~~~~\spbq{\fq_1 \fq_2} = \spbq{\fq_2 \fq_1}\,,
\ee
and the on-shell factorization of the massive momentum mimics the massless case:
\be
|\fq_i\rangle [\fq_i| =  \sigma \cdot p_i\,(\eta_i)^2\,,~~~~~
|\fq_i] \langle\fq_i| = -\overline{\sigma} \cdot p_i\,(\eta_i)^2\,,
\ee
where $ (\eta_i)^2 =  \eta_i^1 \eta_i^2=\eta_{i,1}\eta_{i,2}$.
Contractions of little-group indices are now automatically taken care of when spinors are multiplied.
We also have similar identities involving the complex masses,
\begin{align}
\begin{aligned}
|\fq_i\rangle \langle \fq_i| &=  \mathbb{1} m_i \,(\eta_i)^2\,,
&
|\fq_i] [\fq_i| &= -\mathbb{1} \overline{m}_i \,(\eta_i)^2\,,\\
\spaq{\fq_i \fq_i} &=- 2m_i \, (\eta_i)^2\,,
&
\spbq{\fq_i \fq_i} &= 2\overline{m}_i \, (\eta_i)^2\,.
\end{aligned}
\end{align}
This gives simple spinor-string identities, for instance
\begin{align}
  \begin{aligned}
 \spaq{\fq_1 \fq_2} \spbq{\fq_2 \fq_1} &= 2 p_1 \cdot p_2 \, (\eta_1)^2 (\eta_2)^2\,, \\
 \spaq{\fq_1 \fq_2}^2 &= -2 m_1 m_2 (\eta_1)^2 (\eta_2)^2\,,\\
  \spbq{\fq_1 \fq_2}^2 &= -2 \overline{m}_1 \overline{m}_2 \, (\eta_1)^2 (\eta_2)^2 \,,  \\
  \spaq{\fq_1 \fq_2} \spaq{\fq_2 \fq_3}  \spaq{\fq_3 \fq_1}&= -2 m_1 m_2 m_3  (\eta_1)^2(\eta_2)^2(\eta_3)^2\,.
\end{aligned}
\end{align}
That is, whenever the $\fq_i$ argument appears exactly twice in a spinor string, we can replace it with familiar invariants. If some $\fq_i$ appears more than twice, then the whole term is zero.

Finally, when computing amplitudes from Feynman rules one can directly use $|\fq_i\rangle$ and $|\fq_i]$ as the (momentum-space) wavefunctions of the external fermions which has two benefits: 1) one can leave unstated which physical state of each fermion is being considered, and 2) the anti-commuting nature of the fermions is compensated for by the $\eta_i^a$ variables, making them even under exchange, similar to the bosons.

\subsection{Majorana states}

We will now analyze the external states that one can construct
using massive helicity spinors.
For real SO$(1,3)$ momenta the reality properties of the spinors are
\be \label{realitySpinor}
 (| p^a \rangle_\alpha)^* = [p_a|_{\dot \alpha} \,,~~~~~
 (| p_a ]^{\dot \alpha})^* =\langle p^a |^{\alpha} \,,
\ee
which is easily seen from the explicit expressions in \eqn{massiveWeyl},
assuming that the massless spinors satisfy
$ (| k \rangle_\alpha)^* = |k]_{\dot \alpha}$.
In terms of the Grassmann-dressed spinors the conjugation properties become
\bea \label{etabardef}
(|\fq_i\rangle_\alpha)^* &=&  [i_a |_{\dot \alpha} \bar{\eta}_{i}^a  \equiv  [ \overline \fq_i|_{\dot \alpha} \nn \\
(|\fq_i]^{\dot \alpha})^* &=&  -\langle i^a |^{\alpha} \bar{\eta}_{i,a}  \equiv  -\langle \overline \fq_i|^{\alpha}
\eea
where we have introduced $\bar{\eta}^a\equiv(\eta_a)^*$.
Without further constraints on the Grassmann parameters,
we can conclude that the above reality properties imply
that we are working with four on-shell states,
encoded by $\eta^a$ and $\bar{\eta}^a$.
This is equivalent to working with a Dirac spinor.

Since we are interested in Majorana fermions
we must impose a reality condition on the Grassmann parameters. For example, we could demand that $\bar{\eta}^a=\eta^a$, which would reduce the physical degrees of freedom to two. However, for later purposes it is more convenient to impose the reality condition using a Grassmann-Fourier transform. For any expression $\Omega$ involving $\bar{\eta}^a_i$ we may convert it to only $\eta^a_i$'s using the Fourier transform
\begin{align}\label{eq:FT}
\Omega(\eta_i^a)={\rm FT}_i\, \Omega(\eta_i^a, \bar \eta_i^a)\,, ~~\text{where}~~ {\rm FT}_i= \int\!\d^2\bar\eta_i\,e^{{\bar\eta}_{i,a}\eta_i^a}
\end{align}
and the Grassmann measures are defined by
$\int\!\d^2\eta_i\,(\eta_i)^2=\int\!\d^2\bar\eta_i\, (\bar \eta_i)^2=1$,
$(\bar\eta_i)^2=\bar{\eta}_i^2\bar{\eta}_i^1=\bar{\eta}_{i,2}\bar{\eta}_{i,1}$
(the complex conjugate of $(\eta_i)^2$).
This implies that
${\rm FT}_i\, \bar{\eta}_i^a=\eta_i^a$, as desired, and the spinors are now Majorana.

With a slight abuse of notation we can now consider a superposition of the left-handed and right-handed massive spinors:\footnote{Here its understood we are considering a direct sum of the $(\frac{1}{2},0)$ and $(0, \frac{1}{2})$ representations of the Lorentz group. When contracting such superposed spinors it is always obvious which contributions are zero.}
\be
| \fq_i \} = \Bigg(\begin{matrix}
| \fq_i \rangle   \\
| \fq_i ]
\end{matrix}\Bigg)
\equiv | \fq_i \rangle + | \fq_i]\,,
~~~~~~~
\{\fq_i | = \Big(\begin{matrix}
\langle \fq_i |   & \,[ \fq_i | \,\,
\end{matrix}\Big)
\equiv \langle \fq_i | +  [ \fq_i | \,,
\ee
which are the appropriate Majorana states for external particles in a scattering amplitude.
We will check two properties that confirm that we have defined a Majorana fermion: firstly, that the spinor $| \fq_i \}$ is real up to a similarity transform. Indeed, one can check that the Majorana reality condition $\Psi=\Psi^{C}$ is satisfied,
\be
\Psi=\left(\begin{matrix}
\, | \fq_i \rangle \,  \\
\, | \fq_i] \,
\end{matrix}\right)= {\rm FT}_i\, \gamma^0 C \left(\begin{matrix}
\, | \fq_i \rangle \,  \\
\, | \fq_i] \,
\end{matrix}\right)^*\equiv\Psi^{C}\,,
\ee
where  $\gamma^0$ and the charge-conjugation matrix $C=\gamma^1\gamma^3={\rm diag} (\epsilon^{\dot \alpha \dot \beta}, -\epsilon_{\alpha \beta})$ are given in the Weyl basis.

Secondly, one can check the completeness relation in four-component notation, which should confirm that we have a complete set of states:
\be \label{completenessRel}
 |\fq_i\}\{{-}\fq_i | =
\left(\begin{matrix}
{-} |  \fq_i \rangle  \langle  \fq_i  | \phantom{\Big|}  & |  \fq_i  \rangle  [  \fq_i  |   \\
 {-}  |  \fq_i  ]  \langle  \fq_i  |  \phantom{\Big|}  &  |  \fq_i  ]  [  \fq_i  |
\end{matrix}\right)
\stackrel{\int\!\d^2\eta_i}{\longrightarrow}
\left(\begin{matrix}
-m_i  \phantom{\Big|}  & \phantom{\Big|} \sigma \cdot p_i   \\
 \bar \sigma \cdot p_i    \phantom{\Big|}  & \phantom{\Big|} - \overline{m}_i
\end{matrix}\right)
= \gamma \cdot p_i - m_i P_{\rm L} - \overline{m}_i P_{\rm R}
\ee
where we removed the overall $(\eta_i)^2$ factor by integration, and $P_{\rm L/R} = \frac{1}{2}(1\mp \gamma_5)$ are the Left/Right chiral projectors. Since we are using an all-outgoing formalism, the completeness relation involves flipping the sign of the momentum in the first spinor, which is implemented by the simple rule:
\be
|{-}p^a \rangle = - |p^a \rangle\,,~~~~ |{-}p^a] = |p^a]\,.
\ee
This ensures that the branch cuts in the square roots of the spinors do not interfere with overall sign flips of the momentum.
The right-hand side of \eqn{completenessRel} can be recognized as the numerator of the Feynman propagator for a Majorana fermion with complex mass,
\be
{-}i \frac{\gamma \cdot p - m P_{\rm L} -  \overline{m}  P_{\rm R}}{p^2 -m \overline{m}}\,,
\ee
where the kinetic term is $\frac{i}{2} \overline{\Psi} \gamma \cdot \partial \Psi + \frac{1}{2} \overline{m}  \overline{\Psi} P_{\rm L} \Psi +  \frac{1}{2}m \overline{\Psi} P_{\rm R} \Psi$.
See \App{sec:Lagrangians} for the full Lagrangian.

Using the Majorana spinors we can, for example, compute a color-stripped three-point amplitude between two Majorana fermions and one gluon,
\be
 A_3(1\Psi,2\Psi,3A^{h=\pm})=\{\fq_1| \varepsilon_3^\pm | \fq_2\}  = [ \fq_1 | \varepsilon_3^\pm  | \fq_2 \rangle + \langle \fq_1 | \varepsilon_3^\pm  | \fq_2]=
 \left\{\begin{matrix}
 x  \spaq{\fq_1 \fq_2}~~~~~h=+1 \\
 \frac{1}{m x}  \spbq{\fq_1 \fq_2} ~~~~~~h=-1
\end{matrix}\right.
\ee
where $mx=\sqrt{2} \varepsilon_3^+ \cdot p_1$. Since the spinors now include the Grassmann-odd parameters, we should demand that the color-dressed amplitude is Bose symmetric under exchange of $1\leftrightarrow 2$. Without the color factor the above amplitude is anti-symmetric, which fixes the color factor to also be anti-symmetric $T^A =-(T^A)^{t}$. This corresponds to a real representation of the gauge group, which is expected since the Majorana fermions are real.

Consider three-point amplitudes between two Majorana fermions and a scalar. If the interaction is vector-like $\overline \Psi \Psi \varphi$, then the amplitude is
\be
 A_3(1\Psi,2\Psi,3\varphi) =  \{\fq_1| \fq_2\} = \spaq{\fq_1 \fq_2} +\spbq{\fq_1 \fq_2}
\ee
and if it is axial-like $\overline \Psi \gamma_5 \Psi \tilde \varphi$, we get
\be
 A_3(1\Psi,2\Psi,3\tilde \varphi) =  \{\fq_1|  \gamma_5| \fq_2\} = \spbq{\fq_1 \fq_2}-\spaq{\fq_1 \fq_2}\,.
\ee
If the interactions are chiral $\overline \Psi P_{\rm L}\Psi \phi$, or anti-chiral $\overline \Psi P_{\rm R}\Psi \overline{\phi}$, we get
\bea
 A_3(1\Psi,2\Psi,3\phi) &=&  \{\fq_1|  P_{\rm L} | \fq_2\} =\spaq{\fq_1 \fq_2}\,, \nn \\
  A_3(1\Psi,2\Psi,3\overline{\phi})  &=&  \{\fq_1|  P_{\rm R} | \fq_2\} =\spbq{\fq_1 \fq_2}\,.
\eea
Note that all the color-stripped fermion-scalar amplitudes are symmetric under exchange of $1\leftrightarrow 2$, which means that the Majorana fermions either have to transform in a pseudo-real representation of the gauge group, or we have to introduce additional flavor structure which compensates for the anti-symmetry of the real-representation generators.

\subsection{On-shell $\cN=1$ superspace}

\label{sec:OnShellSuperspace}
Given fermionic charges $Q_\alpha$, $Q^\dagger_{\dot \alpha}$, the ${\cal N}=1$ supersymmetry algebra is defined by
\be
\{Q_\alpha, Q^\dagger_{\dot \alpha} \}=\sigma_{\alpha \dot \alpha}\cdot P\,,~~~\{Q_\alpha, Q_\beta \}=0\,,~~~\{Q^\dagger_{\dot \alpha}, Q^\dagger_{\dot \beta} \}=0\,,
\ee
where $P^\mu$ is the momentum operator
(as there is no R-symmetry manifest, a central charge is not required).
For the situation at hand, where the operators act on a given scattering amplitude, we may replace $Q_\alpha$, $Q^\dagger_{\dot \alpha}$ and $P$ with the total charges of the amplitude. They can then be expressed as sums over the individual charges of the on-shell one-particle states, $Q_\alpha=\sum_i \fq_{i,\alpha}$, $Q^\dagger_{\dot \alpha}=\sum_i\fq^\dagger_{i,\dot \alpha}$, $P=\sum_i p_i$, thus
\be
\{\fq_{i,\alpha}, \fq^\dagger_{i,\dot \alpha} \}=\sigma_{\alpha \dot \alpha}\cdot p_i\,,~~~\{\fq_{i,\alpha}, \fq_{i,\beta} \}=0\,,~~~\{ \fq^\dagger_{i,\dot \alpha}, \fq^\dagger_{i,\dot \beta} \}=0\,,
\ee
and all anti-commutators that mix particle labels $i,j$ are zero.

We can explicitly realize the algebra by using the massive on-shell spinors and Grassmann variables introduced in the previous section,  which gives a {\it holomorphic representation} of the on-shell superspace:
\begin{align}\label{superqonep}
\begin{aligned}
\fq_{i,\alpha} &=  |i^a\rangle_\alpha \eta_{i,a}= |\fq_{i}\rangle_\alpha \,, \\
\fq_{i,\dot \alpha}^\dagger &= [i_a|_{\dot \alpha} \frac{\partial}{ \partial \eta_{i,a}} =  -\int \! \d^2\eta_i\, [\fq_{i}|_{\dot \alpha} \,.
\end{aligned}
\end{align}
In the last equality in \eqn{superqonep} we have traded the $\eta$-derivative for a two-fold $\eta$-integral acting on $[\fq_{i}|=\eta^a_i[i_a|$, which gives an integral operator that acts on a test function in the same way as the derivative operator. In general, we prefer to work with Grassmann integrals rather than derivatives.
Thus we have defined the total supercharges (or supermomenta) as
\be\label{eq:supercharges}
Q = \sum_{i=1}^{n}  |\fq_{i}\rangle\,,~~~~~    Q^\dagger = -\sum_{i=1}^{n}  \int \! \d^2\eta_i [\fq_{i}|\,,
\ee
where for convenience we have suppressed the Lorentz spinor indices, and $n$ is the number of external particles for the amplitude.

Alternatively, we can convert to an {\it anti-holomorphic representation} of the on-shell superspace that uses the $\bar \eta$ variables introduced before. To avoid confusion, we call these alternative versions of the supercharges
$\overline{Q}$ and $\overline{Q}^\dagger$:
\be
 \overline{Q}^\dagger =\sum_{i=1}^{n}  [\overline{\fq}_i|\,,~~~~~    \overline{Q}= \sum_{i=1}^{n}  |i^a\rangle \frac{\partial}{ \partial \bar \eta_{i}^a} = -\sum_{i=1}^{n}   \int \! \d^2\bar \eta_i  |\overline{\fq}_i\rangle \,,
\ee
where $[\overline{\fq}_i| \equiv \bar{\eta}_{i}^a  [i_a| $ and $|\overline{\fq}_i\rangle  \equiv  |i^a\rangle \bar{\eta}_{i,a}$ as defined in \eqn{etabardef}. These supercharges are equivalent to the holomorphic ones, except that they act on functions that have been Grassman-Fourier transformed from $ \eta $ to $\bar \eta$ space. They do not constitute additional supercharges, just a redundancy in our description. We will generally choose to work with the holomorphic $ \eta $ variables and, when necessary, the Grassman-Fourier transform can be used to map to $\bar \eta $ variables.

Scattering amplitudes with massive or massless external legs that preserve ${\cal N}=1$ supersymmetry will be annihilated by both $Q$ and $Q^\dagger$, as well as the momentum operator $P$. Since $P$ and $Q$ are multiplicative operators (not differential or integral operators) they should appear as delta functions
(except at three points, as we will see below).
As is conventional, we define the amplitude $A_n$ by factoring out the momentum conservation factor $\delta^4(P)$, but include the Grassmann delta function:
\be
\delta^4(P) A_n = \delta^4(P)  \delta^2(Q) a_n\,,
\ee
where $a_n$ is a polynomial in $\eta_i^a$, such that each monomial term is a little-group invariant and of even degree $2,4, 6, \ldots, 2n-2$. Using the Grassmann delta function identity $\delta(\eta)=\eta$ one can derive an explicit expression for $\delta^2(Q)$:
\be \label{deltaQ}
 \delta^2(Q) = \frac12\sum_{i,j=1}^{n} \spaq{\fq_i \fq_j}\,,
\ee
where it is important to remember that for massive external states terms of the form $\frac12\spaq{\fq_i \fq_i} = -m_i (\eta_i)^2$ are non-vanishing.

If all external states are massless then the sum collapses to the usual expression involving massless spinors and Grassmann variables with only one upper index,
\be
 \frac12\sum_{i,j=1}^{n} \spaq{\fq_i \fq_j}\Big|_{\rm massless} = \sum_{i<j}^{n}\eta_i^1  \spaq{i \, j}\eta_j^1\,.
\ee
The $\eta_i^1$ are no longer charged under the massive little group, and instead they carry the global $U(1)$ R-symmetry charge of the massless ${\cal N}=1$ multiplets.
While it can be important to remember that massive and massless states are quite different in ${\cal N}=1$ theories, the definition of $\delta^2(Q)$ in this section, using the massive spinor-helicity notation parametrized as \eqn{massiveWeyl}, automatically incorporates these subtleties for any configuration of massive and massless legs. So, henceforth when writing $\delta^2(Q)$  we never need to specify which legs are massless.

\subsection{On-shell multiplets for ${\cal N}=1$ SYM with matter}
\label{sec:neq1Multiplets}

As we are interested in ${\cal N}=1$ SYM with generic matter couplings,
here we list the on-shell multiplets to be encountered.
Firstly, we will have two massless on-shell ${\cal N}=1$ vector multiplets:
\begin{equation}\label{eq:vMults}
V^+ =A^+ + \eta^1 \widetilde \lambda^+\,, ~~~~~~V^- = \lambda^- + A^- \eta^1\,,
\end{equation}
where $A^\pm$ denotes the on-shell gluon polarizations
and $\widetilde\lambda^+$, $\lambda^-$ are the on-shell Weyl spinors of the massless gluino.
The $V^+$ multiplet is manifestly bosonic and the $V^-$ multiplet fermionic,
which is a peculiarity of the fact that we are working
with an odd amount of supersymmetry.
In particular, to perform a state sum over
the physical polarizations and spinors we use
a one-dimensional Grassmann integral:
\be
\int\!\d \eta^1( V^+_{\rm L} V^-_{\rm R}+ V^-_{\rm  L} V^+_{\rm R})\,,
\ee
which gives a Grassmann-even expression overall.

For massive matter we assemble the fermions and scalars into
a long chiral ${\cal N}=1$ multiplet:
\begin{equation}\label{eq:longNeq1}
\Phi =  \phi +\eta^{a} \Psi_a+(\eta)^2\overline \phi \,,
\end{equation}
which satisfies $\Phi={\rm FT}\,\overline{\Phi}$,
assuming that $\Psi_a$ is an on-shell Majorana spinor.
The multiplet thus has two bosonic and two fermionic degrees of freedom, and this is the smallest four-dimensional representation of massive supersymmetric matter. When considering several matter multiplets we add a flavor index to the on-shell multiplet, $\Phi^A$, and it is understood that the components all carry the same flavor index.
The sum over the physical states is taken care of by a two-fold Grassmann integration,
\be
\int\!\d^2 \eta \,  \Phi_{\rm L} \Phi_{\rm R}\,,
\ee
and the resulting answer is little-group invariant.

In the massless limit the long chiral multiplet $\Phi$
can be decomposed into two short chiral multiplets:
\be\label{eq:shortNeq1}
\Phi =\Phi^{-} + \Phi ^{+} \eta^2\,.
\ee
The on-shell Majorana spinor $\Psi_a$ is separated into its chiral on-shell components:
$\Psi_a \rightarrow (- \psi^-, \widetilde   \psi^+)$, so
\begin{align}
\Phi^- &=  \phi -\eta^{1} \psi^-\,,
& \Phi^+ &=  -\widetilde  \psi^+ +\eta^{1}\overline \phi\,.
 \end{align}
where have identified the helicities and little group indices as: $1=+$ and $2=-$.
While these short multiplets are interesting to study separately when calculating amplitudes, ultimately they need to be assigned to a real representation of the gauge group (effectively giving back the long multiplet),
otherwise the theory has a one-loop gauge anomaly.

\section{$\cN=1$ supersymmetric amplitudes}
\label{sec:susyAmps}

In this section we use the on-shell formalism developed in \Sec{sec:neq1Setup} to obtain three- and four-point $\cN=1$ supersymmetric amplitudes.
These will subsequently be used to compute generalized unitarity cuts.
All of our amplitudes are consistent with the supersymmetric Lagrangian given in \App{sec:Lagrangians}.

\subsection{Three-point amplitudes}
\label{sec:3ptAmps}

We begin with the massless MHV amplitudes
for three $\cN=1$ vector multiplets.
For massless particles the spinors in \eqn{massiveWeyl} reduce to
\begin{equation}
  |i^a \rangle = \varepsilon^{ab} |i_b  \rangle=
   \left( {\begin{array}{c}
   0 \\
     |i \rangle \\
   \end{array} } \right) \,,~~~~~~
    |i^a ] = \varepsilon^{ab} |i_b ]  = \left( {\begin{array}{c}
       |i] \\
      0 \\
   \end{array} } \right)\,,
\end{equation}
which implies that $|\fq_i\rangle= |i \rangle \eta_{i}^1$
and $|\fq_i]= -|i] \eta_{i,1}$
using the definitions in \eqn{eq:grassmannSpinors}.
The three-point amplitudes are given by familiar expressions:
\begin{subequations}\label{eq:3ptVectorAmps}
  \begin{align}
  \label{eq:MHVvec}
  A_3(1V^- ,2V^- ,3V^+ )&=i \delta^2(Q) \frac{\spa{1}.{2}^3}{\spa{1}.{2}\spa{2}.{3}\spa{3}.{1}}\,,\\
  \label{eq:MHVbarVec}
  A_3(1V^+ ,2V^+ ,3V^- )&=-i \overline{\delta^2(Q)\eta_1^1\eta_2^1\eta_3^1} \frac{\spb{1}.{2}^3}{\spb{1}.{2}\spb{2}.{3}\spb{3}.{1}}
  =-i \delta(Z_3)\frac{\spb{1}.{2}^3}{\spb{1}.{2}\spb{2}.{3}\spb{3}.{1}}.
  \end{align}
\end{subequations}
The long bar denotes CPT conjugation:
we take the complex conjugate (with $(\eta_a)^*=\bar{\eta}^a$)
and then Grassmann-Fourier transform with respect each of the external legs \eqref{eq:FT}.
This gives rise to the degree-one expression
$Z_3=\spb{1}.{2}\eta_3^1+\spb{2}.{3}\eta_1^1+\spb{3}.{1}\eta_2^1$
in the Grassmann variables,
which conserves supermomentum ($Q=0$) without the need for an
explicit factor $\delta^2(Q)$.

For two long chiral multiplets of equal complex mass $m$ that interact with a vector multiplet we have two non-zero three-point  amplitudes:
\begin{subequations}\label{eq:3ptAmps}
\begin{align}
\label{eq:MMp}
A_3(1\Phi ^{(A)},2\Phi ^{(B)},3V^+ )&=ix \delta^{AB} \delta^2(Q)\,,\\
\label{eq:MMm}
A_3(1\Phi ^{(A)},2\Phi ^{(B)},3V^- )&=
-i\frac{\delta^{AB}}{mx}\delta^2(Q)\!\left(\overline{m}\,\eta_3^1+[3{\fq_1}]\right)=
i\frac{\delta^{AB}}{x} \overline{\delta^2(Q) \eta^1_3}\,,
\end{align}
\end{subequations}
where $A$, $B$ are flavor indices.
The prefactor $x$ is given by
$m x = \langle q|1|3]/\spaq{3\,q}= -\sqrt{2}(p_1\cdot\varepsilon_3^+)= i\,A(\phi,\phi,A^+)$,
where $q^\mu$ is an arbitrary null momentum  and the last equality defines it in terms of a standard scalar-gluon amplitude.
The first amplitude containing $V^+$ is fixed
by the requirement it have overall Grassmann weight 2;
the second amplitude containing $V^-$ is
derived from the first via CPT conjugation,
and is also eliminated by $Q^\dagger$ \eqref{eq:supercharges}.
The kinematic (non-flavor) part of \eqref{eq:MMm} has full Bose symmetry,
yet we have chosen to eliminate $\eta_2^a$ inside the bracket
using supermomentum conservation $Q=0$ to give a more compact expression.

The last non-vanishing massive three-point amplitude is the one between three long chiral multiplets of complex masses $m_1,m_2,m_3$, respectively:
\be
\label{eq:MMM}
A_3(1\Phi ^{(A)},2\Phi ^{(B)},3\Phi ^{(C)})=i\,T^{ABC}
\delta^2(Q)\!\left[1+\frac1{m_3}\!
\left([\fq_1 \fq_2]\!-\!\overline{m}_1(\eta_2)^2\!-\!\overline{m}_2(\eta_1)^2\right)\right],
\ee
where $(\eta_i)^2\equiv\eta_{i}^1\eta_{i}^2$
and $T^{ABC}$ is an arbitrary-flavor
tensor that controls the interaction.
The Grassmann weight-4 part of the amplitude
is derived from the simpler weight-2 part via CPT conjugation;
we have eliminated $\eta_{3}^a$ inside the bracket using $Q=0$.

Despite the apparent factors of mass in the denominators,
all amplitudes are well behaved in the massless limit.
As explained in \Sec{sec:neq1Multiplets},
the long chiral multiplet decomposes into two short chiral multiplets:
$\Phi =\Phi^{-} + \Phi ^{+} \eta^2$.
From \eqn{eq:3ptAmps}
we get the following two massless amplitudes:
\begin{subequations}\label{eq:3ptAmpsMeq01}
\begin{align}
\label{eq:MMpMeq0}
A_3(1\Phi^{-(A)},2\Phi^{+(B)},3V^+ )&=i\delta^{AB} \delta(Z_3) \frac{\spb{2}.{3}}{\spb{1}.{2}}\,,\\
\label{eq:MMmMeq0}
A_3(1\Phi^{-(A)},2\Phi^{+(B)},3V^- )&=
i\delta^{AB}\delta^2(Q)\frac{\spa{1}.{3}}{\spa{1}.{2}}\,,
\end{align}
\end{subequations}
which have lower Grassmann degree by one unit compared to the massive amplitudes since we have extracted and removed an overall factor of $\eta^2_2$.

The massive three-chiral amplitude (\ref{eq:MMM}) has non-uniform Grassmann degree,
so it splits into two sectors in the massless limit.
The two massless amplitudes are
\begin{subequations}\label{eq:3ptAmpsMeq02}
\begin{align}
\label{eq:MMM_Meq0_1}
A_3(1\Phi^{+(A)},2\Phi^{+(B)},3\Phi^{+(C)})&=i\,T^{ABC}
\delta(Z_3)\,, \\
\label{eq:MMM_Meq0_2}
A_3(1\Phi^{-(A)},2\Phi^{-(B)},3\Phi^{-(C)})&=i\,T_{ABC} \delta^2(Q)\,.
\end{align}
\end{subequations}
Here we have lowered the flavor indices in the all-minus amplitude to emphasize that in the massless limit the two cubic tensors may be taken to be distinct, though related by complex conjugation $(T_{ABC})^*=T^{ABC}$.

\subsection{Obtaining four-point amplitudes}
\label{sec:bcfw}

We build up four-point $\cN=1$ amplitudes by sewing together the three-point amplitudes derived above,
essentially using a primitive version of BCFW recursion
\cite{Britto:2004ap,Britto:2005fq}.
(A similar procedure was also used in \rcites{Herderschee:2019ofc,Herderschee:2019dmc}.)
Usually one should worry about four-point contact
terms that might get lost,
both when sewing three-point amplitudes together
and when applying BCFW recursion
(encoded in the large-$z$ scaling behavior).
However, in this case any such ambiguity must correspond to the freedom of adding a local dimensionless
supersymmetric four-point function to our result,
or equivalently the freedom of adding unconstrained quartic interactions to a supersymmetric and renormalizable ${\cal N}=1$ Lagrangian.
The ${\cal N}=1$ SYM Lagrangian has no such freedom,
hence a supersymmetric and dimensionless
four-point function that has consistent factorization limits
will give a correct four-point amplitude.

Consider sewing together two three-point amplitudes of chiral multiplets.
Assuming that only one internal chiral multiplet
(of flavor 5) contributes,
\bea
\label{eq:3px3p}
{\rm Sew}_{12|34}&=&-i \int\!\d^2\eta_5  \, A_3(1\Phi ^{(1)},2\Phi ^{(2)},5\Phi ^{(5)}) \times A_3(-5\Phi ^{(5)},3\Phi ^{(3)},4\Phi ^{(4)})\\
&=& i T^{125}T^{534}
\delta^2(Q) \int\!\d^2\eta_5 \,\delta^2(Q_{125}) \left[1+\frac1{m_5} \left([\fq_1 \fq_2]-\overline{m}_1(\eta_2)^2-\overline{m}_2(\eta_1)^2\right)\right] \nn \\
&&\hskip5.4cm \!\!\!\times \left[1+\frac1{m_5} \left([\fq_3 \fq_4]-\overline{m}_3(\eta_4)^2-\overline{m}_4(\eta_3)^2\right)\right].\nn
\eea
We have used $\delta(A)\delta(B)=\delta(A+B) \delta(B)$
to extract the overall supermomentum delta function;
the leftover delta function has the argument
$Q_{125}=|\fq_1\rangle+|\fq_2\rangle+|\fq_5\rangle$.
As $\eta_5^a$ appears only inside $\delta^2(Q_{125})$
we simply need to evaluate
\bea
\int\!\d^2\eta_5 \,\delta^2(Q_{125})=
\int\!\d^2\eta_5\,\frac{1}{2}\spaq{\fq_5 \fq_5}= -m_5\,.
\eea
So the sewing is given by
\begin{align}
\begin{aligned}
{\rm Sew}_{12|34}= i\frac{T^{125}T^{534}\delta^2(Q) }{m_5} &\left(m_5+ [\fq_1 \fq_2]-\overline{m}_1(\eta_2)^2-\overline{m}_2(\eta_1)^2\right)\\
\times &\left(m_5+ [\fq_3 \fq_4]-\overline{m}_3(\eta_4)^2-\overline{m}_4(\eta_3)^2 \right)\,.
\end{aligned}
\end{align}
The result has terms of Grassmann degree 2, 4 and 6.

The terms of degree 2 and 4 are manifestly local after expanding out the product, while the terms of degree 6 have a spurious $1/m_5$ pole:
\bea
\text{degree 2}:&&~~\delta^2(Q) m_5\,, \\
\text{degree 4}:&&~~\delta^2(Q)\big( [\fq_1 \fq_2]+[\fq_3 \fq_4]-\overline{m}_1(\eta_2)^2-\overline{m}_2(\eta_1)^2 -\overline{m}_3(\eta_4)^2-\overline{m}_4(\eta_3)^2\big) \,, \nn \\
\text{degree 6}:&&~~ \frac{\delta^2(Q)}{m_5}\big([\fq_1 \fq_2]-\overline{m}_1(\eta_2)^2-\overline{m}_2(\eta_1)^2\big)\big([\fq_3 \fq_4]-\overline{m}_3(\eta_4)^2-\overline{m}_4(\eta_3)^2 \big)\,.\nn
\eea
The degree-2 and 6 terms are related by CPT invariance,
which is not apparent since the complex conjugate mass $\overline{m}_5$ is not present in either expression.
We can remedy this by using the on-shell condition of the intermediate leg: $m_5\overline{m}_5= s=(p_1+p_2)^2$.
This gives an expression that instead has a spurious $s$ pole:
\be \label{d2Q_Fourier}
\text{degree 6}:~\frac{\delta^2(Q)\overline{m}_5}{s}([\fq_1 \fq_2]-\overline{m}_1(\eta_2)^2-\overline{m}_2(\eta_1)^2)([\fq_3 \fq_4]-\overline{m}_3(\eta_4)^2-\overline{m}_4(\eta_3)^2 )= \overline{\delta^2(Q) m_5}\,.
\ee
Using the Grassmann-Fourier transform one can show this is the CPT conjugate of $\delta^2(Q) m_5$ for unrestricted four-point kinematics.
We promote the result of the sewing to
an amplitude which has only an s-channel contribution:
\be
A_4(1\Phi ^{(1)},2\Phi ^{(2)}|3\Phi ^{(3)},4\Phi ^{(4)}) = \frac{{\rm Sew}_{12|34}}{s-\mu_5^2}\,,
\ee
where $\mu_5^2 = m_5 \overline{m}_5$.
Other channels can be added when needed,
but the above result already constitutes a fully consistent
partial amplitude.

Next we sew together two three-point
amplitudes where one external leg is a massless $V^+$ vector
and the remaining ones are massive chiral multiplets:
\begin{align} \label{sewing2}
&-i  \int\!\d^2 \eta_5 A_3(1\Phi ^{(1)},2\Phi ^{(2)},5\Phi ^{(3)})A_3(-5\Phi ^{(3)},3\Phi ^{(3)},4V^+ )\nn \\
&\qquad=-i\frac{\langle q|3|4]}{\spa{q}.{4} m_3} \delta^2(Q) \int\!\d^2 \eta_5  \delta^2(Q_{125}) \left[1+\frac1{m_3} \left([\fq_1 \fq_2]-\overline{m}_1(\eta_2)^2-\overline{m}_2(\eta_1)^2\right)\right] \nn \\
&\qquad=-i \frac{[4|13|4]}{2 p_1 \cdot p_4} \delta^2(Q)  \left[1+\frac{\overline{m}_3}{s} \left([\fq_1 \fq_2]-\overline{m}_1(\eta_2)^2-\overline{m}_2(\eta_1)^2\right)\right] \\
&\qquad = - i \frac{\delta^{2}(Q) ([4|13|4]+\overline{m}_1 [\fq_2 4] [4 \fq_3]+\overline{m}_2 [\fq_3 4] [4 \fq_1]+\overline{m}_3 [\fq_1 4] [4 \fq_2])}{2 p_1 \cdot p_4}\,.\nn
\end{align}
For simplicity we have set $T^{123}=1$ and the mass of particle 5 is in this case $m_3$.
In the last step of \eqn{sewing2} some non-trivial simplifications have brought it to a form that can be promoted to unrestricted kinematics. It is easiest to use an ansatz for this step, where the ansatz is constrained by crossing symmetry of the massive legs. The denominator of the sewing can be recognized as the $t$-channel pole, and the color-ordered amplitude is
\be
A_4(1\Phi ^{(1)},2\Phi ^{(2)},3\Phi ^{(3)},4V^+ ) \!= \!- i \delta^{2}(Q) \frac{ [4|13|4]\!+\!(\overline{m}_1 [\fq_2 4] [4 \fq_3]+{\rm cyclic}(123))}{(s-\mu_3^2)(t-\mu_1^2)}\,,
 \ee
where $\mu_i^2=m_i \overline{m}_i$ is the invariant mass square of the internal particles.

As our final example, consider sewing together the massless vector of the three-point amplitudes eqs. (\ref{eq:MMp}) and (\ref{eq:MMm}).  The algebraic steps are not enlightening for this calculation, and it is instead straightforward to write down a very simple ansatz of the correct form. The result has to be local and antisymmetric in legs $1\leftrightarrow 2$ and $3\leftrightarrow 4$, which gives
\begin{align}
\begin{aligned}
&-i \sum_{h=\pm 1}\int\!\d \eta_5^1 A_3(1\Phi ^{(1)},2\Phi ^{(1)},5V^h )A_3(-5V^{-h},3\Phi ^{(3)},4\Phi ^{(3)} ) \\
&\qquad=i \delta^2(Q)\big( [\fq_1 \fq_3]- [\fq_2 \fq_3]- [\fq_1 \fq_4]+ [\fq_2 \fq_4]\big)\,.
\end{aligned}
\end{align}
From this we can define an $s$-channel partial amplitude:
\be
 A_4(1\Phi ^{(1)},2\Phi ^{(1)} | 3\Phi ^{(3)},4\Phi ^{(3)} )= i \delta^2(Q)\frac{ [\fq_1 \fq_3]- [\fq_2 \fq_3]- [\fq_1 \fq_4]+ [\fq_2 \fq_4]}{s}\,,
\ee
which is fully gauge invariant without the need to add other channels.

\subsection{Four-point amplitudes}
\label{sec:fourPoints}

Let us now summarize the four-point color-ordered amplitudes.
The massless  MHV amplitude is given by the familiar Parke-Taylor formula:
\begin{align}
  A_4(1V^-, 2V^-, 3V^+, 4V^+)=
  i \delta^2(Q) \frac{\spa{1}.{2}^3}{\spa{1}.{2}\spa{2}.{3}\spa{3}.{4}\spa{4}.{1}}\,.
\end{align}
Any amplitude (four-point or otherwise) with a single external massive multiplet is zero.
The amplitudes with two equal-flavor chiral multiplets of complex mass $m$ and two vector multiplets are given by\footnote{With
a real mass $m=\overline{m}$ these amplitudes can be obtained by projecting the $\cN=1$ multiplets from a four-point $\cN=4$ SYM amplitude on the Coulomb branch, as demonstrated in \rcites{Herderschee:2019dmc,Herderschee:2019ofc}.}
\begin{subequations}
 \begin{align}
  \label{sector1}
 A_4(1\Phi,2V^+,3\Phi,4V^+) &= - i \frac{\overline{m} \, \delta^{2}(Q)  \spb{2}.{4}^2}{(s- \mu^2)(t- \mu^2)}\,,\\
 \label{sector2}
 A_4(1\Phi,2V^-,3\Phi,4V^+) &= -i \frac{  \delta^{2}(Q)  \langle2|3|4] \big(\spb{4}.{\fq_1}-\spb{4}.{\fq_3}\big)    }{(s- \mu^2)(t- \mu^2)}\,,\\
 \label{sector3}
 A_4(1\Phi,2V^-,3\Phi,4V^-) &= - \frac{i}{2} \frac{  m \, \delta^{2}(Q)  \spa{2}.{4} \big( [\fq_1|-[\fq_3|\big) \big(|\fq_1]-|\fq_3]\big) }{(s- \mu^2)(t-\mu^2)}\,,
 \end{align}
\end{subequations}
where $\mu^2= m \overline m$.
The other orderings are obtained by massive versions of the BCJ relations,
e.g.~$ A_4(1\Phi,3\Phi,2V,4V) = (s- \mu^2)/u\,A_4(1\Phi,2V, 3\Phi,4V)$
(see \Sec{sec:bcjRels}).
These amplitudes are consistent with the non-zero component
amplitudes that can also be derived from the Lagrangian in \App{sec:Lagrangians}:\footnote{Amplitudes given here are correct up to overall signs.}
\begin{subequations}
\begin{align}
\label{comp1}
A_4(1\Psi^a,2A^+,3\Psi^b,4A^+)&= -i   \frac{\overline m \spa{1^a}.{3^b} \spb{2}.{4}^2}{(s-\mu^2)(t-\mu^2)}\,,  \\
A_4(1\phi,2A^+,3\bar{\phi},4A^+)&= i   \frac{\mu^2  \spb{2}.{4}^2}{(s-\mu^2)(t-\mu^2)}\,,  \\
A_4(1\Psi^a,2\lambda^+,3{\phi},4A^+)&=-i   \frac{ \overline m  \langle 1^a 2\rangle \spb{2}.{4}^2}{(s-\mu^2)(t-\mu^2)}\,, \\
A_4(1{\phi},2\lambda^+,3{\phi},4\lambda^+)&=-i   \frac{ \overline m \spa{2}.{4} \spb{2}.{4}^2}{(s-\mu^2)(t-\mu^2)}\,,
\end{align}
\end{subequations}
where $\Psi^a$ is the on-shell Majorana spinor.
Using the $\cN=1$ chiral multiplet $\Phi$ given in \eqn{eq:longNeq1}
one can confirm these all come from \eqn{sector1}.
For \eqn{sector2} the full set of component amplitudes is
\begin{subequations}
\begin{align}
A_4(1\phi,2A^-,3\bar {\phi},4A^+)&=-i   \frac{ \langle 2| 3 | 4]^2 }{(s-\mu^2)(t-\mu^2) }\,,  \\
A_4(1\Psi^a,2A^-,3\Psi^b,4A^+)&=i   \frac{\langle 2| 3 | 4]  (\spb{1^a}.{4} \spa{3^b}.{2}+\spb{3^b}.{4} \spa{1^a}.{2} ) }{(s-\mu^2)(t-\mu^2) } \,, \\
A_4(1\Psi^a,2\lambda^-,3\bar {\phi},4A^+)&=-i   \frac{ \langle2|3|4] \langle1^a|2|4]  }{(s-\mu^2)(t-\mu^2) } \,, \\
A_4(1\Psi^a,2A^-,3{\phi},4\lambda^+)&=i  \frac{ \langle2|3|4] [1^a|4|2\rangle  }{(s-\mu^2)(t-\mu^2)}\,,  \\
A_4(1\Psi^a,2\lambda^-,3\Psi^b,4\lambda^+)&=-i  \frac{ \langle2|3|4] (  [1^a| 4 | 3^b \rangle +  \langle 1^a| 4 | 3^b] )}{(s-\mu^2)(t-\mu^2) } \,, \\
A_4(1\bar{\phi},2\lambda^-,3\phi,4\lambda^+)&=i  \frac{ \langle2|3|4]}{s-\mu^2} \,.
\end{align}
\end{subequations}
The component amplitudes coming from \eqn{sector3} are CPT conjugates of the ones in \eqn{comp1}.

Next we consider the mass configuration $\langle m_1,m_2,m_3, 0 \rangle$
with three distinct-flavor chiral multiplets and one vector multiplet.
Following our discussion in \Sec{sec:bcfw},
the color-ordered superamplitudes are
\begin{align} \label{positiveV}
A_4(1\Phi^{(1)},2\Phi^{(2)},3\Phi^{(3)},4V^+)\!&=\!- i\delta^{2}(Q) \frac{[4|13|4]\!+\!(\overline{m}_1 [\fq_2 4] [4 \fq_3]\!+\!{\rm cyclic}(123))}{(s- \mu^2_3)(t- \mu^2_1)}\,,\\
 A_4(1\Phi^{(1)},2\Phi^{(2)},3\Phi^{(3)},4V^-) &= i\delta^{2}(Q)  \frac{m_1 [\fq_1|2|4\rangle+\mu_1^2 \langle 4 \fq_2 \rangle  +{\rm cyclic}(123)}{(s- \mu^2_3)(t- \mu^2_1)} \nn \\
&\qquad + i \frac{\overline{\delta^{2}(Q)\eta_4^1} \langle 4|13|4 \rangle}{(s- \mu^2_3)(t- \mu^2_1)} \,,
\end{align}
where for simplicity have set the flavor tensor $T^{123}=1$.
In the last term one should apply CPT conjugation through the Grassmann-Fourier transform,
which results in a degree-5 polynomial in the Grassmann variables.
This piece is the CPT conjugate of the first term in \eqn{positiveV}.

With four massive chiral multiplets there are more possibilities depending on the combination of flavors.
First consider a generic mass configuration
$\langle m_1,m_2,m_3,m_4 \rangle$ with four chiral multiplets,
which may or may not be distinct.
Assuming that the $s$-channel amplitude only contains a single internal massive chiral multiplet of complex mass $m_5$ then this part of the superamplitude is
\begin{align} \label{S_channel_Amp}
\begin{aligned}
&A_4(1\Phi^{(1)},2\Phi^{(2)}|3\Phi^{(3)},4\Phi^{(4)})=
i \frac{T^{125}T^{534}}{s-\mu_5^2} \Big[  m_5 \, \delta^{2}(Q)  - \overline m_{5} \, \overline{\delta^{2}(Q)}\,\\
&\qquad\qquad\qquad\!\!\!\!
+\delta^{2}(Q) \big(\spbq{\fq_1 \fq_2} {+} \spbq{\fq_3 \fq_4} {-} \overline{m}_1(\eta_2)^2{-} \overline{m}_2(\eta_1)^2 {-} \overline{m}_3(\eta_4)^2{-}\overline{m}_4(\eta_3)^2  \big)
 \Big]\,,
\end{aligned}
\end{align}
where $(\eta_i)^2=\eta_i^1 \eta_i^2$ and $\overline{\delta^{2}(Q)}$ denotes the CPT conjugate of the supermomentum delta function --- see \eqn{d2Q_Fourier} for an explicit expression.

Let us now specialize to the mass configuration
$\langle m_1,m_2,m_2,m_1 \rangle$
with two distinct chiral multiplets.
The $s$-channel amplitude
$A_4(1\Phi^{(1)},2\Phi^{(2)}|3\Phi^{(2)},4\Phi^{(1)})$
is given by \eqn{S_channel_Amp} after setting $m_3=m_2$ and $m_4=m_1$.
However, we can now also have a gluon exchange in the $t$-channel.
As it corresponds to a gauge-invariant object
we call it the {\it $t$-channel amplitude}:
 \begin{eqnarray} \label{T_channel_Amp}
 A_4(2\Phi^{(2)},3\Phi^{(2)}|4\Phi^{(1)}, 1\Phi^{(1)})  &=& i \frac{ \delta^{2}(Q) }{t} \Big(\spbq{\fq_1 \fq_3}+\spbq{\fq_2 \fq_4}-\spbq{\fq_1 \fq_2} -\spbq{\fq_3 \fq_4}   \Big)\,.
\end{eqnarray}
The full color-ordered amplitude is given by the sum of the two channels:
\begin{align}
\begin{aligned}
&A_4(1\Phi^{(1)},2\Phi^{(2)},3\Phi^{(2)},4\Phi^{(1)})\\
&\qquad=  A_4(1\Phi^{(1)},2\Phi^{(2)}|3\Phi^{(2)},4\Phi^{(1)}) +A_4(2\Phi^{(2)},3\Phi^{(2)}|4\Phi^{(1)}, 1\Phi^{(1)}) \,.
\end{aligned}
\end{align}
By superimposing the $s$- and $t$-channel partial amplitudes
we can obtain any other external mass configuration.
For example, the color-ordered amplitude with four
identical-flavored chiral multiplets and mass configuration  $\langle m_1,m_1,m_1,m_1 \rangle$
is given by the sum of two vector contributions:
 \begin{align}
  \begin{aligned}
 A_4(1\Phi^{(1)}, 2\Phi^{(1)},3\Phi^{(1)},4\Phi^{(1)})  = i \delta^{2}(Q)  \Big[&\frac1{t} \Big( \spbq{\fq_1 \fq_3}+\spbq{\fq_2 \fq_4}   -\spbq{\fq_1 \fq_2} -\spbq{\fq_3 \fq_4}\Big)\\
 + &\frac{1 }{s} \Big( \spbq{\fq_1 \fq_3}+\spbq{\fq_2 \fq_4} -\spbq{\fq_2 \fq_3} -\spbq{\fq_1 \fq_4}  \Big)  \Big]\,.
  \end{aligned}
\end{align}
We now proceed to consider the BCJ tree amplitude relations.

\subsection{Color-kinematics duality and BCJ relations}
\label{sec:bcjRels}

Any four-point amplitude with at least one massless vector can be shown to obey the rules of color-kinematics duality.
An easy way to demonstrate this is by noting that the massive BCJ relations hold for the color-ordered amplitudes. For example,
\bea
A_4(1V,3V,2V,4V) &=& \frac{s}{u} A_4(1V,2V,3V,4V)\,, \nn \\
A_4(1\Phi,3\Phi,2V,4V) &=& \frac{s- \mu^2_3}{u} A_4(1\Phi,2V, 3\Phi,4V)\,, \\
A_4(1\Phi,3\Phi,2\Phi,4V) &=& \frac{s- \mu^2_3}{u- \mu^2_2} A_4(1\Phi,2\Phi, 3\Phi,4V)\,. \nn
\eea
With four identical-flavor chiral multiplets one can also
easily spot a kinematic Jacobi identity.
Defining the numerator with vector exchange in the $s$-channel as
\be
 n_V(1\Phi^1, 2\Phi^1,3\Phi^1,4\Phi^1) = i \delta^{2}(Q)  \Big( \spbq{\fq_1 \fq_3}+\spbq{\fq_2 \fq_4} -\spbq{\fq_2 \fq_3} -\spbq{\fq_1 \fq_4}  \Big) \,,
\ee
the Jacobi identity is
\be\label{1flavorID}
n_V(1\Phi^1, 2\Phi^1,3\Phi^1,4\Phi^1) +n_V(3\Phi^1,1\Phi^1, 2\Phi^1,4\Phi^1) +n_V( 2\Phi^1,3\Phi^1,1\Phi^1,4\Phi^1)\!=\!0.
\ee

In the special case of an $SO(3)$ flavor group
(which may be broken by the mass terms)
there exists a bonus kinematic Jacob identity for amplitudes with four chiral multiplets.
Defining the numerators for the chiral exchange as
\begin{align}
\begin{aligned}
n_\Phi(1\Phi^1, 2\Phi^2&,3\Phi^2,4\Phi^1) = i T^{123}T^{321} \Big[  m_3 \, \delta^{2}(Q)  - \overline m_{3} \, \overline{\delta^{2}(Q)}\,    \\
 &+\delta^{2}(Q) \big(\spbq{\fq_1 \fq_2} {+} \spbq{\fq_3 \fq_4} {-} \overline{m}_1(\eta_2)^2{-} \overline{m}_2(\eta_1)^2 {-} \overline{m}_2(\eta_4)^2{-}\overline{m}_1(\eta_3)^2  \big)
 \Big]\,,
\end{aligned}
\end{align}
where $T^{123}T^{321}= \epsilon^{123}\epsilon^{321}=-1$
and $m_3$ is the mass of the third flavor multiplet
(not external particle three),
the identity is
\begin{align}
\begin{aligned}\label{3flavorID}
&n_\Phi(1\Phi^1, 2\Phi^2,3\Phi^2,4\Phi^1)-n_\Phi(1\Phi^1,3\Phi^2,2\Phi^2,4\Phi^1) \\
&\qquad=i \delta^{2}(Q) (\spbq{\fq_1 \fq_3} +\spbq{\fq_2 \fq_4}-\spbq{\fq_1 \fq_2} - \spbq{\fq_3 \fq_4})
= n_V(2\Phi^2,3\Phi^2,4\Phi^1,1\Phi^1)\,.
\end{aligned}
\end{align}
The last expression is clearly identifiable as the numerator of the vector exchange diagram.

\section{Diagrammatic identities for $\cN=1$ SYM with matter}
\label{sec:diagIden}

Having reviewed the on-shell formalism and tree-level amplitudes with
$\cN=1$ supersymmetries we now proceed to loop level.
Our construction of color-kinematics-satisfying ``color-dual'' numerators
for $\cN=1$ SYM theories with matter largely follows the discussion of $\cN=2$ SQCD in
\rcites{Johansson:2014zca,Johansson:2017bfl}.
In that case there exists a useful additional set of constraints imposable on the color-dual numerators,
which follow from the observation that $\cN=2$ SQCD with $N_f=1$ matter flavors in
the adjoint representation of the gauge group is equivalent to $\cN=4$ SYM.

To extend this concept to an $\cN=1$ SYM theory with matter
let us first discuss the field content of massless $\cN=1$~SQCD from a slightly different point of view.
The $\cN=1$ supersymmetric multiplets can be conveniently projected out from their $\cN=2$ SQCD and $\cN=4$ SYM siblings.
This procedure will allow us to impose strong constraints on the $\cN=1$ SQCD amplitudes at one and two loops,
which will in turn significantly simplify the computation at hand.

\subsection{Supersymmetric decomposition}
\label{sec:susydecomp}

The idea behind supersymmetric decomposition is that we can recycle known supersymmetric amplitudes when constructing new amplitudes with less supersymmetry. This works because most of the necessary ingredients are already present in the maximally supersymmetric $\cN=4$ SYM theory,
albeit packaged so as to manifest the supersymmetry.
Furthermore, the amplitudes with a high degree of supersymmetry will have properties that makes them easier to calculate, hence they will in general be determined before those of lower supersymmetry.

Here we elaborate on two decomposition routes
for obtaining $\cN=1$ SYM amplitudes starting from $\cN=4$ SYM.
The first uses $\cN=2$ SYM as an intermediate step:
\be
(\cN=4)   \xrightarrow{SU(4)\rightarrow SU(2)\times U(1)} (\cN=2) \xrightarrow{SU(2)\rightarrow U(1)} (\cN=1)\,,
\ee
and in the second one directly obtains $\cN=1$ SYM but with more manifest flavor symmetry,
\be
(\cN=4)   \xrightarrow{SU(4)\rightarrow U(1)\times SU(3)}  (\cN=1)\,.
\ee
We will first explain the decomposition in terms of the asymptotic states, which will then lead to decomposition rules for cubic loop graphs.

Starting from the massless on-shell $\cN=4$ SYM superfield\footnote{The Grassmann variables $\Neta^I$ are charged under the $SU(4)$ R symmetry and are distinct from the previously introduced $\eta^a$ variables that transform in the $SU(2)$ little group.}
\begin{equation}
  \cV_{\cN=4} = A^+ + \psi_I^+ \Neta^I  + \frac{1}{2}\phi_{IJ}  \Neta^I \Neta^J  + \frac{1}{3!} \psi_{IJK}^- \Neta^I\Neta^J\Neta^K  + A^- \Neta^1\Neta^2\Neta^3\Neta^4 \,, \\
\end{equation}
we can decompose it into $\cN=2$ components by breaking the R-symmetry as $SU(4)\rightarrow SU(2)\times U(1)$. The result is
\be \label{neq2decomp}
\cV_{\cN=4}=  V^{+}_{\cN=2} +   \Phi_{\cN=2} \,\Neta^3 +\overline{\Phi}_{\cN=2} \,  \Neta^4 +   V^{-}_{\cN=2} \Neta^3\Neta^4
\ee
where the $\cN=2$ components are
\begin{equation}
  \begin{aligned}
    V^{+}_{\cN=2} = & A^+ +  \psi_a^+ \Neta^a +  \phi_{12} \Neta^1\Neta^2 \,,~~~~~~ V^{-}_{\cN=2} =  \phi_{34} + \psi_{a34}^- \Neta^a +  A^- \Neta^1\Neta^2 \,,  \\
    \Phi_{\cN=2} =  & \psi_3^+ + \phi_{a3} \Neta^a+ \psi_{123}^- \Neta^1\Neta^2  \,,~~~~\overline{\Phi}_{\cN=2} = \psi_4^+ + \phi_{a4}\Neta^a + \psi_{124}^-\Neta^1\Neta^2  \,,
  \end{aligned}
\end{equation}
$a=1,2$ is the $SU(2)$ R-symmetry and the hypermultiplet has non-zero $U(1)$ charge.
There is no obstruction in considering the hypermultiplet to be masssive.
In the most general case, we can assign a complex mass
$m$ to $\Phi_{\cN=2}$ and mass $\overline m$ to $\overline \Phi_{\cN=2}$.

It is convenient to represent the supersymmetric decomposition graphically using Feynman-like particle lines. After packaging the two helicity components of the vector multiplet into a single real field, $\cV_{\cN=2}= V^{+}_{\cN=2} +   V^{-}_{\cN=2} \Neta^3\Neta^4 $, we can represent \eqn{neq2decomp} as
\begin{equation} \label{graphicalNeq2}
  \tikzsetnextfilename{decomp4Vector}
    \begin{tikzpicture}
       [
         line width=1pt,
         baseline={([yshift=-0.5ex]current bounding box.center)}
       ]
       \draw[gluon] (0,0) -- node[above=0.1] {\scriptsize$\cN=4$} (0.7,0);
       \node at (1,0) {$=$};
       \draw[gluon] (1.3,0) -- node[above=0.1] {\scriptsize$\cN=2$} (2.1,0);
       \node at (2.4,0) {$+$};
       \draw[quark] (2.7,0) -- node[above=0.1] {\scriptsize$\cN=2$} (3.4,0);
       \node at (3.7,0) {$+$};
       \draw[aquark] (4,0) -- node[above=0.1] {\scriptsize$\cN=2$} (4.7,0);
     \end{tikzpicture}
\end{equation}
where the curly lines are vector multiplets (curly $\cV_{\cN}$) and arrowed lines represent the hypermultiplet as a complex pair. As shown in ref.~\cite{Johansson:2017bfl}, this graphical decomposition can be straightforwardly promoted to off-shell internal lines of diagrams describing multiloop scattering processes. For off-shell lines, this equation represents a constraint that can be imposed on the off-shell states, which is compatible with the on-shell states, and thus compatible with unitarity.

As is relevant to this paper, the $\cN=2$ components can be further decomposed into $\cN=1$ components by breaking the R-symmetry as $SU(2)\rightarrow U(1)$:
\begin{equation}
  \begin{aligned}
V^{+}_{\cN=2} & =  V^{+}+  \Phi_{1}^+ \Neta^2  \,,   ~~~~~~ V^{-}_{\cN=2} = \Phi^{1}_- +  V^{-}\Neta^2\,,  \\
\Phi_{\cN=2} & =   \Phi_{2}^+ +  \Phi^{3}_- \Neta^2   \,, ~~~~~~   \overline{\Phi}_{\cN=2} =  \Phi_{3}^+ +  \Phi^{2}_- \Neta^2\,,
  \end{aligned}
\end{equation}
where the $\cN=1$ components are
\begin{equation}
  \begin{aligned}
    V^{+} = & A^+ + \psi_1^+  \Neta^1  \,,~~~~~~~ V^{-}=  \psi_{234}^- + A^- \Neta^1 \,,  \\
    \Phi_{A-1}^+=  & \psi_A^+ + \phi_{1A} \Neta^1 \,,~~~~~  \Phi^{A-1}_-= \frac{1}{2} \epsilon^{ABC} (\phi_{BC} + \psi_{1BC}^-  \Neta^1) \,,
  \end{aligned}
\end{equation}
and $A,B,C$ are $SU(3)$ flavor indices. 
As before, it is convenient to combine the two helicity components of the vector multiplet into a real field 
\be \label{realNeq1vector}
\cV_{\cN=1}= V^{+} +  V^{-} \theta \,,
\ee
where $\theta =\Neta^2\Neta^3\Neta^4$ is a Grassman-odd variable that we will use to conveniently describe the external states of the two-loop vector amplitude constructed in later sections. 

We can now represent the $(\cN=2)\rightarrow (\cN=1)$ supersymmetic decomposition graphically as
\begin{equation} \label{eq:decompNeq2}
  \begin{aligned}
    \tikzsetnextfilename{decompNeq2Vector}
    &\begin{tikzpicture}
       [
         line width=1pt,
         baseline={([yshift=-0.5ex]current bounding box.center)}
       ]
       \draw[gluon] (0,0) -- node[above=0.1] {\scriptsize$\cN=2$} (0.7,0);
       \node at (1,0) {$=$};
       \draw[gluon] (1.3,0) -- node[above=0.1] {\scriptsize$\cN=1$} (2.1,0);
       \node at (2.4,0) {$+$};
       \draw[rquark] (2.7,0) -- (3.4,0);
       \node at (3.7,0) {$+$};
       \draw[raquark] (4,0) -- (4.7,0);
     \end{tikzpicture}\\
    \tikzsetnextfilename{decompNeq2Hyper}
    &\begin{tikzpicture}
       [
         line width=1pt,
         baseline={([yshift=-0.5ex]current bounding box.center)}
       ]
       \draw[quark] (0,0) -- node[above=0.1] {\scriptsize$\cN=2$} (0.7,0);
       \node at (1,0) {$=$};
       \draw[gquark] (1.3,0) -- (2.1,0);
       \node at (2.4,0) {$+$};
       \draw[baquark] (2.7,0) -- (3.4,0);
     \end{tikzpicture}
  \end{aligned}
\end{equation}
where curly lines are vector multiplets ${\cal V}_\cN$ and the flavors of chiral multiplets are represented by red, green and blue lines. 
Similar to \eqn{graphicalNeq2}, we will impose these two equations for internal off-shell lines, and demonstrate that this is consistent with unitarity at the two-loop level. 

As an alternative to the above supersymmetric decomposition route, we can directly decompose the $\cN=4$ superfield into $\cN=1$ on-shell superfields by breaking $SU(4)\rightarrow U(1)\times SU(3)$, where the $SU(3)$ is a flavor symmetry:
\begin{align}
\cV_{\cN=4} &=  V^{+} +  \Phi_{A}^+ \Neta^{A}+   \Phi_{AB}^-  \Neta^A\Neta^B+   V^{-}  \Neta^2\Neta^3\Neta^4\,,
\end{align}
and $\Phi_{AB}^- =  \frac{1}{2} \epsilon_{ABC}  \Phi^{C}_- $.
This makes it clear that the chiral multiplets belong to the ${\bf 3}$ and $\bar {\bf 3}$ of $SU(3)$. 

For massive $\cN=1$ matter,
we should instead consider the long chiral multiplets introduced in \Sec{sec:neq1Multiplets}, \eqn{eq:longNeq1}.
We obtain these by constructing real matter representations,
thus breaking $SU(3)\rightarrow SO(3)$:
\be
\Phi^{(A)}= \Phi^{A}_- +\Phi_{A}^+\eta^2 \,,
\ee
where $\eta^2$ is the Grassmann variable that previously appeared in the massive supermultiplets. Graphically this implies that the long chiral multiples are symmetric combinations of arrowed graphs
\begin{equation}\label{eq:longToShort}
  \begin{aligned}
    \tikzsetnextfilename{lineToArrow1}
    &\begin{tikzpicture}
      [
        line width=1pt,
        baseline={([yshift=-0.5ex]current bounding box.center)}
      ]
      \draw[rline] (0,0) -- (0.7,0);
      \node at (1,0) {$=$};
      \draw[rquark] (1.3,0) -- (2.1,0);
      \node at (2.4,0) {$+$};
      \draw[raquark] (2.7,0) -- (3.4,0);
    \end{tikzpicture}\\
    \tikzsetnextfilename{lineToArrow2}
    &\begin{tikzpicture}
      [
        line width=1pt,
        baseline={([yshift=-0.5ex]current bounding box.center)}
      ]
      \draw[gline] (0,0) -- (0.7,0);
      \node at (1,0) {$=$};
      \draw[gquark] (1.3,0) -- (2.1,0);
      \node at (2.4,0) {$+$};
      \draw[gaquark] (2.7,0) -- (3.4,0);
     \end{tikzpicture}\\
    \tikzsetnextfilename{lineToArrow3}
    &\begin{tikzpicture}
       [
         line width=1pt,
         baseline={([yshift=-0.5ex]current bounding box.center)}
       ]
       \draw[bline] (0,0) -- (0.7,0);
       \node at (1,0) {$=$};
       \draw[bquark] (1.3,0) -- (2.1,0);
       \node at (2.4,0) {$+$};
       \draw[baquark] (2.7,0) -- (3.4,0);
    \end{tikzpicture}
  \end{aligned}
\end{equation}
In principle one could also define $\cN=1$ matter multiplets that
are antisymmetric combinations of the chiral halfs,
however such multiplets give rise to chiral gauge anomalies
and thus are not consistent at the quantum level.
Similarly, in a massive theory only the symmetric long chiral multiplet
are consistently defined. 

The supersymmetric decomposition should also be done for the interactions.
It is easiest to intuitively understand them by drawing all consistent cubic interactions, without elaborating on the details of the Lagrangian interaction terms (see \App{sec:Lagrangians} for details). The following three-point vertices of $\cN=1$ SYM will appear in the $(\cN=4)\rightarrow(\cN=1)$ decomposition:
\begin{equation}
  \tikzsetnextfilename{threePointDec1}
  \stackrel{\cN=4}{\gTreeTri[scale=1.4,all=gluon]{}}
  ~~\longrightarrow ~~
  \tikzsetnextfilename{threePointDec2}
  \stackrel{\cN=1}{\gTreeTri[scale=1.4,all=gluon]{}}\,,~
  \tikzsetnextfilename{threePointDec3}
  \gTreeTri[scale=1.4,eA=gluon,eB=rquark,eC=raquark]{}\,,~
  \tikzsetnextfilename{threePointDec4}
  \gTreeTri[scale=1.4,eA=gluon,eB=gquark,eC=gaquark]{}\,,~
  \tikzsetnextfilename{threePointDec5}
  \gTreeTri[scale=1.4,eA=gluon,eB=bquark,eC=baquark]{}\,,~
  \tikzsetnextfilename{threePointDec6}
  \gTreeTri[scale=1.4,eA=rquark,eB=gquark,eC=bquark]{}\,,~
  \tikzsetnextfilename{threePointDec7}
  \gTreeTri[scale=1.4,eA=raquark,eB=gaquark,eC=baquark]{}.
\end{equation}
The first vertex on the right-hand side is a pure vector multiplet interaction
(e.g. pure $\cN=1$ SYM), the next three represent a vector multiplet interacting with massless chiral multiplets of three different flavors (e.g. massless $\cN=1$ SQCD), and the last three represent Yukawa-type self-interactions among the chiral multiplets.
The flavor structures appearing in these interactions correspond to $\delta_A^B$,
$\epsilon^{ABC}$ and  $\epsilon_{ABC}$ tensors,
as inherited from $\cN=4$ SYM.
More general flavor tensors can in principle be obtained
by consistently replacing $\epsilon^{ABC} \rightarrow T^{ABC}$
(with some care) in the diagrams computed.

For massive matter we must consider long chiral multiplets with unarrowed lines,
and hence the $\cN=1$ SYM interactions will graphically appear as
\begin{equation}
  \tikzsetnextfilename{threeUndir1}
  \gTreeTri[scale=1.4,all=gluon]{}\,,~
  \tikzsetnextfilename{threeUndir2}
  \gTreeTri[scale=1.4,eA=gluon,eB=rline,eC=rline]{}\,,~
  \tikzsetnextfilename{threeUndir3}
  \gTreeTri[scale=1.4,eA=gluon,eB=gline,eC=gline]{}\,,~
  \tikzsetnextfilename{threeUndir4}
  \gTreeTri[scale=1.4,eA=gluon,eB=bline,eC=bline]{}\,,~
  \tikzsetnextfilename{threeUndir5}
  \gTreeTri[scale=1.4,eA=rline,eB=gline,eC=bline]{}\,.
\end{equation}
In particular, these are the cubic interactions of $\cN=1^*$ SYM, which is a well-known mass-deformation of $\cN=4$ SYM. For equal masses of the chiral multiplets the theory has an $SO(3)$ flavor symmetry. 

While $\cN=1^*$ SYM and $\cN=4$ SYM are directly related by adding mass parameters to the Lagrangian (see \App{sec:Lagrangians}), supersymmetric decomposition seeks to impose a more constraining relation that also works for individual contributions of a given scattering process. At this point we note that supersymmetric decomposition is more subtle for mass-deformed theories. The decomposition identities may not lift from the massless to the massive $\cN=1$ theory in a straightforward manner, if all matter is massive. The reason is that we require the $\cN=1$ vector multiplet to be massless and a chiral multipet to be massive, so they cannot straightforwardly fit into an $\cN=4$ multiplet.
We also cannot get around the problem (at intermediate steps) by considering a Higgs phase of the $\cN=1$ vector multiplet,
because that would change the degrees of freedom (it would have to eat a chiral multiplet). This problem does not happen for massive $\cN=2$ SYM  amplitudes, as obtained from supersymmetric decomposition~\cite{Johansson:2017bfl}, since the $\cN=2$ vector multiplet can become massive on the Coulomb branch without changing the degrees of freedom of the multiplet. We will discuss this subtlety more in the proper context of the two-loop amplitude that we compute in \Sec{sec:2Lcuts}.

The goal now is to use the derived supersymmetric decomposition identities for states and vertices in more complicated diagrams, such as one and two-loop amplitudes. However, first we will discuss further diagrammatic identities that come from imposing color-kinematics duality for $\cN=1$ SYM theories.

\subsection{Color-kinematics duality --- general considerations}

The idea of color-kinematics duality is to express an $L$-loop $D$-dimensional gauge theory amplitude in terms of only cubic graphs,
\begin{equation}
  \label{eq:colorKinDualAmp}
  \cA_m^{(L)}=i^{L-1}g_{\text{YM}}^{m+2L-2}\sum_{i\in\{\text{cubic graphs}\}}\int\!\frac{\d^{LD}\ell}{(2\pi)^{LD}}\frac{1}{S_i} \frac{n_i c_i}{D_i}\,,
\end{equation}
with color factors $c_i$, kinematic numerators $n_i$, propagator denominators $D_i$ and symmetry factors $S_i$. Then impose that the kinematic numerators fulfil the same Lie-algebraic relations as the color factors, such as the three-term relations coming from Jacobi identities,
\begin{align} \label{CKduality}
  c_i = c_j - c_k \quad \Leftrightarrow \quad n_i = n_j - n_k\,.
\end{align}
Kinematic numerators that obey color-kinematics duality are called BCJ numerators, or sometimes color-dual numerators. While \eqn{eq:colorKinDualAmp} has several ingredients that depend on the graph, the only a priori undetermined quantities are the kinematic numerators $n_i$. The quantities $c_i$, $D_i$ and $S_i$ are computed by standard Feynman-diagrammatic methods and are considered trivial for our purposes.  Due to the richness of algebraic identities between BCJ numerators --- possibly supplemented by other relations such as supersymmetric decomposition --- the number of independent numerators can be observed to be surprisingly small. Any subset of numerators that can be used to generate all others are called \emph{master numerators}. In later sections we will see how they can be used to greatly simplify the construction of gauge theory multi-loop amplitudes, which is a sufficient reason for studying color-kinematics duality.

A second central reason for finding BCJ numerators is that they can be used to construct gravitational amplitudes through the double copy. Replacing $c_i \rightarrow \tilde n_i$ and $g_\text{YM} \rightarrow \kappa/2$ (the gravitational coupling) in \eqn{eq:colorKinDualAmp} gives the gravitational amplitude
\begin{equation}
  \label{eq:doubleCopyAmp}
  \cM_m^{(L)}=i^{L-1}\Big(\frac{\kappa}{2}\Big)^{m+2L-2}\sum_{i\in\{\text{cubic graphs}\}}\int\!\frac{\d^{LD}\ell}{(2\pi)^{LD}}\frac{1}{S_i} \frac{n_i \tilde n_i}{D_i}\,,
\end{equation}
where $n_i$ and $\tilde n_i$ are sets of cubic-diagram numerators that may come from two different gauge theories. At least one set of numerators must manifestly obey the duality \eqn{CKduality}, but both sets must come from gauge theories that in principle admit the construction of such numerators.

The reason for imposing constraints on the numerators is that the gravitational amplitude must be invariant under a linearized diffeomorphism transformation, $\delta g_{\mu\nu} = \partial_\mu\varepsilon_\nu+ \partial_\nu\varepsilon_\mu$, applied to any external graviton state. Since in the double copy an external graviton state is obtained as a (symmetrized) tensor product of two vector polarizations, a linearized diffeomorphism transformation can be obtained by sending $\varepsilon_\mu\rightarrow p_\mu$ in one set of numerators, say $\tilde n_i$:\footnote{Since the graviton state is symmetric in the Lorentz indices, the transformation should also be applied to the other copy, $n_i\big|_{\varepsilon_\nu\rightarrow p_\nu}$, and then the two transformations may be added. }
\be
\sum_{i}\int\!\frac{\d^{LD}\ell}{(2\pi)^{LD}}\frac{1}{S_i} \frac{n_i (\tilde n_i\big|_{\varepsilon_\mu\rightarrow p_\mu}) }{D_i} = 0\,.
\ee
This identity is mapped through the double copy, using $n_i \rightarrow c_i$, to the corresponding Ward identity of the gauge theory,
\be
\sum_{i}\int\!\frac{\d^{LD}\ell}{(2\pi)^{LD}}\frac{1}{S_i} \frac{c_i (\tilde n_i\big|_{\varepsilon_\mu\rightarrow p_\mu}) }{D_i} = 0\,.
\ee
So from this consideration we can determine all the algebraic identities that the numerators $n_i$ must obey,
and they will simply correspond to the necessary Lie-algebra identities that the color factors $c_i$ must obey in order to guarantee gauge invariance of the gauge theory amplitude.

We use a diagrammatic notation to represent both the kinematical numerators~$n_i$ and color factors~$c_i$.
When required the (internal and external) legs may be dressed with momentum labels denoting the functional dependence of the corresponding numerator.
Furthermore, many of the diagrammatic identities are to be applied to (tree-level) subgraphs.
Meaning that the numerators, whose subgraphs are related by such identities, obey the corresponding functional constraint.

\subsection{Necessary color and numerator identities}

Here we give all color and numerator identities relevant for amplitudes in a
matter-coupled $\cN=1$ SYM theory.
The color identities are necessary for the gauge amplitude to be gauge invariant,
and the numerator identities are necessary
for the double copy to be diffeomorphism invariant. 

The Jacobi and commutation relations in an arbitrary matter
representation impose three-term numerator identities:
\begin{equation}\label{eq:jacsComs}
  \begin{aligned}
    f^{a_1a_2b}f^{ba_3a_4}&= f^{a_4a_1b}f^{ba_2a_3}
    - f^{a_2a_4b}f^{ba_3a_1}\\
    \tikzsetnextfilename{JacobiIdDiag1}
    \Leftrightarrow \gTreeS[scale=1.2,all=gluon,eLA=1,eLB=2,eLC=3,eLD=4]{} &=
    \tikzsetnextfilename{JacobiIdDiag2}
    \gTreeT[scale=1.2,all=gluon,eLA=1,eLB=2,eLC=3,eLD=4]{}
    \tikzsetnextfilename{JacobiIdDiag3}
    - \gTreeU[scale=1.2,all=gluon,eLA=1,eLB=2,eLC=3,eLD=4]{}\,,\\
    f^{ba_2a_1}T^b_{i_3\bar\imath_4}&= T^{a_2}_{i_3\bar\jmath}T^{a_1}_{j\bar\imath_4}-
    T^{a_1}_{i_3\bar\jmath}T^{a_2}_{j\bar\imath_4} = [T^{a_2},T^{a_1}]_{i_3\bar\imath_4}\\
    \tikzsetnextfilename{CommIdDiag1}
    \Leftrightarrow \gTreeS[scale=1.2,all=gluon,eC=rquark,eD=raquark,eLA=1,eLB=2,eLC=3,eLD=4]{} &=
    \tikzsetnextfilename{CommIdDiag2}
    \gTreeT[scale=1.2,all=gluon,eC=rquark,eD=raquark,iA=raquark,eLA=1,eLB=2,eLC=3,eLD=4]{}
    \tikzsetnextfilename{CommIdDiag3}
    - \gTreeU[scale=1.2,all=gluon,eC=rquark,eD=raquark,iA=raquark,eLA=1,eLB=2,eLC=3,eLD=4]{}\,,\\
    \tikzsetnextfilename{CommIdDiag4}
    \gTreeS[scale=1.2,all=gluon,eC=rline,eD=rline,eLA=1,eLB=2,eLC=3,eLD=4]{} &=
    \tikzsetnextfilename{CommIdDiag5}
    \gTreeT[scale=1.2,all=gluon,eC=rline,eD=rline,iA=rline,eLA=1,eLB=2,eLC=3,eLD=4]{}
    \tikzsetnextfilename{CommIdDiag6}
    - \gTreeU[scale=1.2,all=gluon,eC=rline,eD=rline,iA=rline,eLA=1,eLB=2,eLC=3,eLD=4]{}\,,
  \end{aligned}
\end{equation}
where we have illustrated both the case of arrowed (complex) matter and unarrowed (real) matter. We will illustrate both of these cases side by side in what follows, as it is pertinent to the understanding of the massless chiral multiplet (complex) and the long massive chiral multipet (real) of $\cN=1$ SYM. 

Given that we have cubic matter self-interactions we can have generic
Clebsch-Gordan coefficients $C_{ijk}$ controlling the gauge group structure.
This leads to an additional invariance relation for the $C_{ijk}$ tensor
that is also a three-term identity.
The corresponding numerator relation is illustrated as follows:
\begin{equation}\label{eq:clebschGordan}
  \begin{aligned}
    T^{a_1}_{i_2\bar\jmath}C_{ji_3i_4} &= T^{a_1}_{i_4\bar\jmath}C_{ji_2i_3}
    -T^{a_1}_{i_3\bar\jmath}C_{ji_4i_2}\\
    \tikzsetnextfilename{ThreeTermDiag1}
  \Leftrightarrow\gTreeS[scale=1.2,all=gluon,eB=rquark,eC=bquark,eD=gquark,iA=raquark,eLA=1,eLB=2,eLC=3,eLD=4]{} &=
  \tikzsetnextfilename{ThreeTermDiag2}
  \gTreeT[scale=1.2,all=gluon,eB=rquark,eC=bquark,eD=gquark,iA=gquark,eLA=1,eLB=2,eLC=3,eLD=4]{}
  \tikzsetnextfilename{ThreeTermDiag3}
  - \gTreeU[scale=1.2,all=gluon,eB=rquark,eC=bquark,eD=gquark,iA=baquark,eLA=1,eLB=2,eLC=3,eLD=4]{}\,,\\
  \tikzsetnextfilename{ThreeTermDiag4}
  \gTreeS[scale=1.2,all=gluon,eB=rline,eC=bline,eD=gline,iA=rline,eLA=1,eLB=2,eLC=3,eLD=4]{} &=
  \tikzsetnextfilename{ThreeTermDiag5}
  \gTreeT[scale=1.2,all=gluon,eB=rline,eC=bline,eD=gline,iA=gline,eLA=1,eLB=2,eLC=3,eLD=4]{}
  \tikzsetnextfilename{ThreeTermDiag6}
  - \gTreeU[scale=1.2,all=gluon,eB=rline,eC=bline,eD=gline,iA=bline,eLA=1,eLB=2,eLC=3,eLD=4]{}\,,
  \end{aligned}
\end{equation}
where we have assumed that the vertex is formed by three differently flavored matter multiplets that are all in the same representation of the gauge group.
This mild assumption corresponds to the situation that we encounter in supersymmetric decomposition of  $\cN=4$ SYM into  $\cN=1$ SYM.
For more general chiral matter the Clebsch-Gordan three-term identity is similar, but flavor and arrow assignments may differ. 

To complete the discussion on numerator identities imposed by color-kinematics duality, we should also discuss the symmetry properties of the distinct cubic vertices. For the pure vector multiplet interaction we always impose antisymmetry under a flip of two edges, which is dual to the antisymmetry property of $f^{abc}$.
For the interactions involving matter, in principle we have the choice of imposing symmetric or antisymmetric flip properties, however we will by convention always assume the antisymmetry of two edges. 
Diagrammatically this can be expressed as
\begin{equation} \label{vertexsigns}
  \tikzsetnextfilename{threePointFlip1}
  \gTreeTri[scale=1.4,all=gluon,eLA=3,eLB=1,eLC=2]{}
  \tikzsetnextfilename{threePointFlip2}
  = - \, \gTreeTri[scale=1.4,all=gluon,eLA=3,eLB=2,eLC=1]{}\,,~~~~~
  \tikzsetnextfilename{threePointFlip3}
   \gTreeTri[scale=1.4,eA=gluon,eB=rline,eC=rline,eLA=3,eLB=1,eLC=2]{}= \varmp \,
   \tikzsetnextfilename{threePointFlip4}
   \gTreeTri[scale=1.4,eA=gluon,eB=rline,eC=rline,eLA=3,eLB=2,eLC=1]{}\,,~~~~~
   \tikzsetnextfilename{threePointFlip5}
    \gTreeTri[scale=1.4,eLA=3,eLB=1,eLC=2]{}
    \tikzsetnextfilename{threePointFlip6}
    = \varmp \,  \gTreeTri[scale=1.4, eLA=3,eLB=2,eLC=1]{}\,,
\end{equation}
where the parentheses around the plus sign remind us that
such a sign is allowed for an appropriate gauge-group representation.
The minus sign in the second identity corresponds to working with gauge group
generators that satisfy $T^{a}_{i\bar\jmath}=-T^{a}_{\bar\jmath i}$.
This is compatible with a complex or real representation of the gauge group.\footnote{
For a pseudoreal/symplectic representation the generators are symmetric, and in that case a plus sign has to be chosen in the second identity in \eqn{vertexsigns}.}
The sign of the pure matter interaction is inherited by assuming that the Clebsch-Gordan coefficients $C_{ijk}$ are totally antisymmetric.
This is convenient because we will sometimes assume the matter transforms
in the adjoint representation, hence $C_{ijk}$ becomes $f^{abc}$. 

These relations form the complete set of identities that numerators of a color-kinematics-dual amplitude must fulfill.
The corresponding color-tensor identities are sufficient to guarantee gauge invariance of an $\cN=1$ gauge theory amplitude, from which it follows that the described numerator identities are sufficient to guarantee diffeomorphism invariance of the gravitational double-copy amplitude  that is constructed from the same numerators.

\subsection{Optional numerator relations}

In refs.~\cite{Johansson:2014zca,Johansson:2017bfl} it was observed that an optional \emph{two-term identity} for graphs with complex matter is also helpful for obtaining simple color-dual numerators in $\cN=2$ SQCD.
We will use equivalent identities for graphs with internal matter loops in $\cN=1$ SYM,
involving both complex and real matter.
Diagrammatically the numerator identities can be represented as
\begin{equation}\label{eq:twoTerm}
  \begin{aligned}
    \tikzsetnextfilename{optionalIDDiag1}
    \gTreeS[scale=1.2,eA=rquark,eB=raquark,eC=rquark,eD=raquark,iA=gluon]{} &=
    \tikzsetnextfilename{optionalIDDiag2}
    \gTreeT[scale=1.2,eA=rquark,eB=raquark,eC=rquark,eD=raquark,iA=gluon]{}\,,\\
    \tikzsetnextfilename{optionalIDDiag3}
    \gTreeS[scale=1.2,eA=rline,eB=rline,eC=rline,eD=rline,iA=gluon]{} &=
    \tikzsetnextfilename{optionalIDDiag4}
    \gTreeT[scale=1.2,eA=rline,eB=rline,eC=rline,eD=rline,iA=gluon]{}
    \tikzsetnextfilename{optionalIDDiag5}
    -\gTreeU[scale=1.2,eA=rline,eB=rline,eC=rline,eD=rline,iA=gluon]{}\,.
  \end{aligned}
\end{equation}
The three-term identity we already exposed in \eqn{1flavorID} when considering massive four-point superamplitude numerators. For massless multiplets the three-term identity follows from the two-term identity, and vice versa, if one assumes that the real matter is built out of complex matter components. However, for massive matter the long chiral multiplet cannot be decomposed while preserving supersymmetry, and so {\it a priori} it is not obvious that three-term identity should follow from the massless two-term identity. It is worth mentioning that a similar three-term identity was used in ref.~\cite{Carrasco:2020ywq} to describe numerators for massive scalars.

Note that these identities do not hold in general for the corresponding color factors,
so they are not explicitly required by color-kinematics duality.
However, for appropriate gauge-group representations the color factor identity
may exist and then one may find that enhanced symmetries are present.
For example, if one has a single massless chiral multiplet that transforms in the adjoint then $\cN=1$ SYM enhances to $\cN=2$ SYM,
and the above identities will correspond to Fiertz identities
necessary for enhanced supersymmetry~\cite{Chiodaroli:2013upa}.
If the chiral multiplet remains massive, but still in the adjoint representation,
the theory corresponds to a mass deformation of $\cN=2$ SYM
that we may call $\cN=1^{**}$ SYM ---
see \App{massiveFromSUSY} for the Lagrangian.

Extending the same ideas to graphs involving three different flavors of matter, we propose a new set of optional numerator identities:
\begin{equation}\label{eq:twoTermFl}
  \begin{aligned}
    \tikzsetnextfilename{optionalIDNewDiag1}
    \gTreeS[scale=1.2,eA=gquark,eB=gaquark,eC=baquark,eD=bquark,iA=gluon]{} &=
    \tikzsetnextfilename{optionalIDNewDiag2}
    \gTreeT[scale=1.2,eA=gquark,eB=gaquark,eC=baquark,eD=bquark,iA=rquark]{}\,,\\
    \tikzsetnextfilename{optionalIDNewDiag3}
    \gTreeS[scale=1.2,eA=gline,eB=gline,eC=bline,eD=bline,iA=gluon]{} &=
    \tikzsetnextfilename{optionalIDNewDiag4}
    \gTreeT[scale=1.2,eA=gline,eB=gline,eC=bline,eD=bline,iA=rline]{}
    \tikzsetnextfilename{optionalIDNewDiag5}
    -\gTreeU[scale=1.2,eA=gline,eB=gline,eC=bline,eD=bline,iA=rline]{}\,.
  \end{aligned}
\end{equation}
The second identity appeared in \eqn{3flavorID} when considering the massive four-point superamplitudes. 

As before one can show that the massless three-term identity follows from the two-term identity, and vice versa, if one assumes that the real matter is built out of complex matter components. 
The same warning applies:
the corresponding color-factor identity does not hold in general,
but it may hold for special gauge group representations. For example, for three massless chiral multiplets that transform in the adjoint, the theory will have enhanced symmetry as it becomes equivalent to $\cN=4$ SYM. In that case the color identity is simply the Jacobi identity of the adjoint generators, and the numerator identity is equivalent to Fiertz identities~\cite{Chiodaroli:2013upa} that guarantee the enhanced supersymmetry.
If the chiral multiplets remain massive, but still in the adjoint, the theory corresponds to a mass deformation of $\cN=4$ SYM that is usually called $\cN=1^{*}$ SYM. 

Which combination of the optional identities~(\ref{eq:twoTerm}) and~(\ref{eq:twoTermFl}) can be imposed consistently on general $\cN=1$ SYM numerators remains to be seen, but in this paper we confirm that up to two loops
we have found a four-vector integrand with massive matter that obeys both the necessary and optional numerator identities, up to a minor defect that will be discussed in \Sec{sec:2Lcuts}.
This non-trivial confirmation of the optional identities suggest that they may be useful for more general calculations at higher loops and multiplicities.

\subsection{Diagram and flavor symmetries of $\cN=1$ SYM numerators}\label{sec:flavor}

When working with loop-level numerators it is best assumed that they
fulfill the same automorphism symmetries as the corresponding cubic graphs,
which are labeled by (loop) momenta and particle types.
In physics language we refer to such constraints as imposing (manifest)
\emph{crossing symmetry} on the graphs,
as this will guarantee the standard crossing symmetry of the full amplitude
after summing over all graphs and integrating over the loop momenta.
The main practical advantage of having crossing-symmetric numerators is
that we have fewer graphs to work with,
and also fewer consistency checks to perform as most such checks
are related by trivial relabellings. 

An example of a crossing symmetry constraint for the double-box (db) diagram is
\begin{equation} \label{crossingconstraint}
  \tikzsetnextfilename{boxboxSymm1}
  n_{\mathrm{db}}(1234;\ell_1,\ell_2) = \gBoxBox[all=gluon,eLA=1,eLB=2,eLC=3,eLD=4,iLA=$\ell_1\uparrow$,iLD=$\uparrow\ell_2$]{}
  \tikzsetnextfilename{boxboxSymm2}
  =\gBoxBox[all=gluon,eLA=2,eLB=1,eLC=4,eLD=3,iLA=$\ell_1\downarrow$,iLD=$\downarrow\ell_2$]{}
  =n_{\mathrm{db}}(2143;-\ell_1,-\ell_2)\,.
\end{equation}
There are similar (and more complex) constraints for other graphs,
with and without matter, planar and non-planar diagrams. 

Next, there are isomorphism identities that relate two different graphs,
which arise due to the existence of a symmetry that is not manifest.
In our case it is the flavor symmetry
that may or may not be broken by the mass spectrum of the chiral matter.
Regardless of whether it is broken or not,
we can impose identities that transform from one flavor
to a second flavor if the mass is also changed.
We distinguish between two cases, \emph{overall} permutations of flavor assignments and permutations of flavor for individual \emph{subloops}.
For the former, some examples are
\begin{equation}\label{eq:flavorSym}
  \begin{aligned}
    \tikzsetnextfilename{flavorSymDiag1}
    \gBoxBox[all=gluon,iA=rline,iB=rline,iC=rline,iD=rline,iE=rline,iF=rline]{
      \node at (0.9,1.2) {};
    }
    &=
    \tikzsetnextfilename{flavorSymDiag2}
    \gBoxBox[all=gluon,iA=gline,iB=gline,iC=gline,iD=gline,iE=gline,iF=gline]{
      \node at (0.9,1.2) {};
    }
    =
    \tikzsetnextfilename{flavorSymDiag3}
    \gBoxBox[all=gluon,iA=bline,iB=bline,iC=bline,iD=bline,iE=bline,iF=bline]{
      \node at (0.9,1.2) {};
    }\,,
    \\
    \tikzsetnextfilename{flavorSymDiag4}
    \gBoxBox[all=gluon,iA=gline,iB=gline,iC=bline,iD=bline,iE=bline,iF=gline,iG=rline]{
      \node at (0.9,1.2) {};
    }
    &=
    \tikzsetnextfilename{flavorSymDiag5}
    \gBoxBox[all=gluon,iA=rline,iB=rline,iC=bline,iD=bline,iE=bline,iF=rline,iG=gline]{
      \node at (0.9,1.2) {};
    }
    =
    \tikzsetnextfilename{flavorSymDiag6}
    \gBoxBox[all=gluon,iA=rline,iB=rline,iC=gline,iD=gline,iE=gline,iF=rline,iG=bline]{
      \node at (0.9,1.2) {};
    }\,,
    \\
    \tikzsetnextfilename{flavorSymDiag7}
    \gTriTri[scale=0.8,all=gluon,iA=rline,iB=rline,iC=rline,iE=rline,iF=rline,iG=rline]{
      \node at (0.9,1.2) {};
    }
    &=
    \tikzsetnextfilename{flavorSymDiag8}
    \gTriTri[scale=0.8,all=gluon,iA=gline,iB=gline,iC=gline,iE=gline,iF=gline,iG=gline]{
      \node at (0.9,1.2) {};
    }
    =
    \tikzsetnextfilename{flavorSymDiag9}
    \gTriTri[scale=0.8,all=gluon,iA=bline,iB=bline,iC=bline,iE=bline,iF=bline,iG=bline]{
      \node at (0.9,1.2) {};
    }\,,
    \\
    \tikzsetnextfilename{flavorSymDiag10}
    \gTriTri[scale=0.8,all=gluon,iA=gline,iB=gline,iC=gline,iE=bline,iF=bline,iG=bline]{
      \node at (0.9,1.2) {};
    }
    &=
    \tikzsetnextfilename{flavorSymDiag11}
    \gTriTri[scale=0.8,all=gluon,iA=rline,iB=rline,iC=rline,iE=bline,iF=bline,iG=bline]{
      \node at (0.9,1.2) {};
    }
    =
    \tikzsetnextfilename{flavorSymDiag12}
    \gTriTri[scale=0.8,all=gluon,iA=gline,iB=gline,iC=gline,iE=bline,iF=bline,iG=bline]{
      \node at (0.9,1.2) {};
    }\,.
  \end{aligned}
\end{equation}
The latter, subloop-type, flavor identities are only relevant for graphs with two separate closed matter loops, e.g.
\begin{equation}\label{eq:flavorSym2}
  \tikzsetnextfilename{flavorSymDiag13}
  \gTriTri[scale=0.8,all=gluon,iA=rline,iB=rline,iC=rline,iE=rline,iF=rline,iG=rline]{}=
  \tikzsetnextfilename{flavorSymDiag14}
  \gTriTri[scale=0.8,all=gluon,iA=gline,iB=gline,iC=gline,iE=bline,iF=bline,iG=bline]{}\,.
\end{equation}
So the numerators are insensitive to whether
the matter in the two loops are of the same or of different types.  
There also exist corresponding automorphism/isomorphism identities for graphs dressed with arrows (complex matter).
These are straightforward generalizations of the given ones, so we refrain from further details.

The two-loop numerators that we will present in the next section fulfill all overall flavor symmetry relations.
On thee other hand, some of the subloop flavor symmetries are broken by a sign flip
\begin{equation}\label{eq:brokenFlavor}
  \begin{aligned}
  \tikzsetnextfilename{brokenFlavorDiag1}
  \gBoxBubB[all=gluon,iA=rline,iB=rline,iC=rline,iD=rline,iF=rline,iG=rline]{} &= -
  \tikzsetnextfilename{brokenFlavorDiag2}
  \gBoxBubB[all=gluon,iA=bline,iB=bline,iC=bline,iD=bline,iF=gline,iG=gline]{}\,,\\
  \tikzsetnextfilename{brokenFlavorDiag3}
  \gPentaTad[all=gluon,iA=rline,iB=rline,iC=rline,iD=rline,iE=rline,iG=rline]{} &= -
  \tikzsetnextfilename{brokenFlavorDiag4}
  \gPentaTad[all=gluon,iA=bline,iB=bline,iC=bline,iD=bline,iE=bline,iG=gline]{}\,,\\
  \tikzsetnextfilename{brokenFlavorDiag5}
  \gBoxTadA[all=gluon,iA=rline,iB=rline,iC=rline,iD=rline,iG=rline]{} &= -
  \tikzsetnextfilename{brokenFlavorDiag6}
  \gBoxTadA[all=gluon,iA=bline,iB=bline,iC=bline,iD=bline,iG=gline]{}\,.
  \end{aligned}
\end{equation}
The sign flip can be explained by the decomposition identities discussed in the next subsection.
Each pair of numerators in the above equations need to add up to a $\cN=2$ numerator.
For the $\cN=2$ representation that we chose, these numerators turn out to be zero.
Hence, the $\cN=1$ numerators differ by a sign, also given that they cannot be chosen to be vanishing due to other constraints.

We will also discuss in Sec.~\ref{sec:2Lcuts} that a slight deformation of the massive cuts is needed to find a representation that fulfills the sub-loop flavor symmetry for the bow-tie graph given in \eqn{eq:flavorSym2}.

\subsection{Decomposition of $\cN=2$ SQCD diagrams}\label{sec:decomp}

Let us now finish the discussion of supersymmetric decompositions 
that we started in \Sec{sec:susydecomp}.
Recycling the two-loop calculation in ref.~\cite{Johansson:2017bfl}, we demand that the known numerators for $\cN=2$ SQCD, presented in the same reference, decompose into their $\cN=1$ SYM components according to \eqn{eq:decompNeq2}.
For example, the decomposition identity of an $\cN=2$ SYM double-box graph
with purely vectorial content reads
\begin{equation}\label{eq:decompPure}
  \begin{aligned}
    \tikzsetnextfilename{decompIDsDiag1}
    \gBoxBox[all=gluon]{
      \node at (0.9,1.2) {$\cN=2$};
    }
    =
    \tikzsetnextfilename{decompIDsDiag2}
    \gBoxBox[all=gluon]{
      \node at (0.9,1.2) {$\cN=1$};
    }
    &+
    \tikzsetnextfilename{decompIDsDiag3}
    \gBoxBox[all=gluon,iA=rquark,iB=rquark,iC=rquark,iD=rquark,iE=rquark,iF=rquark]{
      \node at (0.9,1.2) {$\cN=1$};
    }
    +
    \tikzsetnextfilename{decompIDsDiag4}
    \gBoxBox[all=gluon,iA=rquark,iB=rquark,iF=rquark,iG=rquark]{
      \node at (0.9,1.2) {$\cN=1$};
    }
    +
    \tikzsetnextfilename{decompIDsDiag5}
    \gBoxBox[all=gluon,iC=rquark,iD=rquark,iE=rquark,iG=raquark]{
      \node at (0.9,1.2) {$\cN=1$};
    }\\
    \tikzsetnextfilename{decompIDsDiag6}
    &+\gBoxBox[all=gluon,iA=raquark,iB=raquark,iC=raquark,iD=raquark,iE=raquark,iF=raquark]{
      \node at (0.9,1.2) {$\cN=1$};
    }
    +
    \tikzsetnextfilename{decompIDsDiag7}
    \gBoxBox[all=gluon,iA=raquark,iB=raquark,iF=raquark,iG=raquark]{
      \node at (0.9,1.2) {$\cN=1$};
    }
    +
    \tikzsetnextfilename{decompIDsDiag8}
    \gBoxBox[all=gluon,iC=raquark,iD=raquark,iE=raquark,iG=rquark]{
      \node at (0.9,1.2) {$\cN=1$};
    }\\
    =
    \tikzsetnextfilename{decompIDsDiag9}
    \gBoxBox[all=gluon]{
      \node at (0.9,1.2) {$\cN=1$};
    }
    &+
    \tikzsetnextfilename{decompIDsDiag10}
    \gBoxBox[all=gluon,iA=rline,iB=rline,iC=rline,iD=rline,iE=rline,iF=rline]{
      \node at (0.9,1.2) {$\cN=1$};
    }
    +
    \tikzsetnextfilename{decompIDsDiag11}
    \gBoxBox[all=gluon,iA=rline,iB=rline,iF=rline,iG=rline]{
      \node at (0.9,1.2) {$\cN=1$};
    }
    +
    \tikzsetnextfilename{decompIDsDiag12}
    \gBoxBox[all=gluon,iC=rline,iD=rline,iE=rline,iG=rline]{
      \node at (0.9,1.2) {$\cN=1$};
    }\,.
  \end{aligned}
\end{equation}
On the two first lines the decomposition is given in terms of massless chiral matter, and on the last line we see that this recombines into long chiral multiplets. We see from this identity that the pure-vector numerator of $\cN=1$ SYM is completely determined in terms of the $\cN=2$ SYM pure-vector numerator and the $\cN=1$ matter numerators.
So we do not need to separately calculate the pure-vector numerator,
and can instead focus only on the $\cN=1$ matter contributions.

Taking this concept a step further,
we can also decompose $\cN=2$ SQCD graphs with closed matter loops.
For example,
\begin{equation}\label{eq:decompMatterBox1}
  \begin{aligned}
    \tikzsetnextfilename{decompIDsMatterDiag1}
    \gBoxBox[all=gluon,iA=quark,iB=quark,iC=quark,iD=quark,iE=quark,iF=quark]{
      \node at (0.9,1.2) {$\cN=2$};
    }
    &=
    \tikzsetnextfilename{decompIDsMatterDiag2}
    \gBoxBox[all=gluon,iA=gquark,iB=gquark,iC=gquark,iD=gquark,iE=gquark,iF=gquark]{
      \node at (0.9,1.2) {$\cN=1$};
    }
    +
    \tikzsetnextfilename{decompIDsMatterDiag3}
    \gBoxBox[all=gluon,iA=baquark,iB=baquark,iC=baquark,iD=baquark,iE=baquark,iF=baquark]{
      \node at (0.9,1.2) {$\cN=1$};
    }
    +
    \tikzsetnextfilename{decompIDsMatterDiag4}
    \gBoxBox[all=gluon,iA=gquark,iB=gquark,iC=baquark,iD=baquark,iE=baquark,iF=gquark,iG=raquark]{
      \node at (0.9,1.2) {$\cN=1$};
    }
    +
    \tikzsetnextfilename{decompIDsMatterDiag5}
    \gBoxBox[all=gluon,iA=baquark,iB=baquark,iC=gquark,iD=gquark,iE=gquark,iF=baquark,iG=rquark]{
      \node at (0.9,1.2) {$\cN=1$};
    }\\
    &=
    \tikzsetnextfilename{decompIDsMatterDiag6}
    \gBoxBox[all=gluon,iA=gline,iB=gline,iC=gline,iD=gline,iE=gline,iF=gline]{
      \node at (0.9,1.2) {$\cN=1$};
    }
    \tikzsetnextfilename{decompIDsMatterDiag7}
    +\gBoxBox[all=gluon,iA=gline,iB=gline,iC=bline,iD=bline,iE=bline,iF=gline,iG=rline]{
      \node at (0.9,1.2) {$\cN=1$};
    }\,,
  \end{aligned}
\end{equation}
where we have used flavor symmetry to freely permute the three different flavors.
The last independent decomposition identity for a double-box topology in $\cN=2$ SQCD is
\begin{equation}\label{eq:decompMatterBox2}
  \begin{aligned}
    \tikzsetnextfilename{decompIDsMatterDiag8}
    \gBoxBox[all=gluon,iA=quark,iB=quark,iF=quark,iG=aquark]{
      \node at (0.9,1.2) {$\cN=2$};
    }
    &=
    \tikzsetnextfilename{decompIDsMatterDiag9}
    \gBoxBox[all=gluon,iA=gquark,iB=gquark,iF=gquark,iG=gaquark]{
      \node at (0.9,1.2) {$\cN=1$};
    }
    +
    \tikzsetnextfilename{decompIDsMatterDiag10}
    \gBoxBox[all=gluon,iA=baquark,iB=baquark,iF=baquark,iG=bquark]{
      \node at (0.9,1.2) {$\cN=1$};
    }
    +
    \tikzsetnextfilename{decompIDsMatterDiag11}
    \gBoxBox[all=gluon,iA=gquark,iB=gquark,iC=raquark,iD=raquark,iE=raquark,iF=gquark,iG=baquark]{
      \node at (0.9,1.2) {$\cN=1$};
    }
    +
    \tikzsetnextfilename{decompIDsMatterDiag12}
    \gBoxBox[all=gluon,iA=baquark,iB=baquark,iC=rquark,iD=rquark,iE=rquark,iF=baquark,iG=gquark]{
      \node at (0.9,1.2) {$\cN=1$};
    }\\
    &=
    \tikzsetnextfilename{decompIDsMatterDiag13}
    \gBoxBox[all=gluon,iA=gline,iB=gline,iF=gline,iG=gline]{
      \node at (0.9,1.2) {$\cN=1$};
    }
    \tikzsetnextfilename{decompIDsMatterDiag14}
    +\gBoxBox[all=gluon,iA=gline,iB=gline,iC=rline,iD=rline,iE=rline,iF=gline,iG=bline]{
      \node at (0.9,1.2) {$\cN=1$};
    }\,,
  \end{aligned}
\end{equation}
and we recognize that the matter diagrams are the same ones
that appeared in \eqn{eq:decompPure} and \eqn{eq:decompMatterBox1}.
If we focus only on the arrowless double-box diagrams
on the last lines of the above decomposition formulae,
we see that we have three independent equations and four unknown $\cN=1$ diagram numerators.
So we can solve for three of them
(the pure $\cN=1$ SYM diagram and the two single-matter-loop $\cN=1$ SQCD diagrams)
leaving one diagram to be determined:
the double box diagram with three distinct matter flavors. 

The decomposition identities work analogously for other graph topologies (planar and non-planar) that have edges shared between different loops, similar to the double box.
We can always decompose $\cN=2$ SQCD diagrams into $\cN=1$ SYM diagrams, and we obtain sufficient independent constraints so that we can always eliminate the pure-vector and single-matter-loop diagrams.
The only diagrams left as unknown quantities are those with more than
one flavor of matter present on the internal lines. 

When diagram topologies have loops that factorize the details
of the supersymmetric decomposition are somewhat different.
Consider the $\cN=2$ SQCD bow-tie diagram with two matter loops:
\begin{equation}\label{eq:decompBowTie}
  \begin{aligned}
    \tikzsetnextfilename{decompIDsBowTieDiag1}
  	\gTriTri[scale=0.8,all=gluon,iA=quark,iB=quark,iC=quark,iE=quark,iF=quark,iG=quark]{
      \node at (1.1,1.3) {$\cN=2$};
	}
    &=
    \tikzsetnextfilename{decompIDsBowTieDiag2}
    \gTriTri[scale=0.8,all=gluon,iA=gquark,iB=gquark,iC=gquark,iE=gquark,iF=gquark,iG=gquark]{
      \node at (1.1,1.3) {$\cN=1$};
	}
    \tikzsetnextfilename{decompIDsBowTieDiag3}
    +\gTriTri[scale=0.8,all=gluon,iA=gquark,iB=gquark,iC=gquark,iE=baquark,iF=baquark,iG=baquark]{
      \node at (1.1,1.3) {$\cN=1$};
	}
    \tikzsetnextfilename{decompIDsBowTieDiag4}
    +\gTriTri[scale=0.8,all=gluon,iA=baquark,iB=baquark,iC=baquark,iE=gquark,iF=gquark,iG=gquark]{
      \node at (1.1,1.3) {$\cN=1$};
	}
    \tikzsetnextfilename{decompIDsBowTieDiag5}
    +\gTriTri[scale=0.8,all=gluon,iA=baquark,iB=baquark,iC=baquark,iE=baquark,iF=baquark,iG=baquark]{
      \node at (1.1,1.3) {$\cN=1$};
	}\\
    \tikzsetnextfilename{decompIDsBowTieDiag6}
    &=\frac{1}{2}\gTriTri[scale=0.8,all=gluon,iA=gline,iB=gline,iC=gline,iE=gline,iF=gline,iG=gline]{
      \node at (1.1,1.3) {$\cN=1$};
	}
    \tikzsetnextfilename{decompIDsBowTieDiag7}
    +\frac{1}{2}\gTriTri[scale=0.8,all=gluon,iA=gline,iB=gline,iC=gline,iE=bline,iF=bline,iG=bline]{
      \node at (1.1,1.3) {$\cN=1$};
	}
    \,.
  \end{aligned}
\end{equation}
On the first line the $\cN=2$ hypermultiplet is expanded in terms of
short $\cN=1$ chiral multiplets;
on the second line they are reassembled into long $\cN=1$ matter multiplets.
This gives an identity between the $\cN=2$ and $\cN=1$ diagrams
(which may seem surprising at first)
but notice that the arrowed $\cN=2$ lines effectively describe a half-hypermultiplet
with the same degrees of freedom as the long $\cN=1$ chiral multiplet.
Note that, using the subloop flavor symmetry eq.~\eqref{eq:flavorSym2}, one can further identify the latter two diagrams.
This means, that the $\cN=1$ bow-tie numerator with two long chiral multiplets are identical to the corresponding $\cN=2$ numerator with two hypermultiplets running in the loops.
Finally, note that in any of the above supersymmetric decompositions the external
states are projected onto the $\cN=1$ vector multiplet states. 

\section{Numerator construction at two loops}
\label{sec:numCons}

In this section we describe the detailed computational setup and techniques
underlying the calculation of a color-dual representation
of the two-loop four-vector $\cN=1$ SYM amplitude with internal matter. 
Having reduced the functional constraints to a set of master numerators
we make a generic ansatz.
Then we give an example of a two-loop cut,
which provides physical information used to constrain the ansatz.
All the information and constraints need then to be imposed on the ansatz.
Finally, some technical details about the actual implementation of these steps is provided.

\subsection{Recap of needed previous results}

In this paper we consider {\it four-vector superamplitudes} of various SYM theories coupled to generic supersymmetric matter;
hence if no further specification is given the reader can assume these are the amplitudes being discussed.
Furthermore, we will assume that the color-dual numerators of the four-vector amplitudes of $\cN=4$ SYM and $\cN=2$ SQCD are known.
This is indeed the case for $\cN=4$ SYM up to four loops~\cite{Bern:2010ue,Bern:2012uf}, and for $\cN=2$ SQCD up to two loops~\cite{Johansson:2014zca,Johansson:2017bfl}.
Not counting the known master numerators of $\cN=4$ and $\cN=2$ SQCD, one can show that for $\cN=1$ SYM there is a single master numerator at the one-loop level, the box diagram with a closed matter.
Color-kinematics duality combined with supersymmetric decomposition relates this matter diagram to all other $\cN=1$ SYM diagrams. 

At two loops two different representations for the $\cN=2$ numerators were presented in~\cite{Johansson:2017bfl}.
For the results calculated here we used the first version given in that paper, as it fulfills all $\cN=2$ versions of the constraints described in the previous section.
For $\cN=1$ SYM at two loops, we find that there are only two master numerators.\footnote{
This count does not include graphs with massless short chiral multiplets.
Such contributions give rise to a chiral gauge anomaly and hence we ignore them for most of this paper.}
These two masters are conveniently chosen to be the three-flavor double-box and penta-triangle as depicted in \Fig{fig:masters}.
We will discuss the derivation and choice of these master numerators in more detail in \Sec{sec:masterNums}.

\subsection{Master numerators at two loops}\label{sec:masterNums}

\begin{figure}[t]
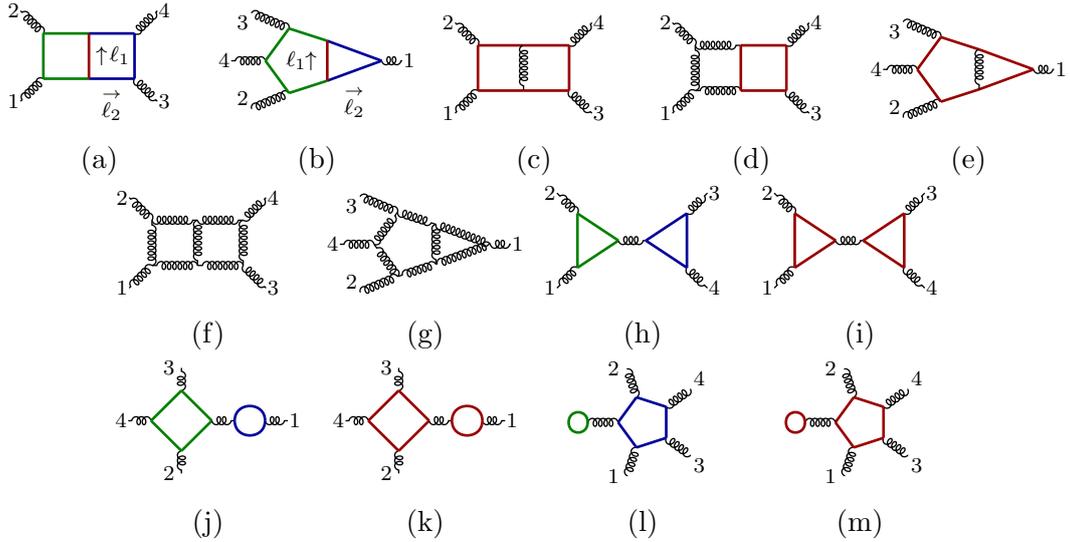

  \centering
  \begin{subfigure}[b]{0.18\textwidth}
    \centering
    \tikzsetnextfilename{allMastersDiag1}
    \gBoxBox[all=gluon,iA=gline,iB=gline,iC=bline,iD=bline,iE=bline,iF=gline,iG=rline,eLA=1,eLB=2,eLC=4,eLD=3,iLG=$\!\uparrow\!\ell_1$,iLE=$\overset{\rightarrow}{\ell_2}$]{}
    \subcaption{}
  \end{subfigure}
  \begin{subfigure}[b]{0.18\textwidth}
    \centering
    \tikzsetnextfilename{allMastersDiag2}
    \gTriPenta[scale=0.9,all=gluon,iA=gline,iB=gline,iC=gline,iD=bline,iE=bline,iF=gline,iG=rline,eLA=2,eLB=4,eLC=3,eLD=1,iLE=$\overset{\rightarrow}{\ell_2}$,iLG=$\ell_1\!\!\uparrow$]{}
    \subcaption{}
  \end{subfigure}
  \begin{subfigure}[b]{0.18\textwidth}
    \centering
    \tikzsetnextfilename{allMastersDiag3}
    \gBoxBox[all=gluon,iA=rline,iB=rline,iC=rline,iD=rline,iE=rline,iF=rline,eLA=1,eLB=2,eLC=4,eLD=3]{}
    \subcaption{}
  \end{subfigure}
  \begin{subfigure}[b]{0.18\textwidth}
    \centering
    \tikzsetnextfilename{allMastersDiag4}
    \gBoxBox[all=gluon,iC=rline,iD=rline,iE=rline,iG=rline,eLA=1,eLB=2,eLC=4,eLD=3]{}
    \subcaption{}
  \end{subfigure}
  \begin{subfigure}[b]{0.18\textwidth}
    \centering
    \tikzsetnextfilename{allMastersDiag5}
    \gTriPenta[scale=0.9,all=gluon,iA=rline,iB=rline,iC=rline,iD=rline,iE=rline,iF=rline,eLA=2,eLB=4,eLC=3,eLD=1]{}
    \subcaption{}
  \end{subfigure}\\
  \begin{subfigure}[b]{0.18\textwidth}
    \centering
    \tikzsetnextfilename{allMastersDiag6}
    \gBoxBox[all=gluon,eLA=1,eLB=2,eLC=4,eLD=3]{}
    \subcaption{}
  \end{subfigure}
  \begin{subfigure}[b]{0.18\textwidth}
    \centering
    \tikzsetnextfilename{allMastersDiag7}
    \gTriPenta[scale=0.9,all=gluon,eLA=2,eLB=4,eLC=3,eLD=1]{}
    \subcaption{}
  \end{subfigure}
  \begin{subfigure}[b]{0.18\textwidth}
    \centering
    \tikzsetnextfilename{allMastersDiag8}
    \gTriTri[scale=0.9,all=gluon,iA=gline,iB=gline,iC=gline,iE=bline,iF=bline,iG=bline,eLA=1,eLB=2,eLC=3,eLD=4]{}
    \subcaption{}
  \end{subfigure}
  \begin{subfigure}[b]{0.18\textwidth}
    \centering
    \tikzsetnextfilename{allMastersDiag9}
    \gTriTri[scale=0.9,all=gluon,iA=rline,iB=rline,iC=rline,iE=rline,iF=rline,iG=rline,eLA=1,eLB=2,eLC=3,eLD=4]{}
    \subcaption{}
  \end{subfigure}\\
  \begin{subfigure}[b]{0.18\textwidth}
    \centering
    \tikzsetnextfilename{allMastersDiag10}
    \gBoxBubB[all=gluon,iA=gline,iB=gline,iC=gline,iD=gline,iF=bline,iG=bline,eLA=2,eLB=4,eLC=3,eLD=1]{}
    \subcaption{}
  \end{subfigure}
  \begin{subfigure}[b]{0.18\textwidth}
    \centering
    \tikzsetnextfilename{allMastersDiag11}
    \gBoxBubB[all=gluon,iA=rline,iB=rline,iC=rline,iD=rline,iF=rline,iG=rline,eLA=2,eLB=4,eLC=3,eLD=1]{}
    \subcaption{}
  \end{subfigure}
  \begin{subfigure}[b]{0.18\textwidth}
    \centering
    \tikzsetnextfilename{allMastersDiag12}
    \gPentaTad[all=gluon,iA=bline,iB=bline,iC=bline,iD=bline,iE=bline,iG=gline,eLA=2,eLB=4,eLC=3,eLD=1]{}
    \subcaption{}
  \end{subfigure}
  \begin{subfigure}[b]{0.18\textwidth}
    \centering
    \tikzsetnextfilename{allMastersDiag13}
    \gPentaTad[all=gluon,iA=rline,iB=rline,iC=rline,iD=rline,iE=rline,iG=rline,eLA=2,eLB=4,eLC=3,eLD=1]{}
    \subcaption{}
  \end{subfigure}
  
  \caption{
    The thirteen master numerators obtained after imposing color-kinematics duality and $SO(3)$ overall flavor symmetry on the $\cN=1$ SYM amplitude.
    Supersymmetric decomposition allows us to remove numerators (c) -- (m) in favor of the previously computed $\cN=2$ SQCD numerators~\cite{Johansson:2017bfl}. 
    The final set of masters consists only of graphs (a) and (b).
  }
  \label{fig:mastersAll}
\end{figure}

Let us now discuss the master numerators obtained after solving the functional
equations that we illustrated (mostly pictorially) in \Sec{sec:diagIden}. 
If we only demand the color-dual numerator relations in \eqns{eq:jacsComs}{eq:clebschGordan},
then combined with the $SO(3)$ \emph{overall} flavor symmetry shown in \eqn{eq:flavorSym}  we can reduce the system to the thirteen masters depicted in \Fig{fig:mastersAll}.

Using additionally the decomposition identities given in \Sec{sec:decomp}, e.g. for the various double-box diagrams in \eqn{eq:decompPure}, \eqn{eq:decompMatterBox1}, and \eqn{eq:decompMatterBox2}, or the bow-tie numerator \eqn{eq:decompBowTie}, we can reduce the numerators (c) - (m) to the master numerators (a) and (b) and known $\cN=2$ SQCD numerators ~\cite{Johansson:2017bfl}. 
The other optional constraints discussed in \Sec{sec:diagIden} are not capable of further reducing this number.

To further reduce the need to solve for the complete master numerator, we use symmetry equations  (e.g. \eqn{crossingconstraint}) to relate different helicity components of a given numerator.
Each numerator has six helicity components,
one for each pair of negative-helicity legs ---
\ie the legs carrying the $V^-$ multiplet given in \eqn{eq:vMults}.
To track these external configurations we introduce the variables
\begin{equation}\label{eq:kappas}
  \hat{\kappa}_{ij} = \frac{\kappa_{ij}}{s_{ij}^3} = \frac{1}{s_{ij}^3}\frac{[12][34]}{\braket{12}\braket{34}}\delta^2(Q)\braket{i\,j}^3 \theta_i \theta_j\,,
\end{equation}
where the Mandelstam invariants are $s_{ij} = (p_i+p_j)^2$, and $i,j\in\{1,2,3,4\}$ denote the two negative-helicity legs. The Grassmann-odd $\theta_i$ were introduced in \eqn{realNeq1vector}, and they serve two purposes: they mark the legs that correspond to negative-helicity vector multiplets, and they make $\hat \kappa_{ij}$ symmetric in $i$ and $j$. Note that $\kappa_{ij}$ is a local variable that carries the same information as the color-ordered $\cN=1$ tree amplitude  $\kappa_{ij} =  i s_{12} s_{23} A^{\rm tree}_{ij}(1,2,3,4)$. 

Both the symmetry equations of the master graph and those of the descendent graphs (via the kinematic numerator relations) can be used to  induce linear relations among coefficients of the different $\hat{\kappa}_{ij} $ components.
For the double-box master, \Fig{fig:mastersAll}(a), we find three independent components,
and we choose to solve for the coefficients of
$\hat{\kappa}_{12}$, $\hat{\kappa}_{13}$, and $\hat{\kappa}_{14}$. For the penta-triangle master, \Fig{fig:mastersAll}(b), we note that there is a kinematic identity that relates it to itself and two double-box master numerators,
\begin{align}
\begin{aligned}
  &n_{\rm (b)}(1234;\ell_1,\ell_2) - n_{\rm (b)}(1432;p_3-\ell_1,p_1-\ell_2)\\
  &-n_{\rm (a)}(1234;\ell_1,\ell_2)+n_{\rm (a)}(1324;p_3-\ell_1,p_1-\ell_2)=0\,.
  \end{aligned}
\end{align}
This identity together with the symmetry of the penta-triangle diagram
imply that it has only one independent component, which is conveniently chosen as $\hat{\kappa}_{12}$. 
Thus, in total the master numerators have four independent $\hat{\kappa}_{ij}$ components, which we paramatrize using an ansatz.  

We keep track of how we have reduced the system to immediately obtain an expression for each numerator (component) as a linear combination of master numerators (components).
Furthermore, it is important to note that there may be several non-trivially inequivalent representations of a given numerator in terms of masters, depending on which ``route'' one takes through the system.
Taking one of them as the value of the numerator, one needs to impose the other representations as conditions on the ansatz.
In practice, this means that some of the Jacobies, commutation relations, and decomposition identities end up constraining the ansatz, together with all other optional constraints.
We were able to find a representation that implements all optional constraints discussed in \Sec{sec:diagIden}, except the subloop flavor symmetries and a tiny defect in the cuts in the form of a single term, as will be discussed in \Sec{sec:2Lcuts}.

\subsection{Ansatz for $\cN=1$ SYM}\label{sec:ansatz}

Now we proceed to make an ansatz for each independent helicity component
of the two master numerators,
working in dimensional regularization with $D=4-2\eps$.
Following \rcites{Johansson:2014zca,Johansson:2017bfl} our ansatz for a given master numerator $n_i$, with a fixed external state configuration $\hat{\kappa}_{jk}$, is given by 
\begin{equation} \label{masterNumAns}
  \begin{aligned}
  n_{i}(1234;\ell_1,\ell_2) &= \sum_{jk} \hat\kappa_{jk}\!\left(\!\sum_m c_{i,jk}^m M_m^{(4)} + i \sum_{m,n}d_{i,jk}^{mn}\,\Xi_n M_m^{(2)}\right. \\
  &\quad\left. + i\sum_{m}e_{i,jk}^m\eps(\mu_1,\mu_2)M_m^{(3)}
  - \sum_{m,n,l}f_{i,jk}^{mn}\Xi_n\epsilon(\mu_1,\mu_2)M_m^{(1)}
  \!\right),
  \end{aligned}
\end{equation}
where $c_{i,jk}^m$, $d_{i,jk}^{mn}$, $e_{i,jk}^m$, and $f_{i,jk}^{mn}$ are the free rational coefficients of the ansatz.

The two sets of kinematic invariants correspond to irreducible parity even and odd variables,
 \begin{align}
    M &= \{s_{12},\, s_{23},\,\ell_i\cdot(p_1{+}p_2),\,\ell_i\cdot(p_2{+}p_3),\,\ell_i\cdot(p_3{+}p_1),\,\ell_i\cdot \ell_j,\,\mu_{ij}\}\,, ~~~(i,j\in\{1,2\})\nn\\
    \Xi &= \{\eps(1,2,3,\ell_1),\,\eps(1,2,3,\ell_2),\,\eps(1,2,\ell_1,\ell_2),\,\eps(2,3,\ell_1,\ell_2),\,\eps(3,1,\ell_1,\ell_2)\}\,.
 \end{align}
Here we assume that the loop momenta $\ell_i$ are $D$-dimensional vectors, but the external momenta $p_i$ are restricted to live in four-dimensional spacetime, where the ${\cN=1}$ theory is well defined.
The extra-dimensional momentum components may explicitly appear in the amplitude as $\mu_{ij}=\bar{\ell}_i\cdot\bar{\ell}_j-\ell_i\cdot\ell_j$, where $\bar{\ell}_i$ is the four-dimensional part of $\ell_i$.
With the mostly-minus Lorentz signature $(+--\cdots-)$ the $\mu_{ii}$ are positive quantities, and can be interpreted as masses $m_i^2=\mu_{ii}$ whenever comparing to unitarity cuts involving massive matter.
The variables $\eps(i,j,k,\ell_l)$ and $\eps(i,j,\ell_k,\ell_l)$ are contractions between the Levi-Civita symbol $\eps_{\mu\nu\rho\sigma}$ and the four-dimensional parts of the vectors that appear in the arguments, e.g.  $\eps(i,j,k,\ell_l)=  \epsilon_{\mu \nu \rho \sigma}p_i^\mu p_j^\nu p_k^\rho \bar \ell_l^\sigma$.
Finally, we include an antisymmetric combination of extra-dimensional components $\epsilon(\mu_1,\mu_2)$.
It is formally defined through a reduction from the six-dimensional Levi-Civita tensor
\begin{equation}
  \epsilon(\mu_1,\mu_2)=\frac{\epsilon^{(6)}(v_1,v_2,v_3,v_4,\ell_1,\ell_2)}{\epsilon(v_1,v_2,v_3,v_4)}=\det(\mu_1,\mu_2)\,,
\end{equation}
where $v_i$ are arbitrary four-dimensional vectors.
Note that it satisfies the non-linear relation $\epsilon(\mu_1,\mu_2)^2=\mu_{11}\mu_{22}-\mu_{12}^2$.

Sets with superscripts, $M^{(n)}$, correspond to all monomials obtained as products of $n$ elements in $M$, and the $m$'th monomial of these sets is denoted by $M^{(n)}_m$. For example, the number of monomials are $|M|= |M^{(1)}| = 14 $, $ |M^{(2)}| = 105 $, $ |M^{(3)}| = 560 $, $ |M^{(4)}| = 2380$.

The ansatz for a fixed $(jk)$-component of a master numerator in~\eqn{masterNumAns} has 3,460 free parameters, after eliminating 75 linear relations among the terms including Levi-Civita objects.
These relations stem from four-dimensional kinematical identities involving the Levi-Civita tensors. Assuming four independent helicity components to solve for, there are a total of 13,840 free parameters in the above ansatz. 

One can further refine the ansatz by imposing so-called \emph{power counting constraints} on individual numerators.
Using a pattern observed in \rcites{Johansson:2014zca,Johansson:2017bfl} we further restrict the ansatz by demanding that no monomial contains more than four powers of loop momenta.
This gives a considerable reduction, bringing down the number of free parameters of a single component to 1,132.

A stronger (heuristic) power counting constraint that has been successfully used in $\cN=4$ SYM calculations~\cite{Bern:2010ue,Carrasco:2011mn,Bern:2012uf,Bern:2017ucb} is to impose power counting constraints on individual one-loop subdiagrams, however as noted in~\rcite{Johansson:2017bfl} this constraint was too strong if applied to every sub-diagram of every numerator of the two-loop $\cN=2$ SQCD amplitude.
Hence we will not use this strong constraint for the $\cN=1$ amplitude since it is related to the latter by the supersymmetric decomposition identities discussed in \Sec{sec:decomp}.
A weaker constraint that worked for the $\cN=2$ amplitude was to demand individual-loop power counting constraints for master numerators and closely related topologies.
However, for the $\cN=1$ amplitude at hand, already individual power counting constraints of double-box numerators clash with themselves or other more favorable constraints and hence we did not impose such power counting.

Finally, we observe that terms containing the extra-dimensional $\mu_{kl}$ terms
require fewer powers of $s_{ij}$ in the denominator of $\hat \kappa_{ij}=\kappa_{ij}/s_{ij}^3$.
The amplitude representation that we give in this paper need only two powers, hence the ansatz for the $\mu_{kl}$ terms can be refined by replacing $\hat \kappa_{ij}$ with $\hat \kappa_{ij} s_{ij}$ and $M^{(n)}$ by $M^{(n-1)}$. 
These constraints decrease the number of free parameters per component to 943, leading to 3,772 free parameters in total. 

\subsection{Two-loop cut construction}

\begin{figure}[t]
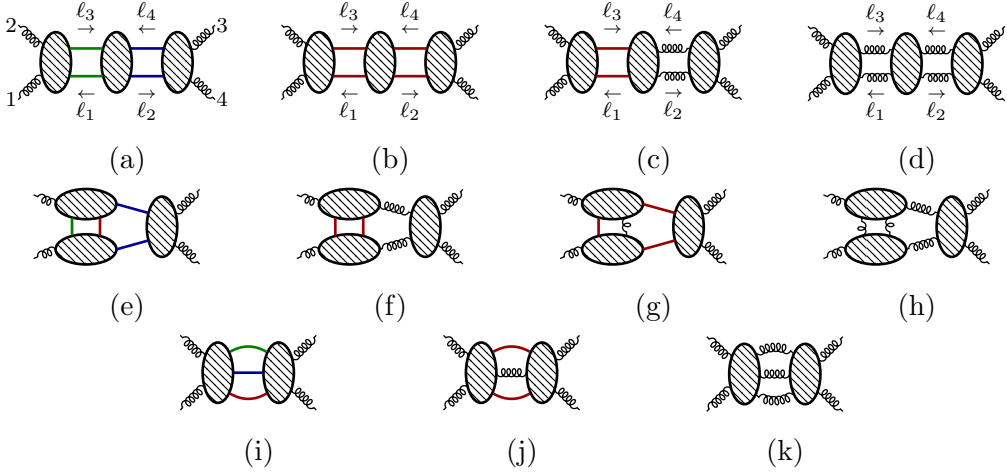

  \centering
  \begin{subfigure}[b]{0.22\textwidth}
    \centering
    \tikzsetnextfilename{allCutsDiag1}
    \cBoxBoxA[all=gluon,iA=gline,iB=bline,iC=bline,iD=gline,
    eLA=$1$,eLB=$2$,eLC=$3$,eLD=$4$,
    iLA=$\underset{\rightarrow}{\ell_3}$,iLB=$\underset{\leftarrow}{\ell_4}$,
    iLC=$\overset{\rightarrow}{\ell_2}$,iLD=$\overset{\leftarrow}{\ell_1}$]{} 
    \subcaption{}
  \end{subfigure}
  \begin{subfigure}[b]{0.22\textwidth}
    \centering
    \tikzsetnextfilename{allCutsDiag2}
    \cBoxBoxA[all=gluon,iA=rline,iB=rline,iC=rline,iD=rline]{} 
    \subcaption{}
  \end{subfigure}
  \begin{subfigure}[b]{0.22\textwidth}
    \centering
    \tikzsetnextfilename{allCutsDiag3}
    \cBoxBoxA[all=gluon,iA=rline,iD=rline]{} 
    \subcaption{}
  \end{subfigure}
  \begin{subfigure}[b]{0.22\textwidth}
    \centering  
    \tikzsetnextfilename{allCutsDiag4}
    \cBoxBoxA[all=gluon]{} 
    \subcaption{}
  \end{subfigure}
  \begin{subfigure}[b]{0.22\textwidth}
    \centering
    \tikzsetnextfilename{allCutsDiag5}
    \cBoxBoxB[all=gluon,iA=gline,iB=bline,iC=bline,iD=rline]{}
    \subcaption{}
  \end{subfigure}
  \begin{subfigure}[b]{0.22\textwidth}
    \centering
    \tikzsetnextfilename{allCutsDiag6}
    \cBoxBoxB[all=gluon,iA=rline,iD=rline]{}
    \subcaption{}
  \end{subfigure}
  \begin{subfigure}[b]{0.22\textwidth}
    \centering
    \tikzsetnextfilename{allCutsDiag7}
    \cBoxBoxB[all=gluon,iA=rline,iB=rline,iC=rline]{}
    \subcaption{}
  \end{subfigure}
  \begin{subfigure}[b]{0.22\textwidth}
    \centering
    \tikzsetnextfilename{allCutsDiag8}
    \cBoxBoxB[all=gluon]{}
    \subcaption{}
  \end{subfigure}
  \begin{subfigure}[b]{0.22\textwidth}
    \centering
    \tikzsetnextfilename{allCutsDiag9}
    \cBoxBoxC[all=gluon,iA=gline,iB=rline,iC=bline]{}
    \subcaption{}
  \end{subfigure}
  \begin{subfigure}[b]{0.22\textwidth}
    \centering
    \tikzsetnextfilename{allCutsDiag10}
    \cBoxBoxC[all=gluon,iA=rline,iB=rline]{}
    \subcaption{}
  \end{subfigure}
  \begin{subfigure}[b]{0.22\textwidth}
    \centering
    \tikzsetnextfilename{allCutsDiag11}
    \cBoxBoxC[all=gluon]{}
    \subcaption{}
  \end{subfigure}
  \caption{A sufficient set of two-loop physical spanning cuts for $\cN=1$ SYM, with different configurations of internal vector and flavored matter multiplets.
  The set of cuts (a) and (e) are sufficient to fully constrain our amplitude; the rest serve as consistency checks.
  Some care needs to be taken in order to fulfill cut (b), see Section~\ref{sec:2Lcuts}.
    The displayed lines are on-shell, and the blobs represent color-ordered tree-level amplitudes, where all orderings and generic flavor assignments are considered, for completeness.
  \label{fig:cuts2L}}
\end{figure}

Physical information about the amplitude is encoded in the generalized unitarity cuts~\cite{Bern:1994zx,Bern:1994cg},
which are constructed by multiplying tree-level amplitudes
and summing over their internal on-shell states.
The two-loop spanning cut topologies, including arbitrarily flavored matter,
are depicted in \Fig{fig:cuts2L}.
The unitarity cuts are directly compared to the ansatz of the amplitude,
and any free parameters in the ansatz are fixed by demanding that
they match the physical information in the cuts.
If some parameters are not fixed by the collection of all spanning cuts, then the contributions of those should ideally correspond to terms that integrate to zero,
or at worst, terms that are part of some scheme dependence of the construction.
In particular, it is advantageous to perform $D$-dimensional unitarity cuts so that all rational terms in the four-dimensional amplitude pick up a logarithmic dependence on the scales of the problem, and thus have a $D$-dimensional branch cut. 

To calculate these cuts using the tree amplitudes in \Sec{sec:susyAmps}
we use an on-shell parametrization of the loop momenta.
Take cut (a) as an example:
we can decompose the four-dimensional parts of the loop momenta $\bar{\ell}_i$
on a basis of external spinors as
\begin{subequations}\label{eq:cutParam}
\begin{align}
\bar{\ell}_1-p_1&=\bar{\ell}_3+p_2=
\alpha_1(p_1-p_2)+
\alpha_2\frac{\braket{23}}{\braket{13}}|1\rangle[2|+
\alpha_3\frac{[23]}{[13]}|2\rangle[1|\,,\\
\bar{\ell}_2-p_4&=\bar{\ell}_4+p_3=
\beta_1(p_4-p_3)+
\beta_2\frac{\braket{23}}{\braket{24}}|4\rangle[3|+
\beta_3\frac{[23]}{[24]}|3\rangle[4|\,,
\end{align}
\end{subequations}
which ensure that $2\bar{\ell}_1\cdot p_{12}=2\bar{\ell}_2\cdot p_{34}=s$,
as the dimensionless parameters $\alpha_i$, $\beta_i$ carry no helicity scaling.
The requirement that $\ell_i^2=\bar{\ell}_i^2-\mu_{ii}=0$
where $\mu_{ii}=m_i\widebar{m}_i$ also enables us to solve for $\alpha_3$ and $\beta_3$.
Appropriate massive spinor variables can then be read off, for example
\begin{subequations}\label{eq:massiveSpinorParam}
\begin{align}
|\bar{\ell}_1^a\rangle&=\left(\begin{matrix}
  \frac{m_1}{\braket{12}}|2\rangle \\
  |1\rangle-\frac{\alpha_1\braket{13}}{\alpha_2\braket{23}}|2\rangle
\end{matrix}\right)\,, &
|\bar{\ell}_1^a]&=\left(\begin{matrix}
  (1+\alpha_1)|1] + \frac{\alpha_2\braket{23}}{\braket{13}}|2] \\
  -\frac{\widebar{m}_1\braket{13}}{\alpha_2[12]\braket{23}}|1]
\end{matrix}\right)\,, \\
|\bar{\ell}_3^a\rangle&=\left(\begin{matrix}
  \frac{m_1}{\braket{12}}|2\rangle \\
  |1\rangle-\frac{(1+\alpha_1)\braket{13}}{\alpha_2\braket{23}}|2\rangle
\end{matrix}\right)\,, &
|\bar{\ell}_3^a]&=\left(\begin{matrix}
  \alpha_1|1] + \frac{\alpha_2\braket{23}}{\braket{13}}|2] \\
  -\frac{\widebar{m}_1\braket{13}}{\alpha_2[12]\braket{23}}|1]
\end{matrix}\right)\,,
\end{align}
\end{subequations}
which reproduces \eqn{eq:cutParam} using
$\bar{\ell}_i=|\bar{\ell}_i^a\rangle[\bar{\ell}_{i,a}|$.
Spinors for $\ell_2$ and $\ell_4$ are related to these by symmetry,
and $|\!-\!\ell^a\rangle=-|\ell^a\rangle$, $|\!-\!\ell^a]=|\ell^a]$.
We are left with 10 independent variables characterizing the cut:
$\alpha_{1,2}$, $\beta_{1,2}$, $m_{1,2,3}$, and $\widebar{m}_{1,2,3}$.
The last two masses $m_3$, $\widebar{m}_3$ are identified with extra-dimensional
components: $m_3\widebar{m}_3=\mu_{11}+\mu_{22}+2\mu_{12}$.

Feeding these decomposed spinors into the tree amplitudes in \Sec{sec:susyAmps}
gives rise to expressions for the cuts depending only on these free parameters,
plus the two Mandelstam variables $s$ and $t$.
For example, the $\hat{\kappa}_{12}$ component takes the form
\begin{equation}
  {\rm Cut}(a)_{\hat{\kappa}_{12}}=
  \frac{2s^2\mu_{11}\mu_{22}}
  {(\ell_1-p_1)^2(\ell_2-p_4)^2(\ell_1+\ell_2)^2}\,.
\end{equation}
By applying the same cut parametrization to the ansatz for each numerator
we have a basis for comparison;
numerical values can be chosen for the free parameters to build up a system
of linear equations.
Due to the constraints discussed in the previous section only the cuts (a) and (e) are required to fully constrain the amplitude;
the others serve as useful consistency checks on the final result, but yield no new information with the exception of a minor discrepancy in cut (b).
We will discuss this issue in the following subsection.

\subsection{Two-loop cuts: merging dimensional regularization and SUSY}
\label{sec:2Lcuts}

It is well known that $\cN=1$ SYM does not have a dimensional uplift to a well-defined supersymmetric theory in $D>4$ dimensions.
This property is sometimes naively generalised to the statement that dimensional regularization is not compatible with supersymmetry.
However, such a statement is too broad to be true in general.
There are certainly some supersymmetric theories ($\cN>1$) that are fully compatible with dimensional regularization in the dimensional reduction scheme.
The dimensional reduction scheme requires that the integrand is constructed for the higher-dimensional theory and then the dimensional reduction back to four dimensions is parametrised with $\epsilon= (4-D)/2$.

For $\cN=1$ SYM such a scheme appears not to be available since the higher-dimensional integrand cannot be constructed following a unique prescription.
Nevertheless, one is allowed to supplement the physical constraints that come from the four-dimensional theory with auxiliary constraints that select terms compatible with a higher-dimensional uplifts.
Such constraints and higher-dimensional terms will then be part of the scheme dependence of the regularization.
The auxiliary constraints should ideally be tied to a higher-dimensional theory, such that the usual theorems and properties of a quantum field theory are built in to the higher-dimensional contributions. 

For the dimensional regularization scheme used in this paper we demand that the following properties are satisfied:
\begin{itemize}
\item supersymmetric decomposition identities to $D>4$ $\cN=2$ SYM obeyed,
\item color-kinematics duality is obeyed for the $D$-dimensional integrand,
\item power counting for the $D$-dimensional integrand is the same as the $D=4$ integrand,
\item higher-dimensional momentum is compatible with massive matter amplitudes.
\end{itemize}
The last point is realized here by matching to cuts in a massive theory.
These cuts include cases where the internal matter are $\cN=1$ vector multiplets and $\cN=1$ chiral long multiplets.

We analytically worked out the cuts (a), (b), and (e) in \Fig{fig:cuts2L}
by sewing together the four-point amplitudes presented in \Sec{sec:fourPoints},
similar to the on-shell recursion we used in \Sec{sec:bcfw}.
As a consistancy check we also numerically evaluated all other cuts on some random phase-space points.
Extra-dimensional $\mu_{ij}$ components present in the ansatz were re-interpreted
as masses when being matched to the cuts.
As the internal vector multiplets are necessarily massless,
they do not yield information about the higher-dimension $\mu_{ij}$ terms on those internal lines.
Yet this limitation is conveniently overcome by our use of the supersymmetric decomposition identities and color-algebra relations:\footnote{
As already observed for $\cN=2$ SQCD in \rcite{Johansson:2017bfl},
three-particle cuts, obtained from gluing two five-point tree amplitudes,
do not impose any further constraints so the iterated two-particle cuts contain all needed information.
We have checked this by numerical evaluation of the three-particle cuts.
}
when these are also imposed, the cuts (a) and (e) fully constrain the amplitude,
with the rest serving only as consistency checks.
As these two cuts contain massive internal lines only,
we gain full access to the higher-dimensional $\mu_{ij}$ terms
\emph{also for graphs with internal vector multiplets}.
Expressions for the (complex) massive helicity spinors $|\ell_i^a\rangle$
and $|\ell_i^a]$ were provided using explicit
on-shell parametrizations of the kinematics,
such as that given for cut (a) in \Eqn{eq:cutParam}.

However, there is a caveat introduced when re-interpreting masses as
extra-dimensional components $\mu_{ij}$ in the context of dimensional regularization.
Consider the next-to-maximal cut obtained by cutting into the spanning cut (b):
\begin{equation}\label{eq:cutExBowTie}
  \tikzsetnextfilename{cutExBowTie1}
  \begin{tikzpicture}
    [line width=1pt,
    baseline={([yshift=-0.5ex]current bounding box.center)},
    scale=1,
    rotate=0,
    font=\scriptsize]
    \fill[blob] (0.9,0.6) ellipse (0.2 and 0.4);
    \draw[gluon] (0.3,0.3) -- (0,0) node[left] {};
    \draw[gluon] (0.3,0.9) -- (0,1.2) node[left] {};
    \draw[gluon] (1.5,0.9) -- (1.8,1.2) node[right] {};
    \draw[gluon] (1.5,0.3) -- (1.8,0) node[right] {};
    \draw[rline] (0.3,0.3) -- (0.3,0.9);
    \draw[rline] (0.3,0.9) -- (0.75,0.9);
    \draw[rline] (1.05,0.9) -- (1.5,0.9);
    \draw[rline] (1.5,0.9) -- (1.5,0.3);
    \draw[rline] (1.5,0.3) -- (1.05,0.3);
    \draw[rline] (0.75,0.3) -- (0.3,0.3);
  \end{tikzpicture}\,.
\end{equation}
Both numerators contributing to this cut are readily determined from the distinct-flavor cuts (a) and (e),
using the supersymmetric decomposition identities (see \Sec{sec:decomp})
and crucially the optional identity \eqref{eq:twoTerm}.
These expressions do not add up to the massive cut as one might naively expect:
\begin{equation}\label{eq:brokenCut}
  \tikzsetnextfilename{ntmefhhc1}
  \begin{tikzpicture}
    [line width=1pt,
    baseline={([yshift=-0.5ex]current bounding box.center)},
    scale=1,
    rotate=0,
    font=\scriptsize]
    \fill[blob] (0.9,0.6) ellipse (0.2 and 0.4);
    \draw[gluon] (0.3,0.3) -- (0,0) node[left] {1};
    \draw[gluon] (0.3,0.9) -- (0,1.2) node[left] {2};
    \draw[gluon] (1.5,0.9) -- (1.8,1.2) node[right] {3};
    \draw[gluon] (1.5,0.3) -- (1.8,0) node[right] {4};
    \draw[rline] (0.3,0.3) -- (0.3,0.9);
    \draw[rline] (0.3,0.9) -- (0.75,0.9);
    \draw[rline] (1.05,0.9) -- (1.5,0.9);
    \draw[rline] (1.5,0.9) -- (1.5,0.3);
    \draw[rline] (1.5,0.3) -- node[below] {$\overset{\rightarrow}{\ell_2}$} (1.05,0.3);
    \draw[rline] (0.75,0.3) -- node[below] {$\overset{\leftarrow}{\ell_1}$} (0.3,0.3);
  \end{tikzpicture}
  =
  \tikzsetnextfilename{ntmefhhc2}
  \left.\frac{1}{2\bar\ell_1\cdot\bar\ell_2-2m^2}\gBoxBox[all=gluon,iA=rline,iB=rline,iC=rline,iD=rline,iE=rline,iF=rline,iLF=$\overset{\leftarrow}{\ell_1}$,iLE=$\overset{\rightarrow}{\ell_2}$,eLA=1,eLB=2,eLC=3,eLD=4]{}\right|_{\textrm{cut}}
  \tikzsetnextfilename{ntmefhhc3}
  +\left.\frac{1}{s}\gTriTri[scale=0.9,all=gluon,iA=rline,iB=rline,iC=rline,iE=rline,iF=rline,iG=rline,iLC=$\overset{\swarrow}{\ell_1}$,iLG=$\overset{\searrow}{\ell_2}$,eLA=1,eLB=2,eLC=3,eLD=4]{}\right|_{\textrm{cut}}
  -2s^2m^2(\hat\kappa_{12}+\hat\kappa_{34})
  \,,
\end{equation}
where
\begin{equation}
  \begin{aligned}
    \tikzsetnextfilename{ntmefhhcres1}
    \left.\frac{1}{2\bar\ell_1\cdot\bar\ell_2-2m^2}\gBoxBox[all=gluon,iA=rline,iB=rline,iC=rline,iD=rline,iE=rline,iF=rline,iLF=$\overset{\leftarrow}{\ell_1}$,iLE=$\overset{\rightarrow}{\ell_2}$,eLA=1,eLB=2,eLC=3,eLD=4]{}\right|_{\textrm{cut}}&=s^2(s \mu_{12}+2\mu_{11}\mu_{22})(\hat\kappa_{12}+\hat\kappa_{34}) + \dots\,,\\
  \tikzsetnextfilename{ntmefhhcres2}
  \left.\frac{1}{s}\gTriTri[scale=0.9,all=gluon,iA=rline,iB=rline,iC=rline,iE=rline,iF=rline,iG=rline,iLC=$\overset{\swarrow}{\ell_1}$,iLG=$\overset{\searrow}{\ell_2}$,eLA=1,eLB=2,eLC=3,eLD=4]{}\right|_{\textrm{cut}}&=-2s^3\mu_{12}(\hat\kappa_{12}+\hat\kappa_{34}) + \dots\,.
  \end{aligned}
\end{equation}
Note that $\mu_{11}=\mu_{22}=-\mu_{12}=m^2$ according to the massive cut conditions.

Interestingly, the same kind of terms are also responsible for a breaking of the sub-loop flavor symmetry of the bow-tie:
\begin{equation}\label{eq:brokenSubLoopFlavor}
  \tikzsetnextfilename{slfs1}
  \gTriTri[scale=0.9,all=gluon,eLA=1,eLB=2,eLC=3,eLD=4,iA=rline,iB=rline,iC=rline,iE=rline,iF=rline,iG=rline,iLC=$\overset{\swarrow}{\ell_1}$,iLG=$\overset{\searrow}{\ell_2}$]{}
  =
  \tikzsetnextfilename{slfs2}
  \gTriTri[scale=0.9,all=gluon,eLA=1,eLB=2,eLC=3,eLD=4,iA=rline,iB=rline,iC=rline,iE=gline,iF=gline,iG=gline,iLC=$\overset{\swarrow}{\ell_1}$,iLG=$\overset{\searrow}{\ell_2}$]{}
  - 2s^3\mu_{12}(\hat{\kappa}_{12}+\hat{\kappa}_{34})
  + 2s^3 i\,\epsilon(\mu_1,\mu_2)(\hat{\kappa}_{12}-\hat{\kappa}_{34})\,.
\end{equation}

Let us try to understand why this problems occur.
The two numerators are related by a color-algebra-like relation
(involving $\mu$-terms, not masses)
\begin{equation}\label{eq:threeTermMatter}
  \tikzsetnextfilename{threeTermMatterProblem1}
  \gTriTri[scale=0.9,all=gluon,iA=rline,iB=rline,iC=rline,iE=rline,iF=rline,iG=rline,eLA=$1$,eLB=$2$,eLC=$3$,eLD=$4$,iLC=$\overset{\swarrow}{\ell_1}$,iLG=$\overset{\searrow}{\ell_2}$]{}
  =
  \tikzsetnextfilename{threeTermMatterProblem2}
  \gBoxBox[all=gluon,iA=rline,iB=rline,iC=rline,iD=rline,iE=rline,iF=rline,eLA=$1$,eLB=$2$,eLC=$3$,eLD=$4$,iLF=$\overset{\leftarrow}{\ell_1}$,iLE=$\overset{\rightarrow}{\ell_2}$]{}
  \tikzsetnextfilename{threeTermMatterProblem3}
  - \gBoxBox[all=gluon,iA=rline,iB=rline,iC=rline,iD=rline,iE=rline,iF=rline,eLA=$1$,eLB=$2$,eLC=$4$,eLD=$3$,iLF=$\overset{\leftarrow}{\ell_1}$,iLC=$\underset{\rightarrow}{\ell_2}$]{}
\end{equation}
where the latter double-box numerator is obtained from the former
by swapping $\ell_2\rightarrow -\ell_2 - p_3 - p_4$,
\ie~$\mu_{12}\rightarrow -\mu_{12}$.
Hence, the term linear in $\mu_{12}$ will change sign such that the equation is fulfilled.
On the other hand, working with masses on the cut kinematics we have
$-\mu_{12}=\mu_{11}=\mu_{22}=m^2$.
So the above color-algebra identity behaves differently as $m^2$
does not change sign under $\ell_2\rightarrow -\ell_2$.
The four-point tree-level amplitude in the cut \eqref{eq:cutExBowTie} fulfills massive-type color-algebra relations, and hence so does the cut.
Numerators constructed using $\mu$-type color-algebra relations are thus not necessarily consistent with these cuts.
Conversely, if we would have used the mass-type version of Eq.~\eqref{eq:threeTermMatter} to determine the bow-tie diagram the cut would be fulfilled.

Such inconsistencies could, in principle, occur for other cuts, diagrams or helicity configurations.
Interestingly, the two places discussed here are the only ones relevant for the amplitude under consideration.
Since we are eventually interested in computing gravitational amplitudes via the double-copy, we prefer having flavor-symmetric numerators.
These are needed for the double-copy of matter contributions, which according to~\rcites{Johansson:2014zca,Johansson:2017bfl} can be promoted to ghosts in order to remove unwanted dilaton- and axion-like multiplets.
Without flavor symmetry this subtraction becomes ambiguous.

Hence, in order to fix eqn.~\eqref{eq:brokenSubLoopFlavor} we deform the massive cuts, by including simple $\mu$-term ans\"{a}tze sitting on top of each of the cubic diagram denominators contributing to the cuts:
\begin{equation}\label{eq:corrections}
  \sum_{i,j} \kappa_{ij} \left(c_{ij}^{(1)} \mu_{11} + c_{ij}^{(2)} \mu_{22} + c_{ij}^{(3)} \mu_{12}+c_{ij}^{(4)}\epsilon(\mu_1,\mu_2)\right)\,.
\end{equation}
All other constraints are kept fixed, which imposes strong cross-talk among the ans\"{a}tze for different diagrams.  Solving the system anew, one finds a one-parameter family of coefficients $c_{ij}^{(k)}$ that satisfy this construction. We choose the remaining parameter such that the deformation take the most symmetric form.
Every correction to a given numerator component can be mapped to the corresponding vacuum topology of a graph ({\it i.e.} in the high-energy limit), such that it is proportional to one of the following:
\begin{equation}
\mu_{11}+\mu_{12}+\mu_{22}\,,~~~~~ \mu_{11}+\mu_{22}\,,~~~~~\mu_{11}\,,~~~~~\mu_{22}\,,
\end{equation}
plus an anti-symmetric term $\epsilon(\mu_1,\mu_2)$, depending on the vacuum topology.

Eventually, a correction in the form of such terms, suitable for a double-copy construction, must be determined by comparing gravitational cuts to double-copied (matter) numerators in $D>4$. Note that for a complete double copy of matter contributions, allowing for the construction of pure supergravity amplitudes, numerators with short multiplets are required.
Furthermore, the dependence of the numerators on the antisymmetric extra-dimensional object $\epsilon(\mu_1,\mu_2)$ is not captured by the massive cuts considered here.
Terms containing the latter will, crucially, no longer necessarily integrate to zero in a double-copied amplitude since $\epsilon(\mu_1,\mu_2)^2=\mu_{11}\mu_{22}-\mu_{12}^2$ is an even function. As a preparation for such studies, we provide a second representation of the numerators in App.~\ref{app:alt} with three free parameters $x,y,z$ which can be used to extract versions that either fulfill a different set of constraints and cuts, or undoes the above cut deformation $y=0$.

\begin{figure}[t]
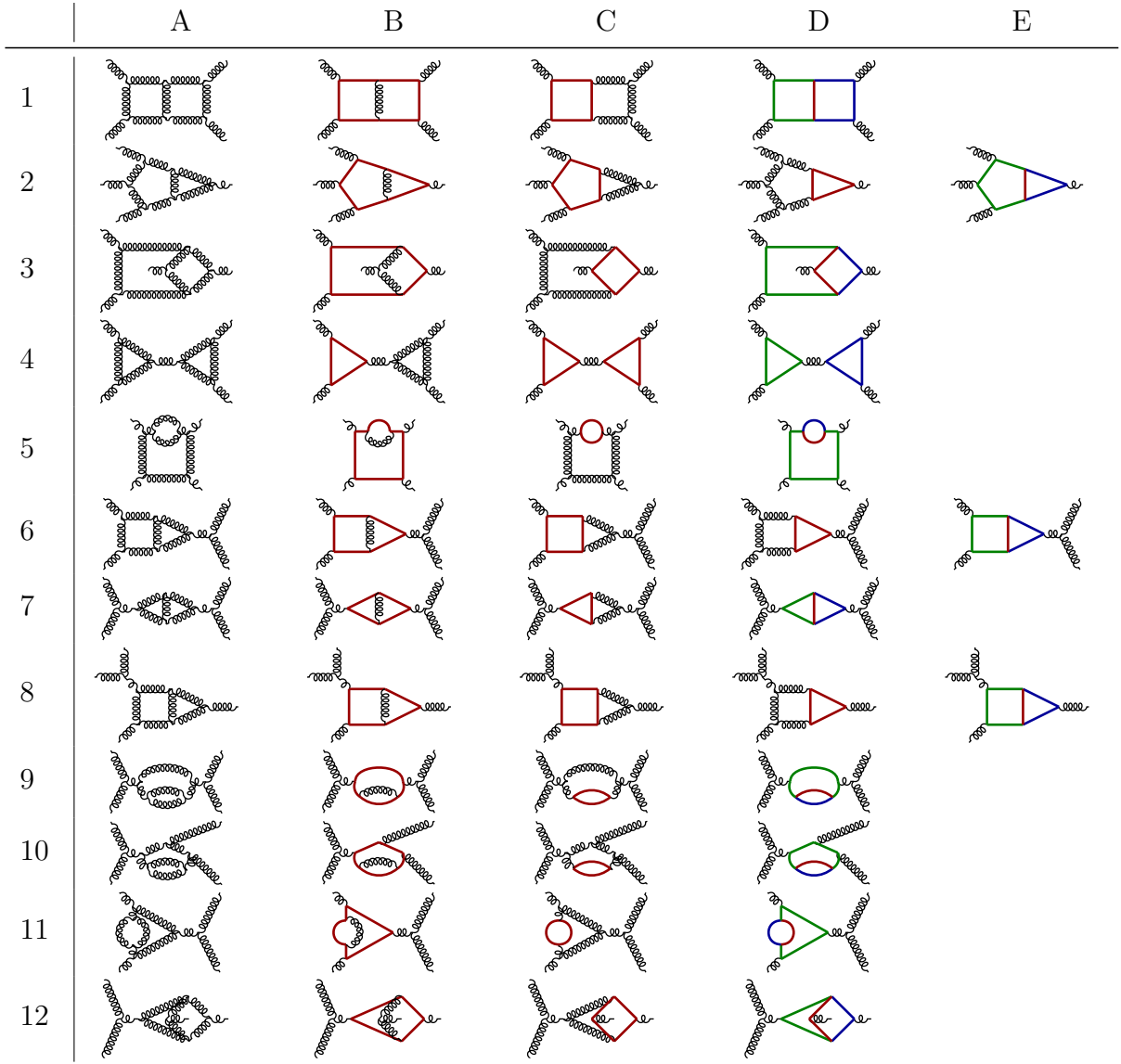

  \centering
  \begin{tabular}{l | c c c c c}
    & A & B & C & D & E\\
    \midrule
    1
    \tikzsetnextfilename{allDiags1}
    & \gBoxBox[scale=0.95,all=gluon]{}
    \tikzsetnextfilename{allDiags2}
    & \gBoxBox[scale=0.95,all=gluon,iA=rline,iB=rline,iC=rline,iD=rline,iE=rline,iF=rline]{}
    \tikzsetnextfilename{allDiags3}
    & \gBoxBox[scale=0.95,all=gluon,iA=rline,iB=rline,iG=rline,iF=rline]{}
    \tikzsetnextfilename{allDiags4}
    & \gBoxBox[scale=0.95,all=gluon,iA=gline,iB=gline,iF=gline,iC=bline,iD=bline,iE=bline,iG=rline]{}\\
    2
    \tikzsetnextfilename{allDiags5}
    &\gTriPenta[scale=0.75,all=gluon]{}
    \tikzsetnextfilename{allDiags6}
    & \gTriPenta[scale=0.75,all=gluon,iA=rline,iB=rline,iC=rline,iD=rline,iE=rline,iF=rline]{}
    \tikzsetnextfilename{allDiags7}
    & \gTriPenta[scale=0.75,all=gluon,iA=rline,iB=rline,iC=rline,iF=rline,iG=rline]{}
    \tikzsetnextfilename{allDiags8}
    & \gTriPenta[scale=0.75,all=gluon,iD=rline,iE=rline,iG=rline]{}
    \tikzsetnextfilename{allDiags9}
    & \gTriPenta[scale=0.75,all=gluon,iA=gline,iB=gline,iC=gline,iF=gline,iD=bline,iE=bline,iG=rline]{}\\
    3
    \tikzsetnextfilename{allDiags10}
    & \gBoxBoxNP[scale=0.85,all=gluon]{}
    \tikzsetnextfilename{allDiags11}
    & \gBoxBoxNP[scale=0.85,all=gluon,iA=rline,iB=rline,iC=rline,iD=rline,iE=rline]{}
    \tikzsetnextfilename{allDiags12}
    & \gBoxBoxNP[scale=0.85,all=gluon,iC=rline,iD=rline,iF=rline,iG=rline]{}
    \tikzsetnextfilename{allDiags13}
    & \gBoxBoxNP[scale=0.85,all=gluon,iA=gline,iB=gline,iE=gline,iC=bline,iD=bline,iF=rline,iG=rline]{}\\
    4
    \tikzsetnextfilename{allDiags14}
    & \gTriTri[scale=0.85,all=gluon]{}
    \tikzsetnextfilename{allDiags15}
    & \gTriTri[scale=0.85,all=gluon,iA=rline,iB=rline,iC=rline]{}
    \tikzsetnextfilename{allDiags16}
    & \gTriTri[scale=0.85,all=gluon,iA=rline,iB=rline,iC=rline,iE=rline,iF=rline,iG=rline]{}
    \tikzsetnextfilename{allDiags17}
    & \gTriTri[scale=0.85,all=gluon,iA=gline,iB=gline,iC=gline,iE=bline,iF=bline,iG=bline]{}\\
    5
    \tikzsetnextfilename{allDiags18}
    & \gBoxBubA[scale=0.85,all=gluon,iC=agluon,iD=agluon]{}
    \tikzsetnextfilename{allDiags19}
    & \gBoxBubA[scale=0.85,all=gluon,iA=rline,iB=rline,iC=rline,iD=agluon,iE=rline,iF=rline,iG=rline]{}
    \tikzsetnextfilename{allDiags20}
    & \gBoxBubA[scale=0.85,all=gluon,iC=rline,iD=rline]{}
    \tikzsetnextfilename{allDiags21}
    & \gBoxBubA[scale=0.85,all=gluon,iA=gline,iB=gline,iE=gline,iF=gline,iG=gline,iC=bline,iD=rline]{}\\
    6
    \tikzsetnextfilename{allDiags22}
    & \gBoxTri[scale=0.85,all=gluon]{}
    \tikzsetnextfilename{allDiags23}
    & \gBoxTri[scale=0.85,all=gluon,iA=rline,iB=rline,iC=rline,iE=rline,iF=rline]{}
    \tikzsetnextfilename{allDiags24}
    & \gBoxTri[scale=0.85,all=gluon,iA=rline,iB=rline,iG=rline,iF=rline]{}
    \tikzsetnextfilename{allDiags25}
    & \gBoxTri[scale=0.85,all=gluon,iC=rline,iE=rline,iG=rline]{}
    \tikzsetnextfilename{allDiags26}
    & \gBoxTri[scale=0.85,all=gluon,iA=gline,iB=gline,iF=gline,iC=bline,iE=bline,iG=rline]{}\\
    7
    \tikzsetnextfilename{allDiags27}
    & \gTriTriB[scale=0.75,all=gluon]{}
    \tikzsetnextfilename{allDiags28}
    & \gTriTriB[scale=0.75,all=gluon,iB=rline,iC=rline,iD=rline,iE=rline]{}
    \tikzsetnextfilename{allDiags29}
    & \gTriTriB[scale=0.75,all=gluon,iB=rline,iE=rline,iF=rline]{}
    \tikzsetnextfilename{allDiags30}
    & \gTriTriB[scale=0.75,all=gluon,iB=gline,iC=bline,iD=bline,iE=gline,iF=rline]{}\\
    8
    \tikzsetnextfilename{allDiags31}
    & \gBoxTriB[scale=0.85,all=gluon]{}
    \tikzsetnextfilename{allDiags32}
    & \gBoxTriB[scale=0.85,all=gluon,iA=rline,iC=rline,iD=rline,iE=rline,iF=rline]{}
    \tikzsetnextfilename{allDiags33}
    & \gBoxTriB[scale=0.85,all=gluon,iA=rline,iC=rline,iG=rline,iF=rline]{}
    \tikzsetnextfilename{allDiags34}
    & \gBoxTriB[scale=0.85,all=gluon,iD=rline,iE=rline,iG=rline]{}
    \tikzsetnextfilename{allDiags35}
    & \gBoxTriB[scale=0.85,all=gluon,iA=gline,iC=gline,iD=bline,iE=bline,iF=gline,iG=rline]{}\\
    9
    \tikzsetnextfilename{allDiags36}
    & \gBubBubB[scale=0.75,all=gluon,iB=agluon,iD=agluon,iE=agluon,iF=agluon,iG=agluon]{}
    \tikzsetnextfilename{allDiags37}
    & \gBubBubB[scale=0.75,all=gluon,iB=rline,iD=rline,iE=rline,iF=agluon,iG=rline]{}
    \tikzsetnextfilename{allDiags38}
    & \gBubBubB[scale=0.75,all=gluon,iB=agluon,iD=agluon,iE=rline,iF=rline,iG=agluon]{}
    \tikzsetnextfilename{allDiags39}
    & \gBubBubB[scale=0.75,all=gluon,iB=gline,iD=gline,iE=bline,iF=rline,iG=gline]{}\\
    10
    \tikzsetnextfilename{allDiags40}
    & \gTriBubB[scale=0.75,all=gluon,iE=agluon,iF=agluon]{}
    \tikzsetnextfilename{allDiags41}
    & \gTriBubB[scale=0.75,all=gluon,iB=rline,iC=rline,iD=rline,iE=rline,iF=agluon,iG=rline]{}
    \tikzsetnextfilename{allDiags42}
    & \gTriBubB[scale=0.75,all=gluon,iE=rline,iF=rline]{}
    \tikzsetnextfilename{allDiags43}
    & \gTriBubB[scale=0.75,all=gluon,iB=gline,iC=gline,iD=gline,iE=bline,iF=rline,iG=gline]{}\\
    11
    \tikzsetnextfilename{allDiags44}
    & \gTriBubC[scale=0.95,all=gluon,iB=agluon,iC=agluon]{}
    \tikzsetnextfilename{allDiags45}
    & \gTriBubC[scale=0.95,all=gluon,iA=rline,iB=rline,iC=agluon,iD=rline,iE=rline,iG=rline]{}
    \tikzsetnextfilename{allDiags46}
    & \gTriBubC[scale=0.95,all=gluon,iB=rline,iC=rline]{}
    \tikzsetnextfilename{allDiags47}
    & \gTriBubC[scale=0.95,all=gluon,iA=gline,iB=bline,iC=rline,iD=gline,iE=gline,iG=gline]{}\\
    12
    \tikzsetnextfilename{allDiags48}
    & \gBoxTriNPB[scale=0.8,all=gluon]{}
    \tikzsetnextfilename{allDiags49}
    & \gBoxTriNPB[scale=0.8,all=gluon,iB=rline,iC=rline,iD=rline,iE=rline]{}
    \tikzsetnextfilename{allDiags50}
    & \gBoxTriNPB[scale=0.8,all=gluon,iC=rline,iD=rline,iF=rline,iG=rline]{}
    \tikzsetnextfilename{allDiags51}
    & \gBoxTriNPB[scale=0.8,all=gluon,iB=gline,iC=bline,iD=bline,iE=gline,iF=rline,iG=rline]{}
  \end{tabular}
  \caption{Graphical representation of all non-vanishing contributions to the ${\cN=1}$ amplitude. We have suppressed certain internal bubble graphs with vanishing numerators, as well as all external bubble or tadpole graphs that can be set to zero in a massless gauge theory. The complete list is given in the ancillary file.}
  \label{fig:allNums}
\end{figure}

\subsection{Technical implementation details}

Having collected expressions for each numerator in terms of masters and written down an ansatz for the master numerators we are ready to constrain the ansatz.
The free parameters are collected in a ($i=3772$)-dimensional vector~$\bc$.
All the constraints, including cuts, are linear and most conveniently collected in matrix form
\begin{equation}\label{eq:matrixConstraints}
  \cB \cdot \bc = \bb\,,
\end{equation}
where $\cB$ is a $i\times j$-matrix with rational entries.
The dimension $j$ corresponds to the number of constraints, and $\bb$ is a constant $j$-dimensional vector with rational entries.

Functional constraints, coming from Jacobi, commutation, decomposition, flavor, and diagram symmetry relations are most efficiently brought into this form using an off-shell parametrization of the kinematic variables. 
Furthermore, we enforced manifest CPT invariance as for example discussed in~\cite{Johansson:2014zca}.

Information from 4-dimensional cuts can be efficiently extracted using algebraic geometry tools~\cite{Zhang:2016kfo}.
Non-linear constraints on the kinematics for a given cut can be encoded in a Groebner basis, which is efficiently found by {\sc Singular}~\cite{DGPS}.
The expressions for both the cut and the ansatz on the cut kinematics can be reduced using simple polynomial division towards this basis.

To reduce the linear system~\eqref{eq:matrixConstraints} we used finite field methods (see e.g.~\cite{vonManteuffel:2014ixa,Peraro:2016wsq}).
It was sufficient to reduce the system modulo two prime numbers: 3037000493, 3036979199.
These are the highest possible prime numbers for a 64-bit implementation\footnote{
We thank Alex Edison and Mao Zeng for sharing their modified version with us.
} of a finite field row-reduction algorithm for sparse matrices in {\sc SpaSM}~\cite{spasm}.
We used a private implementation of a \Cpp \emph{Library Link} from {\sc Mathematica} to have fast memory exchange which allows for an efficient parallelization of the computation.

Even with all constraints implemented the system has 36 free parameters.
An efficient way to find an algebraically appealing representation is by using a version of the Lenstra–Lenstra–Lov\'{a}sz (LLL) algorithm from~\cite{chen2005blas},\footnote{
The algorithm is, for example, implemented in the \emph{Integer Matrix Library (IML)} available on \href{https://cs.uwaterloo.ca/~astorjoh/iml.html}{https://cs.uwaterloo.ca/~astorjoh/iml.html}.}
which determines the smallest (depending on a norm defined on the vector space of parameters $\bc$) minimal-denominator solution of a linear system.

\subsection{Properties of numerator expressions}\label{sec:nums}

The complete set of kinematic numerators for the two-loop $\cN=1$ amplitude is given in computer-readable form in two ancillary files attached to the arXiv submission, including other pertinent information needed for assembling the amplitude. The two versions only differ in their treatment of the $\mu$ terms, and differences can thus be attributed to regularization scheme choices. 

The primary version, given in \texttt{anc\_neq1.dat}, corresponds to the construction discussed in detail in Sec.~\ref{sec:2Lcuts}.
It satisfies:
\begin{itemize}
\item All constraints discussed in Sec.~\ref{sec:diagIden},  namely:
  \begin{itemize}
  \item color-kinematics identities Eqs.~\eqref{eq:jacsComs}, \eqref{eq:clebschGordan}, and~\eqref{vertexsigns};
  \item optional (three-term) numerator relations Eqs.~\eqref{eq:twoTerm} and~\eqref{eq:twoTermFl};
  \item diagram and flavor (overall and subloop) symmetries Sec.~\ref{sec:flavor}, including the slightly modified versions in Eq.~\eqref{eq:brokenFlavor};
  \item supersymmetric decomposition identities Sec.~\ref{sec:decomp};
  \end{itemize}
\item all 4D cuts listed in Fig.~\ref{fig:cuts2L};
\item all deformed massive cuts in Fig.~\ref{fig:cuts2L}, with the $\mu$-terms introduced in Eq.~\eqref{eq:corrections};
\item power counting constraints discussed in the end of Sec.~\ref{sec:ansatz};
\item manifest CPT invariance, see~\rcite{Johansson:2014zca}.
\end{itemize}

The alternative version, given in \texttt{anc\_neq1\_alt.dat}, is similar to primary version, except that it contains three free parameters $x,y,z$. The details are discussed in App.~\ref{app:alt}, here we give a brief summary.
The alternative version satisfies:
\begin{itemize}
\item The majority, but not all, of the constraints discussed in Sec.~\ref{sec:diagIden}:
  \begin{itemize}
  \item color-kinematics identities Eqs.~\eqref{eq:jacsComs}, \eqref{eq:clebschGordan}, and~\eqref{vertexsigns};
  \item optional (three-term) numerator relations Eqs.~\eqref{eq:twoTerm} and~\eqref{eq:twoTermFl} with one exception as discussed in App.~\ref{app:alt};
  \item diagram and flavor (overall and subloop) symmetries Sec.~\ref{sec:flavor}, including the slightly modified versions in Eq.~\eqref{eq:brokenFlavor}, but breaks the bow-tie identity in Eq.~\eqref{eq:flavorSym2};
  \item supersymmetric decomposition identities Sec.~\ref{sec:decomp} with two exceptions for the bow-tie diagram as discussed in App.~\ref{app:alt};
  \end{itemize}
\item all 4D cuts listed in Fig.~\ref{fig:cuts2L};
\item all massive cuts in Fig.~\ref{fig:cuts2L}, without the deformation $(y=1)$, or with the deformation $(y=0)$, introduced in Eq.~\eqref{eq:corrections};
\item power counting constraints discussed in the end of Sec.~\ref{sec:ansatz};
\item manifest CPT invariance, see~\rcite{Johansson:2014zca}.
\end{itemize}
Depending on the choice of free parameters one or several of the exceptions mentioned here are fulfilled.
The free parameter $z$ controls the cut in Eq.~\eqref{eq:brokenCut}, as well as the sub-loop flavor symmetry of the bow-tie diagram in Eq.~\eqref{eq:brokenSubLoopFlavor}, and the color-algebra-like relation in Eq.~\eqref{eq:threeTermMatter}.
The latter two free parameters, $y$ and $x$, multiplies the deformation terms introduced in~\eqref{eq:corrections} and control their exact form. The primary version is recovered for $x=y=z=0$.

For completeness, both ancillary files include vanishing diagrams, either because their numerators vanish, or because they correspond to scaleless integrals: external-leg bubbles or tadpoles.
The latter two contributions cannot be easily constrained by unitarity cuts and hence should be dropped. A reason to still include them is that they are generated by the color-kinematics duality, and as such they carry some non-trivial information on how the full set of constraints affects the global numerator solution, which may be of some interest to the reader. Similar considerations were done for the $\cN=2$ two-loop amplitude in~\rcite{Johansson:2017bfl}, where constrains imposed on the external-leg bubbles and tadpoles influenced the full solution.

The diagrams that are relevant for the integrated answer of the amplitude are depicted in Fig.~\ref{fig:allNums}.
Diagrams not presented there are either zero, integrate to zero, or come with a vanishing color factor.

\subsection{Chiral short matter multiplets}

Two-loop numerators associated to short chiral multiplets have not been considered in this paper, but in principle their four-dimensional integrand-level contributions could be computed. While such chiral contributions would give rise to gauge anomalies, and are thus not desirable in a gauge theory, it turns out that they are important for the double-copy construction of supergravity amplitudes~\cite{Johansson:2014zca}. Short chiral multiplet gives the ability to better control the supergravity matter content, and the double copy of two chiral contribution need not be anomalous~\cite{Johansson:2014zca}. Below we briefly comment on the needed ingredients in their construction. 

\begin{figure}[t]
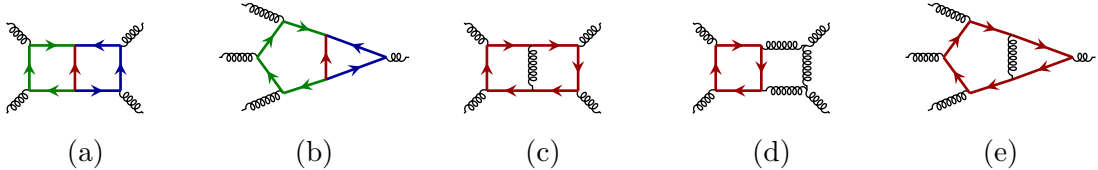

  \centering
  \begin{subfigure}[b]{0.19\textwidth}
    \centering
    \tikzsetnextfilename{mastersDirected1}
    \gBoxBox[all=gluon,iA=gquark,iB=gquark,iC=baquark,iD=baquark,iE=baquark,iF=gquark,iG=raquark]{}
    \subcaption{}
  \end{subfigure}
  \begin{subfigure}[b]{0.19\textwidth}
    \centering
    \tikzsetnextfilename{mastersDirected2}
    \gTriPenta[all=gluon,iA=gquark,iB=gquark,iC=gquark,iD=baquark,iE=baquark,iF=gquark,iG=raquark]{}
    \subcaption{}
  \end{subfigure}
  \begin{subfigure}[b]{0.19\textwidth}
    \centering
    \tikzsetnextfilename{mastersDirected3}
    \gBoxBox[all=gluon,iA=rquark,iB=rquark,iC=rquark,iD=rquark,iE=rquark,iF=rquark]{}
    \subcaption{}
  \end{subfigure}
  \begin{subfigure}[b]{0.19\textwidth}
    \centering
    \tikzsetnextfilename{mastersDirected4}
    \gBoxBox[all=gluon,iA=rquark,iB=rquark,iF=rquark,iG=rquark]{}
    \subcaption{}
  \end{subfigure}
  \begin{subfigure}[b]{0.19\textwidth}
    \centering
    \tikzsetnextfilename{mastersDirected5}
    \gTriPenta[all=gluon,iA=rquark,iB=rquark,iC=rquark,iD=rquark,iE=rquark,iF=rquark]{}
    \subcaption{}
  \end{subfigure}
  \caption{Master numerators for theories with short chiral multiplets denoted by arrowed lines.}
  \label{fig:mastersComplex}
\end{figure}

Numerators coming from short chiral multiplets can be treated analogous to the $\cN=2$ computations in refs.~\cite{Johansson:2017bfl,Kalin:2018thp,Kalin:2019vjc,Kalin:2019qup} after introducing arrows on the matter lines.
We have checked that there are five master numerators -- depicted in figure~\ref{fig:mastersComplex} -- which is substancially larger than for the amplitude with long chiral multiplets.
The chief reason for this is that in the decomposition identities the matter graphs appear twice with a different momentum configuration and hence this equation cannot be used to functionally solve for them.
Instead, the decomposition identity can be used as a constraint for the ansatz. The free parameters can be chosen, such that the relation between the long and short multiplets eq.~\eqref{eq:longToShort} manifestly holds,
i.e. these numerators add up to the four-dimensional part of the long chiral multiplet numerators presented in the previous subsection.

We leave the computation of the extra-dimensional part of these amplitudes, needed for certain double-copy constructions of pure-supergravity amplitudes~\cite{Johansson:2014zca}, for future work.

\section{Conclusions and outlook}
\label{sec:conclu}

In this paper, we have used color-kinematics duality to obtain the integrand for the four-vector two-loop
$\cN=1$ SYM amplitude, coupled to chiral matter multiplets,
in arbitrary representations of the gauge group, with chiral self-interactions (see detailed Lagrangians in App.~\ref{sec:Lagrangians}).
Our starting point was a generalization of
the massive spinor-helicity formalism~\cite{Arkani-Hamed:2017jhn}
and associated on-shell $\cN=1$ superspace
\cite{Herderschee:2019ofc,Herderschee:2019dmc}
to include complex masses (see \Sec{sec:neq1Setup}).
This included a supersymmetric depiction of the long chiral multiplet $\Phi$,
containing a complex scalar and Majorana fermion.
It blends conveniently with the usual massless on-shell formalism which we 
used to describe the $\cN=1$ vector multiplets $V^\pm$.

Equipped with the on-shell formalism we proceeded to obtain tree-level amplitudes.
First we wrote down three-point amplitudes involving various combinations
of long chiral multiplets and massless $\cN=1$ vector multiplets $V^\pm$
using elementary symmetry considerations (\Sec{sec:3ptAmps}).
From these we extracted four-point amplitudes via on-shell recursion (\Sec{sec:bcfw}),
correctly capturing the contact terms, such that it matches the unique supersymmetric $\cN=1$ theory presented in \App{sec:Lagrangians}.
This Lagrangian provided a useful check on the amplitudes;
furthermore we found that the three- and four-point amplitudes satisfy massive versions of the tree-level BCJ amplitude relations, or for pure-matter graphs, the kinematic Jacobi identity~\cite{Bern:2008qj}.

Following the by-now standard procedure for
obtaining loop amplitudes via generalized unitarity,
we sewed together the massive tree-level amplitudes to obtain two-loop cuts.
These provided the physical data needed to constrain an ansatz
for the kinematic color-dual numerators involved in the two-loop amplitude.
An important subtlety here was our re-interpretation of masses
in the tree-level amplitudes as extra-dimensional
components of loop momenta in dimensional regularization.
We found that, while the two can usually be straightforwardly identified,
in one particular instance a difference between the kinematic Jacobi identities
in the massive and extra-dimensional settings
creates tension in the numerator relations.
We resolved this tension by introducing a simple $\mu$-term deformation of the cuts
which we associate with the scheme dependence of dimensional regularization --- see \Sec{sec:2Lcuts} for further details.

Also crucial to finding a two-loop color-dual representation was our use of
\emph{supersymmetric decomposition identities}
(see \Secs{sec:susydecomp}{sec:decomp}).
Using double-box numerators as a simple example,
one can demand that the $\cN=1$ SYM numerators with long chiral matter multiplets (unarrowed lines) satisfy
\begin{align}
\begin{aligned}
  \tikzsetnextfilename{decompConcl1}
  \gBoxBox[all=gluon]{
    \node at (0.9,1.2) {$\cN=2$};
  }
  &=
  \tikzsetnextfilename{decompConcl2}
  \gBoxBox[all=gluon]{
    \node at (0.9,1.2) {$\cN=1$};
  }
  +
  \tikzsetnextfilename{decompConcl3}
  \gBoxBox[all=gluon,iA=rline,iB=rline,iC=rline,iD=rline,iE=rline,iF=rline]{
    \node at (0.9,1.2) {$\cN=1$};
  }
  +
  \tikzsetnextfilename{decompConcl4}
  \gBoxBox[all=gluon,iA=rline,iB=rline,iF=rline,iG=rline]{
    \node at (0.9,1.2) {$\cN=1$};
  }
  +
  \tikzsetnextfilename{decompConcl5}
  \gBoxBox[all=gluon,iC=rline,iD=rline,iE=rline,iG=rline]{
    \node at (0.9,1.2) {$\cN=1$};
  }\,,
  \\
  \tikzsetnextfilename{decompConcl6}
  \gBoxBox[all=gluon,iA=quark,iB=quark,iC=quark,iD=quark,iE=quark,iF=quark]{
    \node at (0.9,1.2) {$\cN=2$};
  }
  &=
  \tikzsetnextfilename{decompConcl7}
  \gBoxBox[all=gluon,iA=rline,iB=rline,iC=rline,iD=rline,iE=rline,iF=rline]{
    \node at (0.9,1.2) {$\cN=1$};
  }
  \tikzsetnextfilename{decompConcl8}
  +\gBoxBox[all=gluon,iA=gline,iB=gline,iC=rline,iD=rline,iE=rline,iF=gline,iG=bline]{
    \node at (0.9,1.2) {$\cN=1$};
  }\,,
  \\
  \tikzsetnextfilename{decompConcl9}
  \gBoxBox[all=gluon,iA=quark,iB=quark,iF=quark,iG=aquark]{
    \node at (0.9,1.2) {$\cN=2$};
  }
  &=
  \tikzsetnextfilename{decompConcl10}
  \gBoxBox[all=gluon,iA=rline,iB=rline,iF=rline,iG=rline]{
    \node at (0.9,1.2) {$\cN=1$};
  }
  \tikzsetnextfilename{decompConcl11}
  +\gBoxBox[all=gluon,iA=gline,iB=gline,iC=rline,iD=rline,iE=rline,iF=gline,iG=bline]{
    \node at (0.9,1.2) {$\cN=1$};
  }\,,
\end{aligned}
\end{align}
where colors are used to represent distinct flavor.
Given prior knowledge of the $\cN=2$ SQCD
double-box numerators on the left-hand side ---
which in \rcite{Johansson:2017bfl} were chosen to satisfy
a decomposition identity to $\cN=4$ SYM similar to the first line above ---
all other $\cN=1$ double-box numerators on the right-hand side are determined
in terms of the one with three-matter interactions.
This observation greatly reduces the number of master numerators for us to solve for;
it also strongly suggests that,
even if one seeks an amplitude without the three-matter interaction
(for example an $\cN=1$ SQCD amplitude)
diagrams of this kind are still useful,
as they allow information to be recycled from the simpler
theories with more supersymmetry.

These identities were supplemented by other useful off-shell relations.
The presence of cubic matter self-interactions with generic Clebsch-Gordan
coefficients $C_{ijk}$ led us to a three-term identity \eqref{eq:clebschGordan},
whose appearance greatly resembles the Jacobi and commutation relations
demanded by color-kinematics duality involving matter.
Following \rcite{Johansson:2017bfl} we also found it useful to impose a two-term
identity between between graphs with complex matter \eqref{eq:twoTerm},
which led to another three-term identity when combined to the full chiral multiplet.
Finally, inspired by the tree-level amplitudes,
we imposed another three-term identity with two pairs of differently flavored
external matter lines.
To summarize: the full set of three-term identities involved in the off-shell
construction are
\begin{align}
\begin{aligned}
  \tikzsetnextfilename{concl4PTIDs1}
  \gTreeS[scale=1.2,all=gluon]{} &=
  \tikzsetnextfilename{concl4PTIDs2}
  \gTreeT[scale=1.2,all=gluon]{}
  \tikzsetnextfilename{concl4PTIDs3}
  -\gTreeU[scale=1.2,all=gluon]{}\,,~~~~ &
  \tikzsetnextfilename{concl4PTIDs4}
  \gTreeS[scale=1.2,all=gluon,eC=rline,eD=rline]{} &=
  \tikzsetnextfilename{concl4PTIDs5}
  \gTreeT[scale=1.2,all=gluon,eC=rline,eD=rline,iA=rline]{}
  \tikzsetnextfilename{concl4PTIDs6}
  -\gTreeU[scale=1.2,all=gluon,eC=rline,eD=rline,iA=rline]{}\,,\\
  \tikzsetnextfilename{concl4PTIDs7}
  \gTreeS[scale=1.2,all=gluon,eB=rline,eC=bline,eD=gline,iA=rline]{} &=
  \tikzsetnextfilename{concl4PTIDs8}
  \gTreeT[scale=1.2,all=gluon,eB=rline,eC=bline,eD=gline,iA=gline]{}
  \tikzsetnextfilename{concl4PTIDs9}
  -\gTreeU[scale=1.2,all=gluon,eB=rline,eC=bline,eD=gline,iA=bline]{}\,,~~~~ &
  \tikzsetnextfilename{concl4PTIDs10}
  \gTreeS[scale=1.2,eA=rline,eB=rline,eC=rline,eD=rline,iA=gluon]{} &=
  \tikzsetnextfilename{concl4PTIDs11}
  \gTreeT[scale=1.2,eA=rline,eB=rline,eC=rline,eD=rline,iA=gluon]{}
  \tikzsetnextfilename{concl4PTIDs12}
  -\gTreeU[scale=1.2,eA=rline,eB=rline,eC=rline,eD=rline,iA=gluon]{}\,,\\
  \tikzsetnextfilename{concl4PTIDs13}
  \gTreeS[scale=1.2,eA=gline,eB=gline,eC=bline,eD=bline,iA=gluon]{} &=
  \tikzsetnextfilename{concl4PTIDs14}
  \gTreeT[scale=1.2,eA=gline,eB=gline,eC=bline,eD=bline,iA=rline]{}
  \tikzsetnextfilename{concl4PTIDs15}
  -\gTreeU[scale=1.2,eA=gline,eB=gline,eC=bline,eD=bline,iA=rline]{}\,.
\end{aligned}
\end{align}
These were supplemented with flavor symmetries,
demanding that graphs with distinct matter loops
and different flavors could always be identified.

Having obtained an integrand for the two-loop four-vector
amplitude of generic ${\cN=1}$ SYM with added chiral multiplets,
we aim to use this result to compute the full integrated ${\cN=1}$ SQCD amplitude
--- an example of particular interest among such theories due to its similarity to QCD.
In \rcite{DelDuca:2017peo} it was found that ---
when tuned to a particular number of chiral multiplets ---
this theory obeys the principle of maximal transcendentality
at next-to-leading logarithmic order in the high-energy limit.
It would be interesting to see how much of
this simplicity survives beyond the high-energy limit,
as the analogously tuned result for $\cN=2$ SQCD was observed to break maximal
transcendentality in a very simple way
\cite{Dixon2008talk,Leoni:2015zxa,Duhr:2019ywc,Kalin:2019vjc}.
Furthermore, for the two-loop four-vector amplitude in $\cN=2$ SQCD for a generic
number of hypermultiplets a particular choice of infrared subtraction motivated by
structure of the integrand~\cite{Kalin:2018thp,Kalin:2019vjc} resulted in a finite
amplitude with enhanced transcendentality behavior~\cite{Duhr:2019ywc,Kalin:2019vjc}.
It would be interesting to see if a similar analysis
is feasible for the $\cN=1$ supersymmetric theory.

Finally, this amplitude provides us with an excellent means to
study a wider class of two-loop supergravity amplitudes in different dimensions ---
and in particular their ultraviolet behavior.
While two-loop Einstein gravity amplitudes are known
to be divergent in five dimensions due to an allowed $R^4$ counterterm~\cite{Goroff:1985th,vandeVen:1991gw},
five-dimensional $\cN=4$ supergravity is finite \cite{Bern:2013qca}
despite such a counterterm not being ruled out by symmetry considerations
\cite{Bossard:2011tq,Bossard:2012xs,Bossard:2013rza,Bern:2010tq}.
An interesting question is therefore whether this ``enhanced cancellation''
persists for theories with $0<\cN<4$ supersymmetries
By double copy with other two-loop $\cN<4$ gauge theory amplitudes ---
such as this one or the two-loop $\cN=2$ amplitude discussed here and in
\rcites{Johansson:2017bfl,Kalin:2018thp,Kalin:2019vjc} ---
we can potentially answer this and other interesting questions.
We emphasize that in order to compute arbitrary matter-coupled gravity theory amplitudes via the double copy one requires gauge-theory amplitudes with short chiral matter multiplets.
Whereas the four-dimensional contributions of these amplitudes can be computed with the same techniques as discussed here, the extra-dimensional pieces, if needed, are not well defined. We leave the computation of these pieces and the study of the double copy for future work.

Finally, we note that with the amplitudes computed here, we are only one step away, using a final superymmetric decomposition, to the non-supersymmetric QCD and pure-gravity amplitudes at two loops. While these amplitudes are well known~\cite{Bern:2017puu,Abreu:2020lyk}, the interesting testbed question is if the current approaches can be streamlined to the point that they become competitive for non-supersymmetric gauge and gravity theories at two loop and beyond. Indeed, we expect that same techniques and ideas presented here will be directly applicable to such computations.

\begin{acknowledgments}

We thank Zvi Bern, Lance Dixon, Alex Edison, Paolo Pichini, Jan Plefka, and Radu Roiban for interesting and useful discussions related to the current work.
We are also greatful to Jakob Andersen for support and help with the \Cpp~graph canonization library \emph{GraphCanon}~\cite{graphCanon}.
This research was funded by the Swedish Research Council under grant 621-2014-5722,  the Ragnar S\"{o}derberg Foundation (Swedish Foundations' Starting Grant),
and the Knut and Alice Wallenberg Foundation under grants KAW 2013.0235, 2018.0116, 2018.0162.
GK was also supported in part by the Knut and Alice Wallenberg Foundation under grant 2018.0441, the US Department of Energy under contract DE--AC02--76SF00515, and is currently funded by the ERC Consolidator Grant “Precision
Gravity: From the LHC to LISA” provided by the European Research Council (ERC) under the European Union’s H2020 research and innovation programme (grant agreement No.
817791). GM’s research is currently funded by the Deutsche Forschungsgemeinschaft
(DFG, German Research Foundation) ---
Projektnummer 417533893/GRK2575 ``Rethinking Quantum Field Theory''.
BV was also supported by the European Research Council under ERC grant \emph{UNISCAMP} (804286).
  
\end{acknowledgments}

\appendix

\section{Grassmann variable conventions}

Here we summarize conventions with respect to massive Grassmann variables $\eta_a$
and their complex conjugates $\bar{\eta}^a\equiv(\eta_a)^*$
(careful: notice that $(\eta^a)^*=-\bar{\eta}_a$).
We define
\begin{align}
|\fq_i \rangle   \equiv  |i^a \rangle \eta_{i,a} \,, &&
|\bar{\fq}_i \rangle   \equiv  |i^a \rangle \bar{\eta}_{i,a} \,, &&
|\fq_i ]   \equiv  |i_a ] \eta_{i}^a \,, &&
|\bar{\fq}_i ]   \equiv  |i_a ] \bar{\eta}_{i}^a\,.
\end{align}
When taking complex conjugates of strings of $\eta_a\!$'s we reverse their order:
e.g. $(\eta_1\eta_2)^*=\bar{\eta}^2\bar{\eta}^1$.
This choice is helpful as it ensures that
\begin{align}
\braket{\fq_i\fq_j}^*=[\bar{\fq}_i\bar{\fq}_j]\,, &&
[\fq_i\fq_j]^*=\braket{\bar{\fq}_i\bar{\fq}_j}\,.
\end{align}
We also adopt the shorthands $(\eta)^2=\eta_1\eta_2=\eta^1\eta^2$ and
$(\bar{\eta})^2=\bar{\eta}^2\bar{\eta}^1=\bar{\eta}_2\bar{\eta}_1$,
which are complex conjugates of each other.

The on-shell supercharges are expressed as
\be
Q_\alpha = \sum_{i=1}^{n}|\fq_{i}\rangle_\alpha\,,~~~~~
Q^\dagger_{\dot{\alpha}} = -\sum_{i=1}^{n}\int\!\d^2\eta_i [\fq_{i}|_{\dot{\alpha}}\,,
\ee
where the Grassmann integration measures are defined so that $\int\!\d^2\eta\,(\eta)^2=1$
and $\int\!\d^2\bar{\eta}\,(\bar{\eta})^2=1$
($\d^2\eta=\d\eta^2\d\eta^1$ and
$\d^2\bar{\eta}=\d\bar{\eta}^1\d\bar{\eta}^2$).
We also have
\be
Q^\dagger = {\rm FT} \, \overline{Q} \,,~~~\text{where}~~~ \overline{Q} =  \sum_{i=1}^{n}  \bar{\eta}_{i}^a  [i_a| \equiv\sum_{i=1}^{n}  [\overline{\fq}_i|\,.
\ee
The supersymmetric delta functions are defined in terms of these as
\begin{align}
\delta^2(Q)&=\frac12Q^\alpha Q_\alpha=Q_2Q_1=\frac12\sum_{i,j}\braket{\fq_i\fq_j}=
\sum_{i<j}\braket{i^aj^b}\eta_{i,a}\eta_{j,b}-\sum_im_i(\eta_i)^2\,,\\
\delta^2(\widebar{Q})&=\frac12\widebar{Q}_{\dot{\alpha}}\widebar{Q}^{\dot{\alpha}}=
\widebar{Q}_1\widebar{Q}_2=
\frac12\sum_{i,j}[\bar{\fq}_i\bar{\fq}_j]=
\sum_{i<j}[i^aj^b]\bar{\eta}_{i,a}\bar{\eta}_{j,b}-\sum_i\bar{m}_i(\bar{\eta}_i)^2\,.
\end{align}

\section{Lagrangians for $\cN=1$ gauge theories}
\label{sec:Lagrangians}

In this Appendix we review the $\cN=1$ supersymmetric gauge theory 
Lagrangians relevant for the amplitudes considered in the main text.
We begin with a generic $\cN=1$ theory coupled to matter multiplets,
and then later specialize to $\cN=1$ SQCD.
We also examine how supersymmetric decomposition manifests on the Lagrangian.

\subsection{Generic $\cN=1$ Yang-Mills theories with matter}

In this section we review the Lagrangian of generic $\cN=1$ supersymmetric Yang-Mills theories with matter.
We begin with pure $\cN=1$ SYM, given by
\begin{equation}\label{eq:pureNeq1L}
 {\cal L}_{{\rm pure}\,\,{\cal N}=1\,{\rm SYM}}={\rm Tr}\Big(-\frac{1}{4} F_{\mu\nu} F^{\mu\nu} + i \widetilde{\lambda}_{\dot \alpha}\,
  \slash \hskip-3mm D^{\dot \alpha \alpha}\lambda_{\alpha } + \mathrm{D}^2\Big)\,,
\end{equation}
where we utilize slash notation $~\slash \hskip-3mm D^{\dot \alpha \alpha} =D^\mu \overline{\sigma}^{\dot \alpha \alpha}_\mu $ with Pauli matrices $\overline\sigma_\mu= \sigma^\mu=(1,\sigma_i)$ and the covariant derivative is $D_\mu=\partial_\mu +\frac{i g}{\sqrt{2}}  {\rm T}^a A_\mu^a $ where ${\rm T}^a $ is to be interpreted as a representation matrix of the field on which it acts. Our conventions are chosen such that ${F_{\mu\nu}^a=\partial_\mu A_\nu^a-\partial_\nu A_\mu^a+\frac{i g}{\sqrt{2}}  A_\mu^b A_\nu^c f^{abc}}$, $f^{abc}={\rm Tr}(T^a[T^b,T^c])$, ${\rm Tr}(T^aT^b)=\delta^{ab}$ where $T^{a}$ denotes the usual Lie algebra generator. The metric signature is $(+---)$ and the gaugino $\lambda_{\alpha}$ is a two-component spinor subjugate to the Majorana condition $\widetilde\lambda= \lambda^C$. We work in the Majorana basis, where $\widetilde\lambda= \lambda^C= \lambda^*$. The pure $\cN=1$ SYM Lagrangian in~\eqn{eq:pureNeq1L} is invariant under the following shifts:
\begin{align}\label{eq:gaugeshifts}
    \begin{aligned}
        \delta_\eps\,\slash\hskip-3mm A_{\alpha\dot{\alpha}}&=
        \sqrt2(\widebar{\eps}_{\dot{\alpha}}\lambda_\alpha+\widetilde{\lambda}_{\dot{\alpha}}\eps_\alpha)\,,\\
        \delta_\eps\lambda_\alpha&=\frac{i}{2\sqrt2}\sigma^\mu_{\alpha\dot{\alpha}}(\widebar{\sigma}^\nu)^{\dot{\alpha}\beta}\eps_\beta F_{\mu\nu}-
        \eps_\alpha \mathrm{D}\,, \\
        \delta_\eps \mathrm{D}&=\frac{i}2\left(\widebar{\eps}_{\dot{\alpha}}\,\slash\hskip-3mm D^{\dot \alpha \alpha}\lambda_\alpha+\eps_\alpha\,\slash\hskip-3mm D^{\dot \alpha \alpha}\widetilde{\lambda}_{\dot{\alpha}}\right)\,.
    \end{aligned}
\end{align}
To~\eqn{eq:pureNeq1L} we add the massless matter Lagrangian, specialising to matter in the adjoint representation,
\begin{align}
    {\cal L}_{\cN=1\text{ matter}}=& \,{\rm Tr}\left(
	    |D_\mu\varphi^A |^2+\widetilde{\psi}^A_{\dot \alpha}\,i\,\slash\hskip-3mm D^{\dot \alpha \alpha}\psi^A_{\alpha }+|\mathrm{F}^A|^2+
    g\widebar{\varphi}^A\lambda^\alpha\psi^A_\alpha-
    g\varphi^A\widetilde{\lambda}_{\dot{\alpha}}\widetilde{\psi}^{A\dot{\alpha}}+
    g\widebar{\varphi}^A\mathrm{D}\varphi^A\,\right).
\end{align}
Taken in combination with the shifts given in~\eqn{eq:gaugeshifts},
this Lagrangian is invariant (up to total derivatives) under
\begin{align}\label{eq:complexShifts}
    \delta_\eps\varphi=\eps^\alpha\psi_\alpha\,, &&
    \delta_\eps\psi_\alpha=
    -i\bar{\eps}^{\dot{\alpha}}\slash\hskip-3mm D_{\alpha \dot \alpha}\varphi+
    \eps_\alpha \mathrm{F}\,, &&
    \delta_\eps \mathrm{F}=-i\bar{\eps}_{\dot{\alpha}}
    \,\slash\hskip-3mm D^{\dot \alpha \alpha}\psi_\alpha+
    g\widebar{\eps}_{\dot \alpha}\widetilde{\lambda}^{\dot{\alpha}}\varphi\,,
\end{align}
where $\delta_\eps\widebar{\mathrm{F}}=(\delta_\eps \mathrm{F})^*$ and (using the Majorana condition)
$\delta_\eps\widetilde{\psi}_{\dot{\alpha}}=(\delta_\eps\psi_\alpha)^*$.
Finally, we include interaction - and mass terms for the chiral multiplets
\begin{align}\label{eq:InteractionLagrangian}
    {\cal L}_{\cN=1\text{ int.}}={\rm Tr}\left(
    \frac{\partial W}{\partial\varphi^A}\mathrm{F}^A-\frac12\frac{\partial^2W}{\partial\varphi^A\partial\varphi^B}\psi^{\alpha A}\psi_\alpha^B\right)+\text{h.c.}\,,
\end{align}
where $W(\varphi^A)$ is the \emph{superpotential} (see e.g.~\cite{Aitchison:2007fn,Weinberg:2000cr}).
It is convenient for us to choose
\begin{equation}\label{eq:SuperPotential}
    W={\rm Tr}\left(\frac{\widebar{m}_A}2\varphi^A\varphi^A+
    \frac{g}3{T}_{ABC}\varphi^A\varphi^B\varphi^C\right)\,,
\end{equation}
where $T_{ABC}$ is a flavor tensor capturing the cubic coupling;
$A,B,C$ are implicitly summed over,
and $m_A=|m_A|e^{i\xi_A}$ is a complex mass.

To obtain a more useful Lagrangian from which one can derive Feynman rules
it might now be expected to solve for $\mathrm{F}^A$ and $\mathrm{D}$.
However, we find it more convenient to simply insert
\begin{equation}
	\mathrm{F}^A=\mathrm{U}^A-m_A\widebar{\varphi}^A\,,
\end{equation}
and leave $\mathrm{D}$ alone.
This diagonalizes the quadratic terms on $\mathrm{U}^A$ and $\varphi^A$. Splitting the terms generated by \eqref{eq:InteractionLagrangian} into mass terms for the chiral component fields and Yukawa-type couplings
gives rise to the following Lagrangians:
\begin{align}\label{eq:Neq1MatterLagrangian}
    {\cal L}_{\cN=1 \text{ m. matter}}&={\rm Tr}\bigg(
    |D_\mu\varphi^A|^2+
    \widetilde{\psi}^{A}_{\dot \alpha}\,
    i\,\slash \hskip-3mmD^{\dot \alpha \alpha}\psi^{A}_{\alpha }
    -\frac12\left(\widebar{m}_A\psi^{\alpha A}\psi^A_{\alpha }+m_A\widetilde{\psi}^A_{\dot \alpha}  \widetilde{\psi}^{\dot \alpha A}\right)\nn\\
    &\quad-|m_A\varphi^A|^2+
    g\widebar{\varphi}^A\lambda^\alpha\psi_\alpha^A-
    g\varphi^A\widetilde{\lambda}_{\dot{\alpha}}\widetilde{\psi}^{\dot{\alpha}A}+
    g\widebar{\varphi}^A\mathrm{D}\varphi^A+|\mathrm{U}^A|^2\bigg)\,,\\
    {\cal L}_{\cN=1 \text{ Yukawa}}&=
    g{T}_{ABC}{\rm Tr}\bigg(\widebar{m}_A\varphi^A\widebar{\varphi}^B\widebar{\varphi}^C-m_A\widebar{\varphi}^A\varphi^B\varphi^C
    -\varphi^A\psi^{\alpha B}\psi_\alpha^C\nn\\
    &\qquad\qquad\qquad+
    \widebar{\varphi}^A\widetilde{\psi}^B_{\dot \alpha}\widetilde{\psi}^{\dot \alpha C}+
    \varphi^A\varphi^B\mathrm{U}^C-\widebar{\varphi}^A\widebar{\varphi}^B\widebar{\mathrm{U}}^C\bigg)\,,
\end{align}
which one may add to the pure $\cN=1$ SYM Lagrangian~\eqn{eq:pureNeq1L} to obtain the full Lagrangian of $\cN=1$ SYM with adjoint massive matter.
The propagators for $\mathrm{D}$ and $\mathrm{U}^A$ are trivial,
and their contributions give rise to four-point contact terms.

The on-shell fermion polarization states in spinor-helicity notation are
\begin{eqnarray}
    \widetilde{\lambda}_{\dot\alpha}(p) &\rightarrow&  [p|_{\dot\alpha}\, ~~~~ \widetilde{\psi}_{\dot\alpha}(p) \rightarrow [p^a|_{\dot\alpha} =
    -\frac{1}{m} \langle p^a |^{\alpha}p_{\alpha \dot\alpha} \,, \nn \\
    \lambda_\alpha(p) &\rightarrow & |p\rangle_\alpha\, ~~~~  \psi_\alpha(p) \rightarrow |p^a\rangle_\alpha =   \frac{1}{\overline m}\, p_{\alpha \dot\alpha} | p^a ]^{\dot \alpha} \,.
\end{eqnarray}
In this convention, $\lambda$ is a negative helicity spinor, and $\tilde \lambda$ is a positive helicity spinor. 

The on-shell multiplets are given by
\begin{eqnarray}
    V^+ &=&A^+ + \widetilde \lambda^+ \eta^1 \,,~~~~~V^- = \lambda^- + A^- \eta^1 \nn \\
    \Phi_{{\cal N}=1\,{\rm long}} &=&  \varphi + \Psi_a\eta^{a}+\overline \varphi (\eta)^{2}  \,,
\end{eqnarray}
where the $SU(2)$ little group indices are labeled by helicity $1=+$ and $2=-$.
The $\Psi_a$ is an on-shell Majorana spinor, which is a superposition of the on-shell Weyl spinors: $\Psi_a \sim \psi_a+\widetilde \psi_a$.
In the massless limit one can decompose the Majorana spinor into its chiral fields, of which only the following survive on-shell:
$\psi \rightarrow \psi_1=- \psi^-$ and $\widetilde \psi \rightarrow \widetilde  \psi_2 = \widetilde \psi^+$, giving the short chiral multiplets
\begin{equation}
    \Phi^+_{{\cal N}=1\,{\rm short}} =  \widetilde  \psi_2+\overline \varphi \eta^{1}\,,
    ~~~~  \Phi^-_{{\cal N}=1\,{\rm short}} =  \varphi + \psi_1 \eta^{1}\,.
\end{equation}
However, these short multiplets need to be assigned to a real representation of the gauge group (effectively giving back the long multiplet),
otherwise the theory has a one-loop gauge anomaly.

\subsection{Massive ${\cal N}=1$ theories from SUSY decomposition }
\label{massiveFromSUSY}

Let us now specify to a particular $\cN=1$ theory with adjoint matter, which amounts to picking a specific form of the superpotential~\eqn{eq:SuperPotential}. The theory we will consider is $\cN=1^*$, which is of particular importance to the supersymmetric decomposition used throughout this paper, as it can be interpreted as a massive deformation of $\cN=4$ SYM. The theory consists of an $\cN=1$ gauge multiplet with three adjoint massive chiral multiplets $\varphi^A$. Its superpotential is given by
\begin{equation}\label{eq:SuperPotentialNeq1st}
W_{\cN=1^*}={\rm Tr}\left(\frac{\widebar{m}_A}2\varphi^A\varphi^A+
\frac{g}3{\eps}_{ABC}\varphi^A\varphi^B\varphi^C\right)\,,
\end{equation}
from which we see that the ${\cal N}=1^*$ Yukawa interaction is proportional to the Levi-Civita tensor ${\eps}_{ABC}$ of $SO(3)$. We decompose the ${\cal N}=1^*$ SYM Lagrangian as before:
\begin{equation}
{\cal L}_{{\cal N}=1^*} = {\cal L}_{{\rm pure}\,\,{\cal N}=1\,{\rm SYM}}  + {\cal L}_{{\cal N}=1\,{\text{m.  matter}}} +  {\cal L}_{{\cal N}=1^*\,{\rm Yukawa}}\,,
\end{equation}
where the pure ${\cal N}=1$ SYM theory is given by~\eqn{eq:pureNeq1L}.
The matter Lagrangian is identical for the 3 flavors of chiral multiplets, and is given by~\eqn{eq:Neq1MatterLagrangian}. 
As indicated by~\eqn{eq:SuperPotentialNeq1st}, the Yukawa interactions are given by
\begin{eqnarray}
{\cal L}_{{\cal N}=1^*\,{\rm Yukawa}} &= &  g  \epsilon_{ABC} {\rm Tr}\Big( \overline{m}_A \varphi^A  \overline{\varphi}^B  \overline{\varphi}^C  - m_A \overline{\varphi}^A  \varphi^B  \varphi^C  - \varphi^A  \psi^{\alpha B}  \psi_{\alpha}^C  \nn \\
				      &&\null ~~~~~~~~~~~~~+ \overline{\varphi}^A \widetilde{\psi}^B_{\dot \alpha}   \widetilde{\psi}^{{\dot \alpha} C} + \varphi^A \varphi^B \mathrm{U}^C- \overline{\varphi}^A \overline{\varphi}^{B} \overline{\mathrm{U}}^C \Big)\,,
\end{eqnarray}
where we may normalize by setting $\epsilon_{123}=1$. 
Much like one can obtain $\cN=1^*$ from a massive deformation of $\cN=4$ SYM, one could also obtain a massive $\cN=1$ theory by deforming $\cN=2$ SYM. Let us call this theory $\cN=1^{**}$ SYM. Its matter content is given by an $\cN=1$ gauge multiplet with a single massive adjoint chiral multiplet. The superpotential is given by
\begin{equation}\label{eq:SuperPotentialNeq1dbst}
	W_{\cN=1^{**}}={\rm Tr}\left(\frac{\widebar{m}}2\varphi\varphi\right)\,,
\end{equation}
from which we see that the Lagrangian is determined entirely by~\eqns{eq:pureNeq1L}{eq:Neq1MatterLagrangian},
\begin{equation}
	{\cal L}_{{\cal N}=1^{**}} = {\cal L}_{{\rm pure}\,\,{\cal N}=1\,{\rm SYM}}  + {\cal L}_{{\cal N}=1\,{\text{m. matter}}}\,,
\end{equation}
and so the theory corresponds to restricting $\cN=1^*$ to a single flavor.

Finally, let us comment on what $\cN=2$ theories look like in our ${\cal N}=1$ language. Note that the pure ${\cal N}=2$ SYM theory is given by
\begin{equation}
	{\cal L}_{{\rm pure}\,\,{\cal N}=2\,{\rm SYM}} =   {\cal L}_{{\rm pure}\,\,{\cal N}=1\,{\rm SYM}}+ {\cal L}_{{\cal N}=1\,{\rm matter}}
\end{equation}
which, eliminating the auxiliary fields ${\rm D}$ and ${\rm F}$ via their equations of motion, yields
\begin{align}
	{\cal L}_{{\rm pure}\,\,{\cal N}=2\,{\rm SYM}}= {\rm Tr}\Big(&-\frac{1}{4} F_{\mu\nu} F^{\mu\nu} +|D_\mu \varphi|^2+ \widetilde{\lambda}_{\dot \alpha A}\,i \,[ \,\slash \hskip-3mm D^{\dot \alpha \alpha},\lambda_{\alpha }^{A}]\nn \\
	&+ g \lambda^{\alpha 1} [\lambda_{\alpha }^{2}, \overline{\varphi}]- g  \widetilde{\lambda}_{\dot \alpha 1} [\widetilde{\lambda}^{\dot \alpha }_{2} ,\varphi] -\frac{g^2}{4}[\overline{\varphi},\varphi]^2\Big)\,,
\end{align}
where we have changed notation for the fermions $\lambda \rightarrow \lambda_1 $, $\psi \rightarrow \lambda_2 $ to make the R-symmetry manifest.

The matter part of ${\cal N}=2^*$ SYM -- the $\cN=2$ theory one obtains via mass deformation of $\cN=4$ SYM -- is
\begin{equation}
	{\cal L}_{{\cal N}=2\,{\rm matter}}= \sum_{A=2}^3 {\cal L}_{{\cal N}=1\,{\rm m. matter}} +  {\cal L}_{{\cal N}=1\,{\rm Yukawa}}\,.
\end{equation}
And thus one may compose
\begin{equation}
{\cal L}_{{\cal N}=2^*} = {\cal L}_{{\rm pure}\,\,{\cal N}=2\,{\rm SYM}}  + {\cal L}_{{\cal N}=2\,{\text{matter}}}\,,
\end{equation}
Changing the $\cN =2$ matter representation to fundamental and increasing the number of matter multiplets to $N_f$ would land on the theory known as ${\cal N}=2$ SQCD.

\subsection{The Lagrangian for ${\cal N}=1$ SQCD}

Another $\cN=1$ gauge theory of particular interest is $\cN=1$ super-QCD (SQCD), which constitutes of $\cN=1$ SYM coupled to $N_f$ fundamental and antifundamental chiral $\cN=1$ multiplets, $\Phi$ and $\widetilde{\Phi}$ respectively. We may use them to compose on-shell massive $\cN=1$ multiplets
\begin{align}
	\Phi_{\cN=1 \text{ long}} &= \varphi + \Psi_a \eta^a +\overline{\widetilde{\varphi}}\, (\eta)^2  \,, \\
	\overline{\Phi}_{\cN=1 \text{ long}} &= \widetilde{\varphi} +\overline{\Psi}^a  {\eta}_a  +\overline{\varphi}\, (\eta)^2\,,
\end{align}
where the Dirac spinors $\Psi$ and $\overline{\Psi}$ are given by
\begin{equation}
    \Psi_a = \left(\begin{matrix}
        \psi \\ \overline{\widetilde{\psi}}
    \end{matrix}\right)\,, \qquad
    \overline{\Psi}^a = \left(\begin{matrix}
        \widetilde{\psi} && \overline{\psi}
    \end{matrix}\right)\,. \qquad
\end{equation}
In the massless limit, these multiplets decompose to on-shell chiral multiplets as follows
\begin{equation}
    \Phi_{\cN=1 \text{ long}} \xrightarrow{m \to 0} \Phi_{\cN=1} + \overline{\widetilde{\Phi}}_{\cN=1}\,{\eta}_2
\end{equation}
where
\begin{align}
    \Phi_{\cN=1} &= \varphi +\psi \, \eta^1  \,, \\
    \overline{\widetilde{\Phi}}_{\cN=1} &= \overline{\widetilde{\psi}} +\overline{\widetilde{\varphi}} \, \eta^1 \,,
\end{align}
and similar relations hold for the conjugate superfield.

The theory's superpotential is given by
\begin{equation}\label{eq:SuperPotentialNeq1SQCD}
    W_{\cN=1 \text{ SQCD}}=\widebar{m}_A\widetilde{\varphi}_A\varphi^{A}\,,
\end{equation}
where $A=1,\dots,N_f$ is the flavor index of the $\cN=1$ matter fields. Note that the superpotential vanishes in the massless limit. Since there are no Yukawa-type couplings, we may decompose the Lagrangian as
\begin{equation}
    {\cal L}_{{\cal N}=1\,{\rm SQCD}} = {\cal L}_{{\rm pure}\,\,{\cal N}=1\,{\rm SYM}}  + {\cal L}_{{\text{matter}}}+ \widetilde{\cal L}_{{\text{matter}}} + {\cal L}_{{\text{mass terms}}}\,,
\end{equation}
where -- upon fixing the auxiliary $\mathrm{F}$-field to $\mathrm{F}^A = \mathrm{U}^A -m_A \overline{\widetilde{\varphi}}^A$ and $\widetilde{\mathrm{F}}_A = \widetilde{\mathrm{U}}_A -m_A \overline{\varphi}_A$ -- the kinetic terms stemming from the fundamental chiral multiplet are given by
\begin{equation}
	{\cal L}_{{\text{matter}}} =|D_\mu \varphi^A|^2 + \widebar{\psi}_{A\,\alpha} i \, \slash \hskip-3mmD^{\dot \alpha \alpha}\psi^{A}_{\alpha } + g\widebar{\varphi}_A  \lambda^\alpha\psi_\alpha^A -g\varphi^A  \widebar{\lambda}_{\dot{\alpha}}\widebar{\psi}^{\dot{\alpha}}_A+g\widebar{\varphi}_A \mathrm{D}\varphi^A+|\mathrm{U}^A|^2\,,
\end{equation}
and the matter terms $\widetilde{\cal L}_{\text{matter}}$ stemming from the antifundamental multiplet are given by the same Lagrangian, where appropriate tildes are added to the component fields of the chiral multiplet. Finally, the mass terms are given by
\begin{align}
	{\cal L}_{\text{mass terms}}= -|m_A\varphi_A|^2 -|m_A\widetilde{\varphi}_A|^2 -\frac{\overline{m}_A}{2}\widetilde{\psi}_A\psi^A - \frac{m_A}{2} \overline{\psi}_A \overline{\widetilde{\psi}}^A\,.
\end{align}

\section{Alternative representations}\label{app:alt}

An alternative way to handle the broken cut (b) in Fig.~\ref{fig:cuts2L}, as discussed in Sec.~\ref{sec:2Lcuts}, is by modifying some of the optional identities we impose on the numerators.
The representation of the amplitude discussed here is attached as an ancillary file, \texttt{anc\_neq1\_alt.dat}, to the arXiv submission.
We start by deforming the decomposition identity of the $\cN=2$ bow-tie diagrams:
\begin{equation}
  \begin{aligned}
    \tikzsetnextfilename{decompBowTie1Again}
  	\gTriTri[scale=0.8,all=gluon,eLA=1,eLB=2,eLC=3,eLD=4,iA=quark,iB=quark,iC=quark,iE=quark,iF=quark,iG=quark,iLC=$\overset{\swarrow}{\ell_1}$,iLG=$\overset{\searrow}{\ell_2}$]{
      \node at (1.1,1.3) {$\cN=2$};
	}
    &=
    \frac{1}{2}
    \tikzsetnextfilename{decompBowTie2Again}
    \gTriTri[scale=0.8,all=gluon,eLA=1,eLB=2,eLC=3,eLD=4,iA=gline,iB=gline,iC=gline,iE=gline,iF=gline,iG=gline,iLC=$\overset{\swarrow}{\ell_1}$,iLG=$\overset{\searrow}{\ell_2}$]{
      \node at (1.1,1.3) {$\cN=1$};
	}
    +\frac{1}{2}
    \tikzsetnextfilename{decompBowTie3Again}
    \gTriTri[scale=0.8,all=gluon,eLA=1,eLB=2,eLC=3,eLD=4,iA=gline,iB=gline,iC=gline,iE=bline,iF=bline,iG=bline,iLC=$\overset{\swarrow}{\ell_1}$,iLG=$\overset{\searrow}{\ell_2}$]{
      \node at (1.1,1.3) {$\cN=1$};
	}\\
    &\quad- z\, s^3 \left[\mu_{12} (\hat{\kappa}_{12}+\hat{\kappa}_{34})
    - i \epsilon(\mu_1,\mu_2) (\hat{\kappa}_{12}-\hat{\kappa}_{34})\right]
    \,,\\
    \tikzsetnextfilename{decompPureBowTie1Again}
    \gTriTri[scale=0.8,all=gluon,eLA=1,eLB=2,eLC=3,eLD=4,iLC=$\overset{\swarrow}{\ell_1}$,iLG=$\overset{\searrow}{\ell_2}$]{
      \node at (1.1,1.3) {$\cN=2$};
    }
    &=
    \tikzsetnextfilename{decompPureBowTie2Again}
    \gTriTri[scale=0.8,all=gluon,eLA=1,eLB=2,eLC=3,eLD=4,iLC=$\overset{\swarrow}{\ell_1}$,iLG=$\overset{\searrow}{\ell_2}$]{
      \node at (1.1,1.3) {$\cN=1$};
    }
    \tikzsetnextfilename{decompPureBowTie3Again}
    +\gTriTri[scale=0.8,all=gluon,eLA=1,eLB=2,eLC=3,eLD=4,iA=rline,iB=rline,iC=rline,iE=rline,iF=rline,iG=rline,iLC=$\overset{\swarrow}{\ell_1}$,iLG=$\overset{\searrow}{\ell_2}$]{
      \node at (1.1,1.3) {$\cN=1$};
    }
    +\tikzsetnextfilename{decompPureBowTie4Again}
    \gTriTri[scale=0.8,all=gluon,eLA=1,eLB=2,eLC=3,eLD=4,iA=rline,iB=rline,iC=rline,iLC=$\overset{\swarrow}{\ell_1}$,iLG=$\overset{\searrow}{\ell_2}$]{
      \node at (1.1,1.3) {$\cN=1$};
    }
    +\tikzsetnextfilename{decompPureBowTie5Again}
    \gTriTri[scale=0.8,all=gluon,eLA=1,eLB=2,eLC=3,eLD=4,iE=rline,iF=rline,iG=rline,iLC=$\overset{\swarrow}{\ell_1}$,iLG=$\overset{\searrow}{\ell_2}$]{
      \node at (1.1,1.3) {$\cN=1$};
    }\\
    \\
    &\quad- 2z\, s^3 \left[\mu_{12} (\hat{\kappa}_{12}+\hat{\kappa}_{34})
    - i \epsilon(\mu_1,\mu_2) (\hat{\kappa}_{12}-\hat{\kappa}_{34})\right]
    \,,
  \end{aligned}
\end{equation}
parametrized by the deformation variable~$z$.
The expression for the broken cut Eq.~\eqref{eq:brokenCut} then gets deformed by
\begin{equation}
  \begin{aligned}
  \tikzsetnextfilename{ntmefhhc1Again}
  \begin{tikzpicture}
    [line width=1pt,
    baseline={([yshift=-0.5ex]current bounding box.center)},
    scale=1,
    rotate=0,
    font=\scriptsize]
    \fill[blob] (0.9,0.6) ellipse (0.2 and 0.4);
    \draw[gluon] (0.3,0.3) -- (0,0) node[left] {1};
    \draw[gluon] (0.3,0.9) -- (0,1.2) node[left] {2};
    \draw[gluon] (1.5,0.9) -- (1.8,1.2) node[right] {3};
    \draw[gluon] (1.5,0.3) -- (1.8,0) node[right] {4};
    \draw[rline] (0.3,0.3) -- (0.3,0.9);
    \draw[rline] (0.3,0.9) -- (0.75,0.9);
    \draw[rline] (1.05,0.9) -- (1.5,0.9);
    \draw[rline] (1.5,0.9) -- (1.5,0.3);
    \draw[rline] (1.5,0.3) -- node[below] {$\overset{\rightarrow}{\ell_2}$} (1.05,0.3);
    \draw[rline] (0.75,0.3) -- node[below] {$\overset{\leftarrow}{\ell_1}$} (0.3,0.3);
  \end{tikzpicture}
  &=
  \tikzsetnextfilename{ntmefhhc2Again}
  \left.\frac{1}{2\ell_1\cdot\ell_2-2m^2}\gBoxBox[all=gluon,iA=rline,iB=rline,iC=rline,iD=rline,iE=rline,iF=rline,iLF=$\overset{\leftarrow}{\ell_1}$,iLE=$\overset{\rightarrow}{\ell_2}$,eLA=1,eLB=2,eLC=3,eLD=4]{}\right|_{\textrm{cut}}
  \tikzsetnextfilename{ntmefhhc3Again}
  +\left.\frac{1}{s}\gTriTri[scale=0.9,all=gluon,iA=rline,iB=rline,iC=rline,iE=rline,iF=rline,iG=rline,iLC=$\overset{\swarrow}{\ell_1}$,iLG=$\overset{\searrow}{\ell_2}$,eLA=1,eLB=2,eLC=3,eLD=4]{}\right|_{\textrm{cut}}\\
  &\quad-2(1-z)s^2m^2(\hat\kappa_{12}+\hat\kappa_{34})
  \,,
  \end{aligned}
\end{equation}
where 
\begin{equation}
  \begin{aligned}
    \tikzsetnextfilename{ntmefhhcres1Again}
  \left.\gBoxBox[all=gluon,iA=rline,iB=rline,iC=rline,iD=rline,iE=rline,iF=rline,iLF=$\overset{\leftarrow}{\ell_1}$,iLE=$\overset{\rightarrow}{\ell_2}$,eLA=1,eLB=2,eLC=3,eLD=4]{}\right|_{\textrm{cut}}&=s^2(s \mu_{12}+2\mu_{11}\mu_{22})(\hat\kappa_{12}+\hat\kappa_{34}) + \dots\,,\\
  \tikzsetnextfilename{ntmefhhcres2Again}
  \left.\gTriTri[scale=0.9,all=gluon,iA=rline,iB=rline,iC=rline,iE=rline,iF=rline,iG=rline,iLC=$\overset{\swarrow}{\ell_1}$,iLG=$\overset{\searrow}{\ell_2}$,eLA=1,eLB=2,eLC=3,eLD=4]{}\right|_{\textrm{cut}}&=-2(1-z)s^3\mu_{12}(\hat\kappa_{12}+\hat\kappa_{34}) + \dots\,.
  \end{aligned}
\end{equation}
Note that we set $\epsilon(\mu_1,\mu_2)=0$ on the cut kinematics.
Simultaneously also the subloop flavor symmetry equation gets deformed in a similar way
\begin{equation}
  \tikzsetnextfilename{slfs1Again}
  \gTriTri[scale=0.8,all=gluon,eLA=1,eLB=2,eLC=3,eLD=4,iA=rline,iB=rline,iC=rline,iE=rline,iF=rline,iG=rline,iLC=$\overset{\swarrow}{\ell_1}$,iLG=$\overset{\searrow}{\ell_2}$]{}
  =
  \tikzsetnextfilename{slfs2Again}
  \gTriTri[scale=0.8,all=gluon,eLA=1,eLB=2,eLC=3,eLD=4,iA=rline,iB=rline,iC=rline,iE=gline,iF=gline,iG=gline,iLC=$\overset{\swarrow}{\ell_1}$,iLG=$\overset{\searrow}{\ell_2}$]{}
  - 2(1-z)s^3\left[\mu_{12}(\hat{\kappa}_{12}+\hat{\kappa}_{34})+i\epsilon(\mu_1,\mu_2)(\hat{\kappa}_{12}-\hat{\kappa}_{34})\right]\,.
\end{equation}
Hence, both of the former equalities are fixed by choosing $z=1$.
On the other side this will not only invalidate the decomposition identity above, but also one of the optional three-term identities:
\begin{equation}
  \begin{aligned}
  \tikzsetnextfilename{threeTermMatterProblem1again}
  \gTriTri[scale=0.9,all=gluon,iA=rline,iB=rline,iC=rline,iE=rline,iF=rline,iG=rline,eLA=$1$,eLB=$2$,eLC=$3$,eLD=$4$,iLC=$\overset{\swarrow}{\ell_1}$,iLG=$\overset{\searrow}{\ell_2}$]{}
  &=
  \tikzsetnextfilename{threeTermMatterProblem2again}
  \gBoxBox[all=gluon,iA=rline,iB=rline,iC=rline,iD=rline,iE=rline,iF=rline,eLA=$1$,eLB=$2$,eLC=$3$,eLD=$4$,iLF=$\overset{\leftarrow}{\ell_1}$,iLE=$\overset{\rightarrow}{\ell_2}$]{}
  \tikzsetnextfilename{threeTermMatterProblem3again}
  - \gBoxBox[all=gluon,iA=rline,iB=rline,iC=rline,iD=rline,iE=rline,iF=rline,eLA=$1$,eLB=$2$,eLC=$4$,eLD=$3$,iLF=$\overset{\leftarrow}{\ell_1}$,iLC=$\underset{\rightarrow}{\ell_2}$]{}\\
  &\quad+ 2z s^3 \left[\mu_{12}(\hat{\kappa}_{12}+\hat{\kappa}_{34})-i\epsilon(\mu_1,\mu_2)(\hat{\kappa}_{12}-\hat{\kappa}_{34})\right]\,.
  \end{aligned}
\end{equation}

It is fair to interpret the leftover term with a new all-matter bow-tie diagram:
\begin{equation}
  \tikzsetnextfilename{ambtAgain}
  \gTriTri[scale=0.9,all=rline,eA=gluon,eB=gluon,eC=gluon,eD=gluon,iD=line,iLC=$\overset{\swarrow}{\ell_1}$,iLG=$\overset{\searrow}{\ell_2}$,eLA=$1$,eLB=$2$,eLC=$3$,eLD=$4$]{}
  =2z s^3 \left[\mu_{12}(\hat{\kappa}_{12}+\hat{\kappa}_{34})-i\epsilon(\mu_1,\mu_2)(\hat{\kappa}_{12}-\hat{\kappa}_{34})\right]\,.
\end{equation}
where the middle line represent a supersymmetric version of a scalar which appears upon dimensionally reducing a 6D vector.
This additional numerator does not talk to any diagram via color-algebra relations or other identities.
Including this new contribution also into the broken cut above would fix it (independently of the value of $z$).
Note that this new diagram integrates to zero.

Furthermore, we added a fudge factor~$(1-y)$ controlling the addition of the deformation terms introduced in Eq.~\ref{eq:corrections},
which themselves contain a single free parameter~$x$.
Setting $z=y=x=0$ one obtains our preferred representation (see Sec.~\ref{sec:2Lcuts} and Sec.~\ref{sec:nums}), whereas choosing $z=y=1$ leads to a representation fulfilling all massive cuts.

\bibliographystyle{JHEP}
\bibliography{references}

\end{document}